\documentclass[aps,prb,twocolumn,showpacs,10pt,nofootinbib,longbibliography]{revtex4-2}
\usepackage{graphicx}
\usepackage{amssymb}
\usepackage{amsmath,color,mathrsfs}
\pdfoutput=1
\usepackage{hyperref}

\usepackage{enumitem}

\usepackage[table]{xcolor}

\usepackage{booktabs}

\usepackage[english]{babel}%
\usepackage[T1]{fontenc}%

\usepackage{cancel}

\graphicspath{{./fig/}}

\newcommand{\be}{\begin{equation}}
\newcommand{\ee}{\end{equation}}

\newcommand{\ps}[1]{ {#1} }

\begin{document}
\makeatother

\title{Dynamical nuclear spin polarization in a quantum dot with an electron spin driven by electric dipole spin resonance}

\author{Peter Stano$^{1,2}$}
\author{Takashi Nakajima$^{1}$}
\author{Akito Noiri$^{1}$}
\author{Seigo Tarucha$^{1,3}$}
\author{Daniel Loss$^{1,3,4}$}

\affiliation{\textsuperscript{1}Center for Emergent Matter Science, RIKEN, 2-1 Hirosawa, Wako-shi, Saitama 351-0198, Japan}
\affiliation{\textsuperscript{2}Institute of Physics, Slovak Academy of Sciences, 845 11 Bratislava, Slovakia}
\affiliation{\textsuperscript{3}RIKEN Center for Quantum Computing, 2-1 Hirosawa, Wako, Saitama, 351-0198 Japan}
\affiliation{\textsuperscript{4}Department of Physics, University of Basel, Klingelbergstrasse 82, CH-4056 Basel, Switzerland}

\date{\today}

\begin{abstract}

We analyze the polarization of nuclear spins in a quantum dot induced by a single-electron spin that is electrically driven to perform coherent Rabi oscillations. We derive the associated nuclear-spin polarization rate and analyze its dependence on the accessible control parameters, especially the detuning of the driving frequency from the electron Larmor frequency. The arising nuclear-spin polarization is related to the Hartmann-Hahn effect known from the NMR literature with two important differences. First, in quantum dots one typically uses a micromagnet, leading to a small deflection of the quantization axes of the electron and nuclear spins. Second, the electric driving wiggles the electron with respect to the atomic lattice. The two effects, absent in the traditional Hartmann-Hahn scenario, give rise to two mechanisms of nuclear-spin polarization in gated quantum dots.
The arising nuclear-spin polarization is a resonance phenomenon, achieving maximal efficiency at the resonance of the electron Rabi and nuclear Larmor frequency (typically a few or a few tens of MHz). As a function of the driving frequency, the polarization rate can develop sharp peaks and reach large values at them. Since the nuclear polarization is experimentally detected as changes of the electron Larmor frequency, we often convert the former to the latter in our formulas and figures. In these units, the polarization can reach hundreds of MHz/s in GaAs quantum dots and at least tens of kHz/s in Si quantum dots. 
We analyze possibilities to exploit the resonant polarization effects for achieving large nuclear polarization and for stabilizing the Overhauser field through feedback.

\end{abstract}

\maketitle

\section{Introduction}

Spin qubits in semiconducting quantum dots \cite{loss_quantum_1998} are pursued as promising qubit hosts \cite{burkard_semiconductor_2021,stano_review_2022}. 
The advantage of semiconducting spin qubits is that they can be controlled electrically. 
For example, single-qubit gates exploit electric dipole spin resonance (EDSR) where an oscillating electric field drives spin rotations through either the material spin-orbit interaction 
\cite{golovach_electric-dipole-induced_2006,nowack_coherent_2007}, or a designed micromagnet \cite{tokura_coherent_2006,pioro-ladriere_electrically_2008}. The manipulation thus proceeds by applying a resonant radio-frequency (rf) field through local gates instead of a global field typical for electron spin resonance (ESR) experiments.

Since the first experiments with gated spin qubits, it has been routinely observed that some form of nuclear spin polarization often accompanies electrical manipulation \cite{ono_nuclear-spin-induced_2004,koppens_control_2005} including a resonant driving \cite{koppens_driven_2006, nowack_coherent_2007}.
Since in semiconductors nuclear spins have a strong impact on spin qubits, limiting their lifetime and coherence \cite{khaetskii_electron_2002,merkulov_electron_2002,coish_hyperfine_2004,petta_coherent_2005,johnson_tripletsinglet_2005,chesi_dephasing_2016},
a lot of research went into understanding the electron-nuclear spin interactions 
(see, for example, the reviews in Refs.~\cite{schliemann_electron_2003,hanson_spins_2007,coish_nuclear_2009,chekhovich_nuclear_2013} and the references therein). 
A possible control of nuclear spins through the arising nuclear polarization received particular attention
\cite{coish_hyperfine_2004,klauser_nuclear_2006,danon_nuclear_2008,klauser_nuclear_2008,rudner_nuclear_2011,gullans_preparation_2013}.
While a certain degree of control was demonstrated \cite{foletti_universal_2009,vink_locking_2009,bluhm_enhancing_2010}, 
overall it remained limited \cite{petta_dynamic_2008,rudner_phase_2010,barthel_relaxation_2012,nichol_quenching_2015} and the large nuclear noise persists as the major issue of III-V materials to be dealt with \cite{nadj-perge_disentangling_2010,shulman_suppressing_2014,malinowski_notch_2016,nakajima_coherence_2020}.\footnote{Nuclear spins are one of the main reasons to switch from element--III-V to element-IV quantum dots. However, even here nuclear spins might still remain as a performance limit of spin qubits made with electrons \cite{camenzind_spin_2022, yoneda_noise-correlation_2022} as well as holes \cite{bosco_fully_2021}.}\footnote{Since our focus is on gated semiconducting dots manipulated electrically, we will make only sporadic comments to works on self-assembled dots accessed optically, where the nuclear-spin-control program continues unabated \cite{gangloff_quantum_2019,bodey_optical_2019,gangloff_witnessing_2021,jackson_quantum_2021,jackson_optimal_2022}.}

We revisit here the nuclear spin polarization induced by an EDSR-driven and coherently precessing electron spin in an isolated quantum dot. 
We consider the coherent regime with the electron spin Rabi frequency large compared to the relevant decay times, either the electron lifetime in the dot or its spin Rabi decay time. This regime of well-defined Rabi rotations (or strong driving) is the essential difference to previous works on this topic \cite{rudner_electrically_2007,danon_nuclear_2008,danon_multiple_2009} which implicitly or explicitly considered the limit of weak driving.\footnote{See especially Footnote [18] in Ref.~\cite{rudner_electrically_2007}} 

The physics' essence is closely related to the Hartmann-Hahn resonance \cite{hartmann_nuclear_1962}, well known from nuclear magnetic resonance (NMR): dynamical nuclear spin polarization (DNSP)\footnote{In line with the literature on spin qubits, we will use the name `dynamical nuclear spin polarization' (DNSP) rather than the `dynamical nuclear polarization' (DNP) used in the NMR community.} arises when the Rabi frequency of the driven electron is equal to the Larmor frequency of nuclear spins.\footnote{Since the effect exists in several flavors, it might be useful to mention further names that are used: The original work, Ref.~\cite{hartmann_nuclear_1962}, considered two different nuclear species, both of which are driven. The spin `cross-polarization' then arises when their Rabi frequencies are equal, a condition called also `double resonance' \cite{pines_protonenhanced_1973}. Ref.~\cite{henstra_nuclear_1988} coined the acronym `NOVEL' for the variant where one of the spins is electronic, being driven, and the other is nuclear, not driven. This is the situation we consider in a quantum dot.} However, important differences preclude using the existing NMR results: (1) since the electron-spin driving is electrical, the electron shifts in space with respect to the atomic lattice; (2) there is often a micromagnet gradient giving dispersion to nuclear Larmor frequencies and spin quantization axes; and (3) unlike the dipole-dipole interaction relevant in NMR, the electron-hyperfine spin-spin interaction is isotropic, preserving the total spin. As an illustration of the difference, if
the electron is driven purely magnetically (ESR) and the spin quantization axes of all nuclei and as well as the electron are collinear, the DNSP effects that we describe would not be present.\footnote{However, we reason that such a highly idealized situation does not describe realistic experiments even if they do not employ micromagnets. The DNSP arises, and our formulas apply also in this scenario; see Sec.~\ref{sec:extensions} for details.}
 
\begin{figure}
\begin{center}
\includegraphics[width=\linewidth]{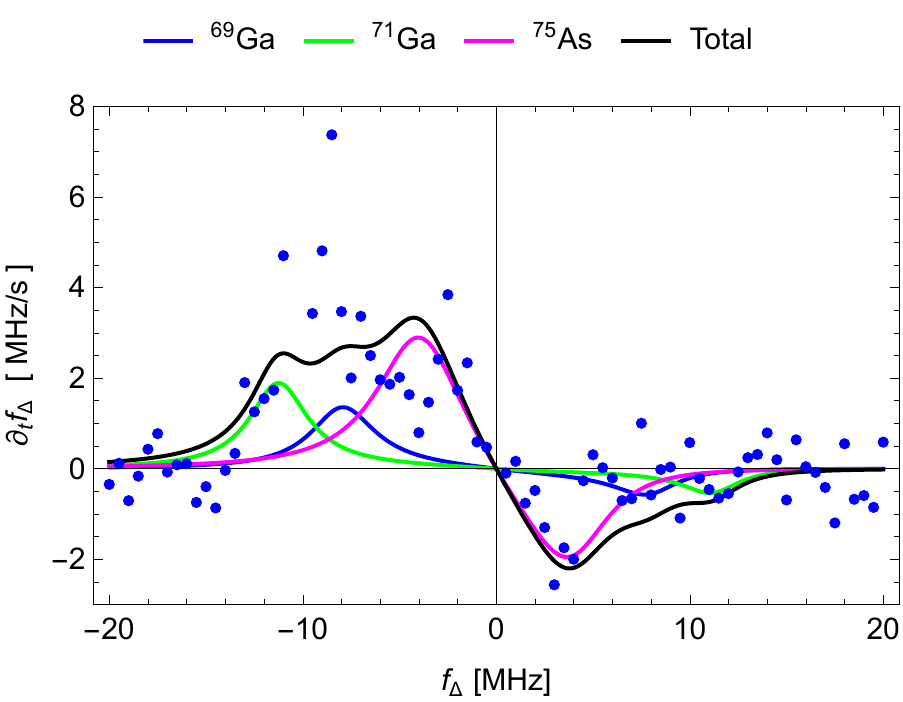}
\end{center}

\caption{\textbf{DNSP in a GaAs quantum dot.}
The measured data (points) show the polarization rate observed in an EDSR-driven single-electron quantum dot. The three color curves plot Eq.~\eqref{eq:mainResult} for the three isotopes of GaAs as given in the plot legend. The following parameters were used in the evaluation of the theory expressions: external field $B=1$ T, Rabi frequency at resonance 6.5 MHz, pulse time $T_\mathrm{pulse}=1$ $\mu$s, cycle time $T_\mathrm{cycle} = 20$ $\mu$s, dot displacement $d=0.5$ nm, dot in-plane size $l=34$ nm, dot out-of-plane size $l_z=10$ nm, longitudinal magnetic field gradient $\nabla_{||}B=1$ T/$\mu$m, and transverse magnetic field gradient $\nabla_\perp B=0.3$ T/$\mu$m. Finally, we used energy density $G_\Sigma$ with a Lorenzian profile and included an additional smearing of $2\pi \times 250$ kHz according to the discussion in Sec.~\ref{sec:extensions}. For better comparison to the data, the theoretically calculated rates were multiplied by 1/2.
\label{fig:data-fit}
}
\end{figure}

The paper focuses on a detailed derivation of the DNSP rate, but it also contains measured data on it (Fig.~\ref{fig:data-fit}). The derivation is presented in Secs.~\ref{sec:definitions}--\ref{sec:rate} with auxiliaries delegated to Appendixes~\ref{app:nuclearDensity}--\ref{app:tables}. The main result is the polarization rate given in Eq.~\eqref{eq:rateDNSP}. It is derived for a generic material (we present theory plots for GaAs and Si), the electron spin 1/2,\footnote{The formula covers also the case of a hole spin, if the hole-spin--nuclear-spin interaction tensor is known. We discuss the hole-spin scenario in Appendix~\ref{app:hole}.} and nuclear spins of arbitrary magnitude and isotopic composition. The effects that we describe here are resonance phenomena and very sharp resonance peaks result in the theory if applied naively. When fitting experimental results, one needs to account for additional `smearing' effects as discussed in Sec.~\ref{sec:extensions}. In Sec.~\ref{sec:feedback}, we analyze the dependence of the polarization rate on the detuning from the resonance to implement feedback to control the nuclei, similarly to previous works along this line \cite{rudner_electrically_2007, danon_nuclear_2008,danon_multiple_2009,vink_locking_2009,bluhm_enhancing_2010,tenberg_narrowing_2015}.

We uncover two mechanisms of the DNSP: the first is due to the electron spatial displacement due to the electric field, the second due to the misalignment of the quantization axes for the electron and the nuclei induced by the micromagnet magnetic-field gradient. 
The two mechanisms coexist and interfere, making the polarization-rate dependence on parameters involved. Nevertheless, in GaAs with the Hartmann-Hahn resonance condition fulfilled, the polarization rate can reach hundreds of MHz per second (we convert the nuclear polarization to the change of the electron precession frequency due to the induced Overhauser field). 

An important question is whether one expects sizable DNSP in natural Si. While the rates are orders of magnitude smaller than in GaAs, the effect might be observable because of longer spin coherence times in Si. We estimate that the rates can reach tens of kHz/s, and even more in smaller dots. On the other hand, our estimates given in Sec.~\ref{sec:restoringForce} suggest that, unlike in GaAs, the arising DNSP does not appreciably affect gate fidelities in Si. 

Concerning the experiment, the measurements were performed by driving a single electron spin in a double dot GaAs sample with a micromagnet using the Pauli spin blockade as the spin detection. While we find the qualitative correspondence to the theory satisfactory, the measured data are noisy and do not show clear resonance peaks. We believe that this is because of strong feedback: the polarized nuclear spins change the DNSP rate by changing the EDSR resonance frequency. It is only through compensating for this effect in the experiment (that is, readjusting the driving frequency to the actual value of the hyperfine field) that polarization rates could have been measured. The compensation precision is limited and, therefore, the correspondence of the theory and measurements is only qualitative concerning the shape of the curve for the DNSP rate. On the other hand, the magnitude of the observed rate aligns with the theory almost without any fitting, using the material constants and parameters of the dot obtained independently.

\newcommand{\gFactor}{g}
\newcommand{\bField}{\mathbf{B}}
\newcommand{\axisEnergy}{\mathbf{z}}
\newcommand{\frequency}{f}
\newcommand{\angularFrequency}{\omega}
\newcommand{\frequencyLarmor}{\frequency}

\newcommand{\spinElectron}{s}
\newcommand{\spinElectronVector}{\mathbf{\spinElectron}}
\newcommand{\gFactorElectron}{\gFactor_e}
\newcommand{\bFieldElectron}{\bField_e}
\newcommand{\axisEnergyElectron}{\axisEnergy_e}
\newcommand{\axisEnergyElectronX}{{\mathbf{x}_e}}
\newcommand{\axisEnergyElectronY}{{\mathbf{y}_e}}
\newcommand{\axisEnergyElectronO}{{\mathbf{o}_e}}
\newcommand{\angularFrequencyElectron}{\angularFrequency_e}
\newcommand{\frequencyLarmorElectron}{\frequencyLarmor_e}

\newcommand{\spinNucleus}{I}
\newcommand{\spinNucleusVector}{\mathbf{\spinNucleus}}
\newcommand{\gFactorNucleus}{\gFactor_n}
\newcommand{\bFieldNucleus}{\bField_n}
\newcommand{\axisEnergyNucleus}{\axisEnergy_n}
\newcommand{\angularFrequencyNucleus}{\angularFrequency_n}
\newcommand{\frequencyLarmorNucleus}{\frequencyLarmor_n}
\newcommand{\fractionIsotope}{\phi_i}
\newcommand{\axisQuadrupolar}{q}
\newcommand{\axisQuadrupolarVector}{\mathbf{\axisQuadrupolar}}
\newcommand{\angularFrequencyQuadrupolar}{\angularFrequency_Q}

\newcommand{\rf}{{\mathrm{rf}}}
\newcommand{\dotShift}{\mathbf{d}}
\newcommand{\axisBFieldRF}{\mathbf{b}}
\newcommand{\angularFrequencyRF}{\angularFrequency_\rf}
\newcommand{\frequencyRF}{\frequency_\rf}
\newcommand{\angleRF}{\phi_\rf}
\newcommand{\angularFrequencyRabi}{\angularFrequency_R}
\newcommand{\frequencyRabi}{\frequency_R}
\newcommand{\angularFrequencyRabiResonant}{\angularFrequency_{RR}}
\newcommand{\frequencyRabiResonant}{\frequency_{RR}}
\newcommand{\frequencyDetuning}{\frequency_{\Delta}}
\newcommand{\angularFrequencyDetuning}{\angularFrequency_{\Delta}}
\newcommand{\angleDetuning}{\gamma}

\newcommand{\angularFrequencyRabiHH}{{\angularFrequency_R^\mathrm{hh}}}

\section{Electron spin coupled to ensemble of nuclear spins}

\label{sec:definitions}

We consider an electron confined in a quantum dot interacting with nuclear spins of the atoms of the semiconducting host. We now list the elements of the problem.

\subsection{Quantum dot}

On top of a homogeneous field of a solenoid coil, a micromagnet fabricated nearby the quantum dot adds an inhomogeneous component, together resulting in a spatially dependent magnetic field $\bField(x,y,z)$. For the DNSP rates studied here, one can neglect the spin-orbit effects (both from the intrinsic spin-orbit interactions and from the magnetic field inhomogeneity) on the electron wave function and take it as separable to the spin part and the orbital part. We take the latter as
\be
\Psi(\mathbf{r},z) =\frac{1}{\sqrt{\pi} l} \exp[-(\mathbf{r}-\mathbf{r}_0)^2/2l^2] \psi(z).
\label{eq:wavefunction}
\ee
The Gaussian form in the in-plane (the 2DEG plane) coordinates $(x,y) \equiv \mathbf{r}$ corresponds to harmonic confinement with the scale $l$ and the minimum at $\mathbf{r}_0$. Together with the effective mass $m$, the length scale $l$ defines the in-plane orbital energy $\hbar^2/ml^2$. The wave-function profile along the coordinate $z$ (out-of-plane), $\psi(z)$, will not be important and is left unspecified except of assigning it a corresponding length scale $l_z$. With that, we define the quantum dot effective volume $V_D=2\pi l^2 l_z$ and the effective number of nuclei within the quantum dot $N_\mathrm{tot} = V_D / v_0$ (see Appendix~\ref{app:nuclearDensity} for the definition of $V_D$ and the definition motivation). Here $v_0 = a_0^3 / 8$ is the volume per atom in a zinc-blende or diamond lattice. $N_\mathrm{tot}$ counts all atomic nuclei, irrespective of their spin. The spin depends on the isotope. Introducing the isotope fractions $\fractionIsotope$, the number of atoms of isotope $i$ in the dot is $N_i = \fractionIsotope N_{\mathrm{tot}}$. The total number of spin-carrying nuclei is large, up to a million in a typical GaAs gated dot and ten thousand in a Si dot with natural isotopic concentrations. 

\subsection{Nuclei}

Concerning atoms, we need to distinguish different isotopes as they differ in their nuclear-spin characteristics. We use the following notation. The atoms within the quantum dot are indexed by subscript $n$. When the individual position of the nucleus is not relevant, we trade the individual index $n$ for the isotope index $i$. (The latter is a function of the former, $i=i(n)$, but we omit the argument for notational clarity.) In GaAs $i \in \{^{69}\mathrm{Ga},^{71}\mathrm{Ga},^{75}\mathrm{As}\}$, while $n$ is an integer going from one to about a million. A quantity $X$ specified for a given nucleus then reads $X_n$ or $X_i$.
For notational clarity, we sometimes omit the nuclear index on the spin operator entirely, $\spinNucleusVector_n$ or $\spinNucleusVector_i$ $\to$ $\spinNucleusVector$. 
There are also quantities that are defined only with the isotope index $i$, for example, the material isotopic fractions $\fractionIsotope$. 

The nuclear spin is coupled to the magnetic field through the Zeeman term,
\be
H_n^Z = -\gFactorNucleus \mu_N \bFieldNucleus \cdot \spinNucleusVector_n.
\label{eq:ZeemanNucleus}
\ee
Here, $\gFactorNucleus$ is the nuclear $g$ factor, $\mu_N$ is the nuclear magneton, $\spinNucleus_n$ is the nuclear spin magnitude (not necessarily 1/2), and $\spinNucleusVector_n$ is the vector of nuclear spin operators. Among these, the $g$ factor and spin magnitude depend only on the isotope, so that the atom index $n$ could be traded for the isotope index $i$. Importantly, the magnetic field $\bFieldNucleus = \bField(x_n,y_n,z_n)$ depends on the location of the atom because of the micromagnet induced gradients. They are parameterized by $\nabla \bField$,  a second-rank tensor defined by $(\nabla \bField)_{ij} = \nabla_i B_j$. While the gradients are small, $l |\nabla \bField| \ll B$, taking them into account is crucial for one of the DNSP mechanisms. Finally, we define the unit vector $\axisEnergyNucleus$ pointing along $\mathrm{sgn}(\gFactorNucleus) \bFieldNucleus$, being the direction of the nuclear spin in the ground state of $H_n^Z$. With that, we rewrite Eq.~\eqref{eq:ZeemanNucleus} as
\be
H_n^Z = - \hbar \angularFrequencyNucleus \spinNucleusVector_n \cdot \axisEnergyNucleus,
\label{eq:ZeemanNucleus2}
\ee
where the angular Larmor frequency $\angularFrequencyNucleus$ is positive independently on the sign of the $g$ factor, a form that will be useful in the derivations below.

\subsection{Electron and its hyperfine interaction with nuclei}

The DNSP arises due to a coupling of the electron and nuclear spins. It takes the form of the Fermi-contact, or hyperfine, interaction,
\be
H_\mathrm{hf} = \sum_n A_n v_0 |\Psi(\mathbf{r}_n,z_n)|^2  \spinNucleusVector_n \cdot \spinElectronVector.
\label{eq:hyperfine}
\ee
Here, $A_n$ is an isotope-dependent constant, and $\spinElectronVector$ is the vector of electron spin operators. We consider a spin one-half, $\spinElectron=1/2$, and use the spin operator $\spinElectronVector = \boldsymbol{\sigma}/2$ with $\boldsymbol{\sigma}$ the Pauli sigma matrices. Once the electron orbital degrees of freedom have been separated and specified by Eq.~\eqref{eq:wavefunction}, the spin is the remaining degree of freedom. It is described by the Hamiltonian \be
H_e^Z = \gFactorElectron \mu_B \bFieldElectron \cdot \spinElectronVector,
\label{eq:ZeemanElectron}
\ee
where $\gFactorElectron$ is the $g$ factor and $\mu_B$ is the Bohr magneton. Equation \eqref{eq:ZeemanElectron} is the analog of Eq.~\eqref{eq:ZeemanNucleus} (the overall sign is opposite due to the opposite electric charge), but there are differences concerning the field $\bFieldElectron$. Namely, in the lowest approximation that we adopted by Eq~\eqref{eq:wavefunction}, it is a sum of two contributions. The first is the spatial average of the magnetic field within the quantum dot,  
\be
\langle \bField \rangle = \int |\Psi(\mathbf{r},z)|^2 \bField(\mathbf{r},z) \,\mathrm{d}\mathbf{r} \,\mathrm{d} z.
\label{eq:bFieldAverage}
\ee
The second is the statistical average of Eq.~\eqref{eq:hyperfine}, the Overhauser field, which we specify introducing polarizations $p_n$,
\be
\langle \spinNucleusVector_n \rangle = p_n \spinNucleus_n \axisEnergyNucleus.
\label{eq:polarization}
\ee 
To make progress, we adopt further approximations. The goal of this paper is to calculate the nuclear spin polarization $p_n$, or its rate of change, the DNSP rate. However, we are not interested in polarizations of individual atoms, which are not observable anyway, but rather in their collective effect on the electron spin. Therefore, we assign all atoms of a given isotope the same polarization 
\be
p_n \to p_i,
\label{eq:reductionToIsotopes}
\ee
drastically reducing the set of unknowns. Compared to this approximation, in reality the nuclei in the center of the dot will be polarized more and on the outskirts less. In the derivations below, we repeatedly average over the nuclei (or over the dot coordinates) in this spirit. The second approximation is to neglect the deflection of the Overhauser field from the average external field concerning the electron Zeeman energy. This deflection is a higher-order effect (in the magnetic field gradients) and neglecting it is in line with using Eq.~\eqref{eq:wavefunction}. We thus write the electron Zeeman term 
\be
H_e^Z = - \hbar \angularFrequencyElectron \spinElectronVector \cdot \axisEnergyElectron,
\label{eq:ZeemanElectron2}
\ee
with $\axisEnergyElectron$ a unit vector along $-\mathrm{sgn}(\gFactorElectron) \langle \bField \rangle$ and the positive Zeeman energy
\be
\hbar \angularFrequencyElectron = |\gFactorElectron \mu_B \langle \bField \rangle | + \sum_i \mathrm{sgn} (\gFactorElectron) p_i \fractionIsotope I_i |A_i|.
\label{eq:frequencyElectron}
\ee
To arrive at this form, we assumed that the polarizations are small, so that the magnitude of the first term is bigger than the second (see Appendix~\ref{app:conversion} for the derivation).

\subsection{EDSR}

The last basic element is the EDSR driving. Applying an oscillating electric field $\mathbf{E}(t) = \mathbf{E}_0 \cos(\angularFrequencyRF t -\angleRF)$ drives the electron in space. The micromagnet-field gradients result in an effective oscillating magnetic field. Since the driving frequency is small compared to the electron orbital confinement energy, $\hbar \angularFrequencyRF \ll \hbar^2/ml^2$, the drive is adiabatic with respect to the electron orbital degrees of freedom and results in a time-dependent displacement of the dot center $\mathbf{r}_0$ by
\be
\dotShift(t) = \frac{e \mathbf{E}(t) l^2}{\hbar^2/ml^2}.
\label{eq:dotShift}
\ee 
The EDSR drive can thus be taken into account by using Eq.~\eqref{eq:wavefunction} with a time-dependent center, $\mathbf{r}_0 \to \mathbf{r}_0 +\mathbf{d}(t)$, in Eqs.~\eqref{eq:hyperfine} and \eqref{eq:bFieldAverage}. The replacement in %
Eq.~\eqref{eq:hyperfine}  will lead to one of the DNSP mechanisms (as we will see below), while in Eq.~\eqref{eq:bFieldAverage}, it gives an effective oscillating magnetic field
\be
\bField_\rf(t) = (\dotShift(t) \cdot \nabla_{\mathbf{r}_0}) \langle \bField \rangle.
\ee
The component of $\bField_\rf(t)$ perpendicular to the average field $\bFieldElectron$  is denoted as
\be
-\gFactorElectron \mu_B \left[ \bField_\rf(t) \right]_{\perp} \equiv - 2\hbar \angularFrequencyRabiResonant \axisBFieldRF \cos(\angularFrequencyRF t - \phi_\rf).
\label{eq:EDSRTerm}
\ee
The equation defines the unit vector $\axisBFieldRF$ and the Rabi angular frequency at resonance $\angularFrequencyRabiResonant$. The Rabi oscillations of the electron due to this term, induced by the electric field, are called EDSR.

All quantities that were defined in this section and will be used in the following are collected for reference in %
Table~\ref{tab:tables}
in Appendix~\ref{app:tables}.

\section{Electron-nuclear spin pair}

We consider DNSP arising in the following repeated experiment. The electron spin is initialized to the ground state of $H_Z^e$ (using the electron reservoir, not nuclei), and then EDSR driven for a fixed time, of order microseconds, at a fixed detuning $\angularFrequencyDetuning = \angularFrequencyRF - \angularFrequencyElectron$ of order tens of $2\pi$ $\times$ MHz. Reference~\cite{noiri_unpublished_nodate} gives a detailed description of these steps and their implementation.\footnote{The regularity of re-initalization of the electron spin is crucial for auto-focusing in experiments such as Ref.~\cite{markmann_universal_2019}. On the other hand, assuming random re-initialization times was important for the description in Ref.~\cite{merkulov_long-term_2010}. In our model, the (ir)regularity of the moments at which the electron spin is initialized is irrelevant (although it matters into what state the electron spin is initialized): The DNSP is happening continuously during the electron Rabi precession.}

We derive the polarizations $p_i$ and the corresponding rates 
\be
\Gamma_i = \partial_t p_i,
\label{eq:rateDefinition}
\ee
proceeding in two steps: First, we consider an isolated nucleus $n$, of the isotope $i$, in contact with a driven electron. We solve for its dynamics. Second, we average the arising polarization rate over the dot, in line with Eq.~\eqref{eq:reductionToIsotopes}. Considering the nuclei polarization rates as independent is a good approximation as long as only a small fraction of the electron spin is transferred to the nuclear ensemble over one experimental cycle (after which the electron is reinitialized).\footnote{Reference~\cite{henstra_theory_2008} went beyond the approximation of independent rates and considered the electron spin being dissipated into the nuclear ensemble as a whole.} This condition is well fulfilled in all our numerical examples and plots.

The restriction to a single nucleus allows us to simplify the notation. We introduce a shorthand notation for the hyperfine coupling (the Knight field) as
\be
J_n(t) = A_n v_0 |\Psi_n(t)|^2,
\label{eq:J}
\ee
where we denoted the time dependence explicitly. 
The Hamiltonian for the electron-nuclear pair is
\be
\begin{split}
H = &
-\hbar\angularFrequencyNucleus \spinNucleusVector \cdot \axisEnergyNucleus
-\hbar\angularFrequencyElectron \spinElectronVector \cdot \axisEnergyElectron\\
&- 2\hbar\angularFrequencyRabiResonant   \spinElectronVector \cdot \axisBFieldRF\cos(\angularFrequencyRF t - \phi_\rf)
+J_n(t) \delta \spinNucleusVector \cdot \spinElectronVector.
\end{split}
\label{eq:electronNuclearPairH}
\ee
The first two terms are the Zeeman energies, Eqs.~\eqref{eq:ZeemanNucleus} and \eqref{eq:ZeemanElectron}, the third is the EDSR-driving term, Eq.~\eqref{eq:EDSRTerm}, and the last is the hyperfine coupling, originating from Eq.~\eqref{eq:hyperfine}. In this term, we subtracted the statistical average, defining
$\delta \spinNucleusVector = \spinNucleusVector - \langle \spinNucleusVector \rangle$, since the average has been included in $\angularFrequencyElectron$. For further convenience, all frequencies in the above equation are defined as positive. Inverting a sign, for example of a $g$-factor, would be reflected by inverting the corresponding unit vector $\axisEnergy$.\footnote{We find that while the $g$-factor signs are not entirely irrelevant as they show up in the formulas below, they do not lead to qualitative differences. Rather, inverting a $g$-factor maps the problem to an equally relevant scenario for all questions that we consider. See especially Sec.~\ref{sec:feedback}.} While the hyperfine coupling $J_n$ is signed, neither the DNSP rate nor the feedback through Eq.~\eqref{eq:frequencyElectron} will depend on the sign.

\section{Polarization rate}
\label{sec:rate}

We now proceed with the derivation of the polarization rate using Eq.~\eqref{eq:electronNuclearPairH}. As already noted, the calculation is related to some results of the NMR and molecular-chemistry literature \cite{weis_solid_2000,weis_electron-nuclear_2006,henstra_theory_2008,jain_off-resonance_2017}. Nevertheless, there are important differences, which we point out on the way. 

\subsection{The origin of the DNSP}

\label{sec:DNSPorigin}

The first step is to gauge away the time-dependent driving, transforming to a rotating reference frame, $\Psi^\prime= U \Psi$. It is useful to transform both the electron and nuclear spins, with the following unitary,
\be
U(t) = \exp ( - i \spinElectronVector \cdot \axisEnergyElectron \angularFrequencyRF t) \exp ( - i \spinNucleusVector \cdot \axisEnergyNucleus \angularFrequencyRF t).
\label{eq:U}
\ee
Adopting the rotating-wave approximation in the third term of Eq.~\eqref{eq:electronNuclearPairH} gives the transformed Hamiltonian
\be
\begin{split}
H^\prime = & 
-(\hbar\angularFrequencyNucleus-\hbar\angularFrequencyRF) \spinNucleusVector \cdot \axisEnergyNucleus
-(\hbar\angularFrequencyElectron-\hbar\angularFrequencyRF) \spinElectronVector \cdot \axisEnergyElectron\\
&-  \hbar\angularFrequencyRabiResonant \, \spinElectronVector \cdot \axisEnergyElectronY
+\delta \spinNucleusVector \cdot J_n^\prime(t) \cdot \spinElectronVector.
\end{split}
\label{eq:Htilde}
\ee
We have defined $\axisEnergyElectronY$ as the vector $\axisBFieldRF$ rotated by angle $\phi_\rf$ around axis $\axisEnergyElectron$, and the transformed hyperfine tensor by the relation
\be
U(t) J_n(t) \, \delta \spinNucleusVector \cdot \spinElectronVector \, U(t)^\dagger = \sum_{ij} [J_n^\prime(t)]_{ij} \delta \spinNucleus_i \spinElectron_j.
\label{eq:Jtilde}
\ee
Here, the differences to the existing derivations can be appreciated. First, were the quantization axes of the electron and the nucleus parallel, which would be the case for a magnetic field constant in space, the transformation $U$ would commute with the spin-spin interaction
\be
[J^\prime_n(t)]_{ij} = J_n(t) \delta_{ij},
\tag{{uniform\,B-field}}
\ee
This result arises because the hyperfine interaction Eq.~\eqref{eq:hyperfine} conserves the total (electron plus nuclear) spin. In this case, the transformation into the rotating frame does not generate time-dependent terms. In our problem, the Knight field $J_n$ is still time-dependent due to the spatial displacement of the electron, making the wave function modulus $|\Psi_n|^2$ in Eq.~\eqref{eq:J} time-dependent. This additional time dependence is the second difference to the existing results. For a typical NMR scenario with two nuclei in the lattice of a crystal or in a molecule, their mutual interaction in the laboratory frame is constant. That would here correspond to an ESR\cite{shchepetilnikov_nuclear_2016} (and not EDSR) driving of the electron, by which the transformed hyperfine tensor would become not only diagonal in spin indexes but also time-independent
\be
[J_n^\prime(t)]_{ij} = J_n(0) \delta_{ij},  \tag{ESR\, and\, uniform\,B-field}
\ee
Under such conditions, the transformed Hamiltonian in Eq.~\eqref{eq:Htilde} would be time-independent and no DNSP effects would arise.\footnote{On the other hand, both ESR \cite{koppens_driven_2006} and EDSR \cite{nowack_coherent_2007} experiments in a gated quantum dot showed signatures of DNSP. In the former, an oscillating electric field probably accompanied the desired oscillating magnetic field.}

Since in the NMR scenarios the laboratory frame $J_n$ is time-independent, a finite DNSP requires either `nonsecular' terms in the exchange tensor, such as $\spinNucleus_x \spinElectron_z$,\cite{henstra_nuclear_1988,weis_solid_2000,henstra_theory_2008,jain_off-resonance_2017} or Rabi-driving also the nuclear spin \cite{weis_electron-nuclear_2006}, where the `secular' exchange term $\spinNucleus_z \spinElectron_z$ allows for spin flips (as in the standard Hartmann-Hahn scenario \cite{hartmann_nuclear_1962}).

Concluding, there are two sources of the time dependence of the transformed hyperfine tensor $J^\prime_n$. One is the noncollinearity of the spin quantization directions and is due to the micromagnet-induced magnetic field gradients. The second is due to the time-dependent spatial oscillations of the electron induced by the EDSR drive. We refer to the two sources as the two mechanisms of the DNSP. Neither is present in the standard NMR scenario, while at least one is necessary for a finite DNSP in a quantum dot in the coherent regime of EDSR.

\subsection{Identification of the secondary resonance}

\begin{figure}
\begin{center}
\includegraphics[width=0.7\linewidth]{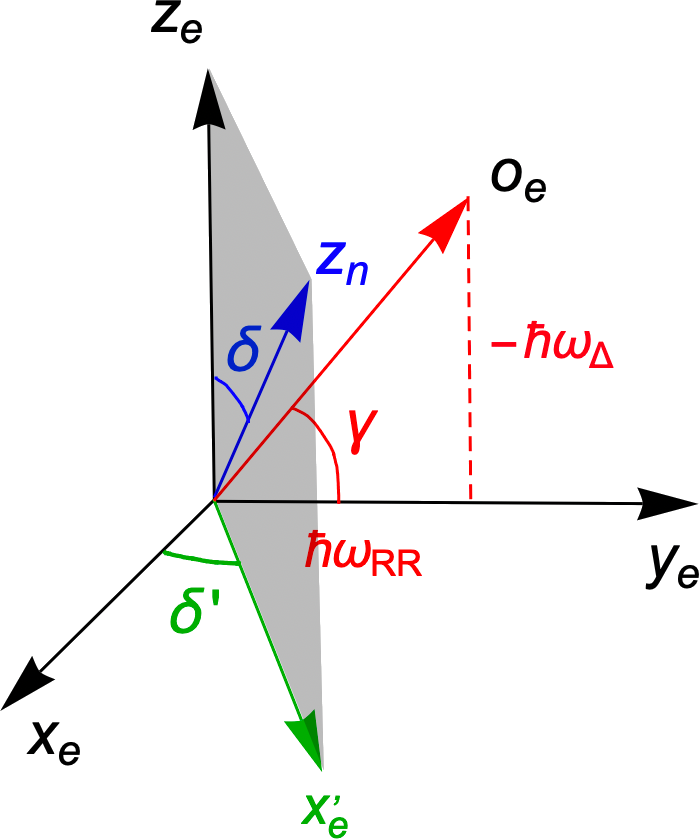}
\end{center}

\caption{\textbf{Angles and directions relevant for EDSR.} 
The electron ground-state spin direction $\axisEnergyElectron$ is along or opposite to the magnetic field $\bFieldElectron$, depending on the electron $g$-factor sign. In the frame rotating with the electron Larmor frequency, the EDSR field is along the vector $\axisEnergyElectronO$ and makes an angle $\pi/2-\gamma$ with the axis of the precession. For the nuclear ground-state spin direction $\axisEnergyNucleus$, apart from relating to the magnetic field by the $g$ factor, it also changes with the position within the dot due to the micromagnet. The rotation $R_{\axisEnergyElectron \to \axisEnergyNucleus}$ that rotates the axis $\axisEnergyElectron$ into the axis $\axisEnergyNucleus$ is parameterized by two Euler angles: $\delta^\prime$ for the initial rotation around $\mathbf{z_e}$, and $\delta$ for the final rotation around the rotated axis $\axisEnergyElectronY^\prime$.
\label{fig:angles}
}
\end{figure}

After explaining the physical origins of the effects and their differences from the Hartmann-Hahn scenario of NMR, we now proceed with straightforward manipulations of Eq.~\eqref{eq:Htilde}. The technical reason for employing the transformation $U$ was to move all time-dependence into the last term of $H^\prime$. Since it is the smallest term, it can be treated perturbatively. To this end, we first diagonalize the unperturbed part by introducing the following unit vectors and angles:
\begin{subequations}
\begin{align}
\axisEnergyElectronY &= R_{\axisEnergyElectron,\phi_\rf} \cdot  \axisBFieldRF,\\
\axisEnergyElectronX &= \axisEnergyElectronY \times \axisEnergyElectron,\\
\axisEnergyElectronO &= \axisEnergyElectron \sin \gamma  + \axisEnergyElectronY \cos \gamma ,\\
\sin\gamma &= -\frac{\angularFrequencyDetuning}{\angularFrequencyRabi},\\
\cos\gamma &= \frac{\angularFrequencyRabiResonant}{\angularFrequencyRabi}.
\end{align}
\end{subequations}
Also, $\angularFrequencyRabi = \sqrt{\angularFrequencyRabiResonant^2 + \angularFrequencyDetuning^2}$ is the (positive) Rabi frequency and $R_{\mathbf{n},\alpha}$ is a  $3\times 3$ matrix corresponding to a rotation around vector $\mathbf{n}$ by angle $\alpha$. The axes and angles are shown in Fig.~\ref{fig:angles}. The Hamiltonian becomes
\be
H^\prime =  
-(\hbar\angularFrequencyNucleus-\hbar\angularFrequencyRF) \spinNucleusVector \cdot \axisEnergyNucleus
-\hbar\angularFrequencyRabi \spinElectronVector \cdot \axisEnergyElectronO
+\delta \spinNucleusVector \cdot J_n^\prime(t) \cdot \spinElectronVector.
\label{eq:Htilde2}
\ee
The unperturbed part of the Hamiltonian (the first two terms) has eigenstates with the nuclear and electron spins parallel or antiparallel to the vectors $\axisEnergyNucleus$ and $\axisEnergyElectronO$. We denote them as $|sj\rangle$,
\be
\label{eq:basisStates}
H^\prime(J^\prime_n=0) |sj\rangle = E_{sj}|sj\rangle,
\ee
where $s$ and $j$ denote the spin eigenvalues: $s\in\{+1/2,-1/2\}$ and $j\in\{+I,+I-1,\ldots,-I\}$ with a general integer or half-integer value for $I$. The corresponding energy is \footnote{Concerning unperturbed energies, we thus include only the collective Overhauser field from all nuclei acting on the electron [entering into $\hbar\angularFrequencyRabi$ through Eq.~\eqref{eq:frequencyElectron}] and neglect the Overhauser field and the Knight field stemming from the last term in Eq.~\eqref{eq:Htilde2}. Reference \cite{weis_electron-nuclear_2006} deals with the scenario where the diagonal part of the hyperfine interaction is strong and needs to be included in the unperturbed energies.} 
\be
E_{sj} = -(\hbar\angularFrequencyNucleus-\hbar\angularFrequencyRF) j -\hbar\angularFrequencyRabi s,
\label{eq:Esj}
\ee
The energy difference between a pair of eigenstates is
\be
E_{sj} -E_{s^\prime j^\prime }=  (\hbar\angularFrequencyNucleus-\hbar\angularFrequencyRF) (j^\prime -j) +\hbar\angularFrequencyRabi (s^\prime - s).
\label{eq:pairEnergyDifference}
\ee
The DNSP (and the Hartmann-Hahn effect) arises if a pair of these eigenstates is Rabi-driven through the time-dependent term in Eq.~\eqref{eq:Htilde2} on resonance with the energy difference Eq.~\eqref{eq:pairEnergyDifference}.\footnote{In contrast, in Refs.~\cite{danon_nuclear_2008,danon_multiple_2009} the energy mismatch is assumed to be compensated by an additional agent, such as an applied source-drain voltage. In Ref.~\cite{rudner_electrically_2007} (Ref.~\cite{yang_collective_2012}), it is the finite linewidth of the electron (hole) spin.}
Since we are interested in transitions that change the nuclear spin, $j^\prime \neq j$, the large energy difference $\hbar \angularFrequencyRF$ must be compensated by a time-dependent term oscillating at a similar frequency. The matrix elements of $J_n^\prime$ are polynomial functions of exponentials $\exp (\pm i \angularFrequencyRF t)$ and, therefore, contain only integer multiples of $\angularFrequencyRF$ as frequencies. The integer one multiple can compensate the driving frequency in Eq.~\eqref{eq:pairEnergyDifference} and what remains is\footnote{This step can be understood as going into a rotating frame effectively undoing the rotation due to the second term of Eq.~\eqref{eq:U}. We include an alternative derivation of the polarization rate using such a frame in Appendix \ref{app:quadrupolar}, see Eq.~\eqref{eq:U-alt}.} 
$$
\hbar\angularFrequencyNucleus(j^\prime -j) +\hbar\angularFrequencyRabi (s^\prime - s).
$$
This difference can become zero only if the electron also flips, $s=-s^\prime$, and we get that a quasi-resonant pair fulfills
\be
s+j = s^\prime+j^\prime.
\label{eq:spinIndexRelation}
\ee
Concluding, the only states that can become quasi-resonant are (we explicitly denote the spin quantization directions in subscripts as a reminder)
\be
| s_{\axisEnergyElectronO} = 1/2, j_{\axisEnergyNucleus} \rangle \leftrightarrow | s_{\axisEnergyElectronO} =-1/2,  (j+1)_{\axisEnergyNucleus}  \rangle,
\label{eq:resonantPair}
\ee
and that happens if the Hartmann-Hahn-like condition, 
\begin{equation}
\label{eq:Hartmann-Hahn}
\hbar \angularFrequencyNucleus \approx \hbar \angularFrequencyRabi,
\end{equation} 
is fulfilled. 

\subsection{Secondary Rabi oscillations}

\label{sec:secondaryRabi}

We depict the result of the preceding analysis by the following Hamiltonian for the electron-nuclear spin pair,
\be
\left( \begin{tabular}{c|cccc}
$H_{s^\prime j^\prime, s j}$ & $\uparrow \uparrow$ & $\uparrow \downarrow$ & $\downarrow \uparrow$ & $\downarrow \downarrow$ \\
\hline
$ \uparrow \uparrow $ & $E_{\uparrow \uparrow}$& $ \cdot$ & $\cdot$ & $\cdot$ \\
$\uparrow \downarrow $ & $\cdot$& $E_{\uparrow \downarrow}$& $Y^\dagger$ & $\cdot$ \\
$\downarrow \uparrow $ & $\cdot$ & $Y $& $E_{\downarrow \uparrow}$& $ \cdot$ \\
$\downarrow \downarrow $ & $ \cdot$ & $\cdot$ & $\cdot$& $E_{\uparrow \uparrow}$ \\
\end{tabular} \right).
\label{eq:Hsjsj}
\ee 
We have used a pictorial notation for the spin states, $\uparrow$ and $\downarrow$ for the electron states $+1/2$ and $-1/2$, and for nuclear states $j+1$ and $j$. In addition to the state energies, given by Eq.~\eqref{eq:Esj}, one needs only one matrix element,
\be
\label{eq:YXI}
Y = \langle \downarrow \uparrow | \delta \spinNucleusVector \cdot J_n^\prime(t) \cdot \spinElectronVector | \uparrow \downarrow \rangle 
     \equiv \langle \downarrow \uparrow | \spinNucleus_+ \spinElectron_- | \uparrow \downarrow \rangle X
     \equiv \mathcal{I}_+ X
     ,
\ee
for the only pair of states that might become quasi-resonant. Here, we have introduced the spin ladder operators $\spinElectron_\pm = \spinElectron_x \pm i \spinElectron_y$ and $\spinNucleus_\pm = \spinNucleus_x \pm i \spinNucleus_y$. All other states are off-resonant, with the Hamiltonian matrix elements negligible with respect to the energy differences. These negligible off-resonant elements are denoted by dots in Eq.~\eqref{eq:Hsjsj}. Therefore, one can focus on the state pair $\{\uparrow \downarrow, \downarrow \uparrow\}$ as an effective two-level system displaying Rabi oscillations.\footnote{Note that these are `secondary' Rabi oscillations, different from the Rabi oscillations of the electron spin itself. The `primary' electron spin Rabi oscillations are taken into account---in the basis corresponding to Eq.~\eqref{eq:Hsjsj}---through the energies only. The appearance of the `secondary' Rabi oscillations in a frame where the `primary' oscillations are already trivial is the essence of the Hartmann-Hahn effect, see Eqs.~(49) and (50) in Ref.~\cite{hartmann_nuclear_1962}.} In Eq.~\eqref{eq:YXI}, we have introduced the abbreviations $X$ and $\mathcal{I}_+$, as parts of the matrix element $Y$ that we calculate below separately.

\begin{figure}
\begin{center}
\includegraphics[width=0.7\linewidth]{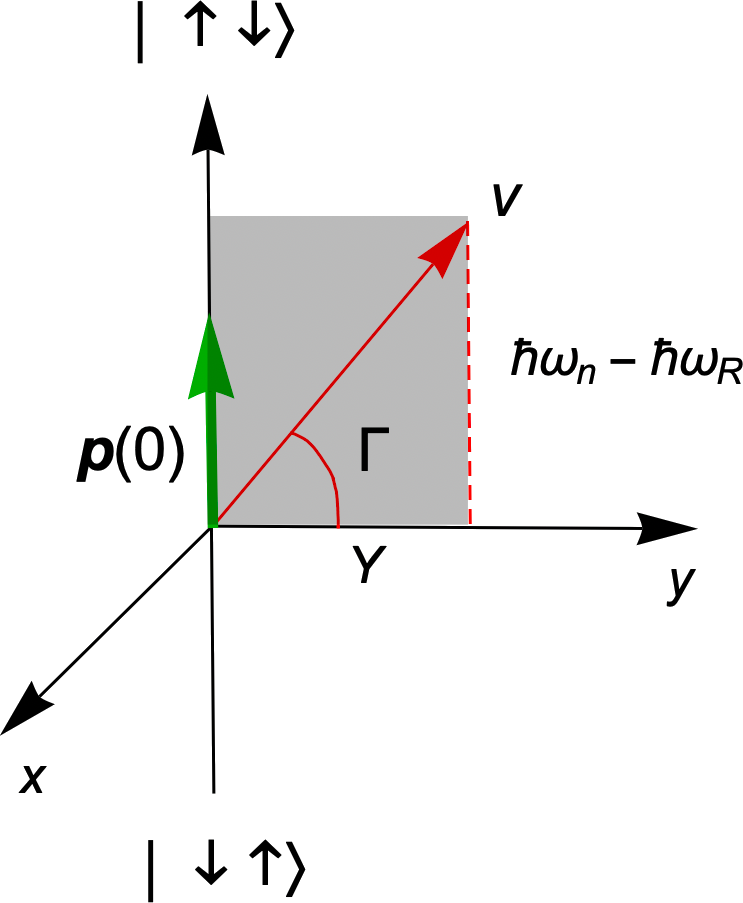}
\end{center}
\caption{\textbf{Bloch sphere for the secondary Rabi oscillations.} 
The electron-nuclear states $|\!\!\uparrow \downarrow\rangle$ and  $|\!\!\downarrow \uparrow \rangle$ define the north and south pole of the Bloch sphere. The in-plane axes are chosen so that the energy vector is in the yz plane, making an angle $\pi/2-\Gamma$ with the $z$ axis. This angle is defined by the matrix element $|Y|$ and the energy difference $\hbar\angularFrequencyNucleus-\hbar\angularFrequencyRabi$. Finally, $\mathbf{p}(0)$ is the initial polarization of the system.
\label{fig:Rabi}
}
\end{figure}

The corresponding $2\times 2$ block of the Hamiltonian can be then treated by the textbook method for the Rabi problem. We define a Bloch sphere spanning the two orthogonal states $\{\uparrow \downarrow, \downarrow \uparrow\}$ which we place on the sphere $z$ axis. There are two parameters important for the Rabi oscillations: the energy difference of the states, which is $E_{\uparrow\downarrow}-E_{\downarrow\uparrow} = \hbar\angularFrequencyNucleus-\hbar\angularFrequencyRabi$, and the magnitude of the matrix element $|Y|$. The phase of $Y$ defines only where to put the in-plane axes, $x$ and $y$, of the Bloch sphere, and is not relevant in the following. The two parameters define the (positive) Rabi frequency
\be
\hbar \angularFrequencyRabiHH = \sqrt{ |Y|^2 + \left| \hbar\angularFrequencyNucleus-\hbar\angularFrequencyRabi\right|^2 },
\label{eq:wHH}
\ee
and the angle $\Gamma$ which will turn out useful,
\begin{subequations}
\begin{align}
\sin \Gamma &= \frac{\hbar\angularFrequencyNucleus-\hbar\angularFrequencyRabi}{\hbar \angularFrequencyRabiHH},\\
\cos \Gamma &= \frac{|Y|}{\hbar \angularFrequencyRabiHH}.
\label{eq:sinGamma}
\end{align}
\end{subequations}
These quantities are shown in Fig.~\ref{fig:Rabi}.

Now we come to a somewhat subtle point concerning the initial state, that is, the system state at the time when the electron enters the dot or its EDSR driving begins. The phase relation of the $\uparrow \downarrow$ and $\downarrow \uparrow$ components of this initial state has contributions not only from the phase difference of the electron spin up and down state, which is controlled, but also from a similar phase difference on the nuclear spin, which is not controlled.
Alternatively viewed, the Hamiltonian in Eq.~\eqref{eq:Hsjsj} induces entanglement between the electron and nuclear spin. This entanglement is lost repeatedly as the electron is repeatedly reinitialized through the reservoir or otherwise. As a consequence, on average (over the experiment cycle repetitions) the initial state in the Bloch sphere in Fig.~\ref{fig:Rabi} can only have a non-zero component along the sphere $z$ axis.\footnote{In the language of Ref.~\cite{muller_coherence_1979}, in our scenario we have `cross-polarization', but no 'coherence transfer'. Our assumption means that we do not consider that the nuclear spin precession and the moments when the electron EDSR rotation starts are synchronized over many cycles. Such a long-time synchronization is essential for nuclear autofocusing \cite{greilich_nuclei-induced_2007,markmann_universal_2019}.} We will thus describe this initial state by a density matrix, parameterized by a vector $\mathbf{p}(t)$, which is at time $t=0$ aligned with the $z$ axis, and its length might be smaller than one.

The dynamics of this `polarization' vector is a simple precession and can be expressed, for example, by
\be
\mathbf{p}(t) = R^{-1}_{\mathbf{x},\pi/2-\Gamma} \cdot R_{\mathbf{z},\angularFrequencyRabiHH t} \cdot R_{\mathbf{x},\pi/2-\Gamma} \cdot \mathbf{p}(0),
\ee
where $\mathbf{x}$ and $\mathbf{z}$ are unit vectors along the Bloch-sphere axes. In addition, it is only the $z$ component that is relevant, as we have just discussed. It is then straightforward to evaluate the previous equation for that component arriving at
\be
p_z(t) = p_z(0) \left( \sin^2 \Gamma + \cos^2 \Gamma \cos \angularFrequencyRabiHH t \right).
\label{eq:pzt}
\ee
This result is the first main ingredient of the DNSP polarization rate.

\subsection{Calculation of the matrix element $\mathcal{I}_+$}

We now look at the initial polarization $p_z(0)$. As explained, it is contributed by the initial polarization of both the electron and the nucleus. The conversion from these two polarizations to $p_z(0)$ is not completely trivial because we consider a general nuclear spin $I$ and because the polarizations need to be weighted by the spin-dependent transition matrix element $\mathcal{I}_+$. The calculation is shown in Appendix~\ref{app:pzI} and gives
\be
\overline{ p_z(0) |\mathcal{I}_+|^2} = I \times \Big( \alpha_I(p_n) p_e - p_n \Big).
\label{eq:pzI}
\ee
Here, $p_e$ is the initial polarization of the electron spin \emph{along the axis $\axisEnergyElectronO$} and $p_n$ is the polarization of the nuclear spin along the external magnetic field. Both of these polarizations are normalized so that the maximal possible polarization corresponds to $p=1$. Finally, $\alpha_I(p_n)$ is a factor of order one, which we calculate in Appendix~\ref{app:pzI}, see Eq.~\eqref{eq:pz03b}. We get $\alpha_I = (2/3)(I+1)$ for $p_n\ll 1$ and $\alpha_I=1$ for $1-p_n\ll1$.  

Equation \eqref{eq:pzI} states that the electron polarization $p_e$ is the source of the nuclear polarization $p_n$. As the latter develops a finite value, the rate diminishes. For nuclear spin $I=1/2$ one has $\alpha_I=1$ for any $p_n$, and the rate is proportional to the difference $p_e-p_n$, a natural result. The proportionality factor $\alpha_I$ differs from one for nuclear spin $I>1/2$.  
In any case, in the majority of experiments the steady state nuclear polarization will be reached once the DNSP rate is balanced by additional decay channels, such as nuclear diffusion, rather than due to the DNSP rate dropping to zero at $p_n = p_e \alpha_I$. 
Therefore the second term in the bracket in Eq.~\eqref{eq:pzI} can usually be dropped.

Let us now elucidate the electron polarization $p_e$, considering two typical experiments. In the first, the initial electron state is along the external magnetic field. Once the driving is turned on, the electron performs Rabi oscillations. This choice is the standard EDSR and means the initial electron polarization is equal to $\axisEnergyElectron \cdot \axisEnergyElectronO = \sin \gamma$. In the second, the electron is `spin-locked', meaning its spin is along $\axisEnergyElectronO$ and the initial polarization is one.\footnote{In this case the system has to be initialized either adiabatically changing the driving frequency \cite{hartmann_nuclear_1962} or using a phase shift in the driving pulse \cite{weis_solid_2000}. Even if it is the former, we are not concerned with the transition period needed to spin-lock the electron. We assume that the transition period is shorter than the time during which the electron remains spin-locked with a constant Rabi frequency. Finally, a more complicated initial polarization when the electron is driven off resonance was considered in Ref.~\cite{jain_off-resonance_2017}.} Summarizing, we get
\begin{subequations}
\begin{align}
p_e & = \sin \gamma & \mathrm{(EDSR)},\label{eq:peForEDSR}\\
p_e & = 1&\mathrm{(spin\,locking)}.
\end{align}
\end{subequations}
While the first choice corresponds to the standard EDSR, the advantage of the second one (concerning possible DNSP) is that the lifetime of the electron spin is longer in the spin-locked state, compared to the lifetime of the Rabi oscillations \cite{redfield_nuclear_1955}. Finally, we note that in both scenarios one can invert the polarization $p_e \to -p_e$, by preparing the electron in the excited state, rather than in the ground state.

\subsection{Calculation of the matrix element $X$}

We now turn to $X$, the second component of the matrix element $Y$. Writing $\delta \spinNucleusVector \cdot {J_n^\prime}(t) \cdot \spinElectronVector$ in the interaction picture
\be
\langle s j | \exp(i E_{sj}t /\hbar) \delta\spinNucleusVector \cdot J_n^\prime(t) \cdot \spinElectronVector \exp(-i E_{s^\prime j^\prime}t /\hbar) |s^\prime j^\prime \rangle,
\label{eq:interactionPicture}
\ee
one can see that the resonant component of $J^\prime_n(t)$ is the one of frequency $2s \times \angularFrequencyRF$. Introducing the Fourier components in this matrix,
\be
\label{eq:Jprime}
J_n^\prime(t) = \sum_{k\in \mathbb{Z} } \exp(i k \angularFrequencyRF t) {J_n^\prime}^{(k)},
\ee
the resonant matrix element in Eq.~\eqref{eq:interactionPicture} would be ${J_n^\prime}(t) \approx {J_n^\prime}^{(2s)} \exp(i2s \angularFrequencyRF t)$. Keeping only this term, essentially the rotating wave approximation, the calculation of the matrix element is a straightforward algebraic exercise and we delegate it to Appendix~\ref{app:derivationX}. The result, after the spatial average over the dot coordinates, is
\begin{subequations}
\label{eq:X2}
\be
\langle |X|^2 \rangle = X_\mathrm{df}^2 + X_\mathrm{sh}^2 +2 \xi X_\mathrm{df} X_\mathrm{sh}, 
\label{eq:Xave}
\ee
where
\begin{align}
X_\mathrm{df} &= \frac{J}{4} \frac{\nabla_\perp B l}{B}\cos \gamma,\label{eq:xdf}\\
X_\mathrm{sh} &= \frac{J}{4} \frac{d}{l} (1+\sin \gamma),\label{eq:xsh}\\
\xi &= \cos( \delta^\prime + \angleRF) \cos( \phi),
\label{eq:xi}
\end{align}
\end{subequations} 
with $J$ being the average $\langle J_n \rangle$, $\nabla_\perp B$ being the magnitude of the gradient of the transverse components of the magnetic field, and the angles $\delta^\prime$ and $\phi$ express the mutual orientation of the magnetic field gradient and the dot displacement (see App.\ref{app:derivationX} for details). As in experiments these directions are difficult to control or even to know, instead of going into its rather tedious analysis, we drop the interference term from Eq.~\eqref{eq:Xave}. We retain only the first two terms:\footnote{The latter mechanism was considered in several previous works on DNSP in quantum dots \cite{rudner_electrically_2007,danon_nuclear_2008,danon_multiple_2009,tenberg_narrowing_2015} starting with Ref.~\cite{laird_hyperfine-mediated_2007}. As we explain in Appendix~\ref{app:laird}, our Eq.~\eqref{eq:xsh} can be thought of as a generalization of these previous works.}
$X_\mathrm{df}$ is due to a deflection (thus `df') of the spin quantization axes of the electron and the nucleus, and requires a finite gradient of the transverse magnetic field component. $X_\mathrm{sh}$ is due to the time dependence of the electron-nuclear spin coupling constant, in turn due to the time dependence of $|\Psi(\mathbf{r}_n)|^2$, in turn due to the physical shifts (thus `sh') of the quantum dot electron with respect to the crystal lattice. 
Equation \eqref{eq:X2} is the second main ingredient for the calculation of the DNSP rates. 

\subsection{Evaluation of the DNSP rate}

We now have all ingredients needed to evaluate the DNSP rate. We define the individual nuclear spin polarization rate by
\be
\Gamma_n = \frac{\langle \overline{p_z(0) - p_z(t)} \rangle }{t},
\ee
where the bar denotes the statistical average over the nuclear spin distribution and the angle brackets the average over the dot coordinates. The overall sign has been chosen to define a positive polarization rate as the decrease of $p_z$, that is a transition of nuclear spin from $\downarrow$ towards $\uparrow$. In other words, a positive polarization rate means that nuclear spins are pumped into their energy ground state, being along or opposite to the external field depending on the nuclear $g$-factor sign. 
Using Eqs.~\eqref{eq:sinGamma} and \eqref{eq:pzt} gives
\be
\Gamma_n = \frac{1}{ \hbar^2} \,\langle \overline{p_z(0) |X \mathcal{I}_+|^2} \rangle \frac{1-\cos \angularFrequencyRabiHH t}{(\angularFrequencyRabiHH)^2t}.
\ee
Next, we approximate the averaging over the nuclear spins by evaluating it separately for the matrix element $X^2$ and the rest,
\be
\Gamma_n = \frac{1}{\hbar^2} \, \overline{ p_z(0) |\mathcal{I}_+|^2 } \,\,  \langle |X^2| \rangle \frac{1-\cos \angularFrequencyRabiHH t}{(\angularFrequencyRabiHH)^2t}.
\label{eq:gamman}
\ee
The two needed results are given in Eqs.~\eqref{eq:pzI} and \eqref{eq:Xave}.

We have arrived at a rate for an `average' nuclear spin, which is not really a rate: it contains time, since it originates from coherent precession expressed by Eq.~\eqref{eq:pzt}. 
We convert it to a time-independent rate\footnote{As a remark, this time-dependence was kept in Ref.~\cite{henstra_theory_2008}, resulting in a non-trivial time-dependence of the polarization. For example, a polarization overshoot seen in the data in Fig.~3 therein could be explained with it.}
 by considering the limit $t\to\infty$ upon which the last factor in Eq.~\eqref{eq:gamman} becomes a delta function of a finite width given by the matrix element $|Y|$. Since for our parameters the latter is several orders of magnitude smaller than other energy smearings that we consider below, we neglect it,
\begin{equation}
\label{eq:promotionToDeltaFunction}
\frac{1-\cos \angularFrequencyRabiHH t}{(\angularFrequencyRabiHH)^2t} \to \pi \delta(\angularFrequencyRabi - \angularFrequencyNucleus).
\end{equation}
We now define the total polarization rate 
\be
\Gamma_{i,\mathrm{tot}} = \sum_{n \in i} \Gamma_n = N_i \int \mathrm{d}\angularFrequencyNucleus \, g(\angularFrequencyNucleus) \Gamma_n(\angularFrequencyNucleus),
\label{eq:GammaTotalAsSum}
\ee
introducing the nuclear frequency density $g(\omega)$ as the fraction of $i$-isotope nuclei with Larmor frequency $\omega$ out of their total number $N_i$. The function is derived in Appendix~\ref{app:nuclearDensity}. We get
\be
\Gamma_{i,\mathrm{tot}} = N_i \frac{\pi}{ \hbar^2} \, \overline{p_z(0) |\mathcal{I}_+|^2} \,\,  \langle|X^2|\rangle g(\angularFrequencyRabi).
\ee
Note the crucial role of the micromagnet, setting the width of the distribution $g(\omega)$: the larger the gradient, the more dispersed the Larmor frequencies of the nuclei in the dot area, and the wider the resonance. Here, the resonance means the electron Rabi frequency $\angularFrequencyRabi$ hitting the peak of the function $g$, which is located at the Larmor frequency of the nuclei in the dot center.

\subsection{Final form of the DNSP rate and its discussion}

We now put together the pieces to present the rate in a user-friendly form. In the course of derivation, we have used several approximations, which are expected to bring an error of order one. Therefore, we neglect small terms, in order to arrive at a simple formula with an appealing physical interpretation:
\begin{subequations}
\label{eq:mainResult}
\be
\partial_t p_i = \frac{\pi}{\hbar^2}\left(X_\mathrm{df}^2 + X_\mathrm{sh}^2 \right) \left(\alpha_I p_e-p_i\right) G_\Sigma\left(\angularFrequencyRabi - \omega_i \right),
\label{eq:rateDNSP}
\ee
where
\begin{align}
X_\mathrm{df} &= \frac{A_i}{4N_{\mathrm{tot}}} \frac{l \nabla_\perp B}{B}\cos \gamma, \label{eq:XM}\\
X_\mathrm{sh} &= \frac{A_i}{4N_{\mathrm{tot}}} \frac{d}{l} (1+\sin \gamma),\label{eq:XT}\\
p_e & =  \pm \left\{\begin{tabular}{c@{\hspace{2cm}}r} $\sin \gamma$ &for EDSR, \\ 1 & for spin locking, \end{tabular} \right.\label{eq:pe}\\ 
G_\Sigma(x) &= \frac{1}{\sqrt{2\pi}\Sigma} \exp\left( -\frac{x^2}{2\Sigma^2}\right),\label{eq:gaussianDensity}\\
\Sigma_\mathrm{\mu M} &= \omega_i \frac{l\nabla_{||}B}{2B}, \label{eq:sigmaMM}\\
\alpha_I & =  \left\{\begin{tabular}{c@{\hspace{2cm}}r} $\frac{2}{3}(I_i+1)$ &for $p_i\approx 0$, \\ 1 & for $p_i\approx 1$, \end{tabular} \right.\\
\tan \gamma &= \frac{\angularFrequencyElectron - \angularFrequencyRF}{\angularFrequencyRabiResonant} .
\end{align}
\end{subequations}
Here, the quantities dependent on the atomic isotope have the subscript $i$, $l$ is the dot in-plane confinement length, typically tens of nanometers, $d$ is the dot displacement magnitude given by Eq.~\eqref{eq:dotShift}, typically below a nanometer. The plus sign for $p_e$ applies if the electron is initially in the ground state of the static field in the laboratory frame (for EDSR) or the rotating frame (for spin-locking). If initially the electron spin is in the excited state, the minus sign applies. Finally, $\nabla_{||}B $ is the magnitude of the longitudinal (along the vector $\langle \bField \rangle$) component of the magnetic-field gradient, and $\nabla_\perp B$ is the magnitude of the gradient of the magnetic field transverse components. For the moment, we assume that $\Sigma = \Sigma_\mathrm{\mu M}$; however, below we list additional sources contributing to $\Sigma$ in Eq.~\eqref{eq:rateDNSP} beyond Eq.~\eqref{eq:sigmaMM}.

Let us make a few comments on the DNSP rate given in Eq.~\eqref{eq:mainResult}, our main result. %
\newcounter{myCounter}
\newcommand{\mySymbol}{\addtocounter{myCounter}{1}\themyCounter.}
\begin{itemize}[leftmargin=*, itemindent=0.5em]
\item[\mySymbol] Equation \eqref{eq:rateDNSP} gives the rate of polarization of isotope $i$. It can be converted to the total `spin-injection' rate by $\Gamma_{i,\mathrm{tot}} =I_i  N_i \partial_t p_i $. Since the hyperfine interaction is spin preserving, this total rate of spin injected into the nuclei is compensated by the opposite change of the electron spin (component along the external magnetic field).
\item[\mySymbol] The nuclear polarization direction is defined as the positive rate corresponding to pumping-in the nuclear spin energy ground state (along the magnetic field if the nuclear $g$ factor is positive).
\item[\mySymbol] Neglecting the saturation effect, meaning dropping $p_i$ from the right-hand side of Eq.~\eqref{eq:rateDNSP}, the DNSP rate has a characteristic shape as a function of the detuning from the electron Rabi resonance, parameterized by $\gamma$ here. Namely, since $\angularFrequencyRabi(\gamma) = \angularFrequencyRabi(-\gamma)$, the DNSP rate is antisymmetric in $\gamma$ in EDSR and symmetric in spin-locking experiments if the 
`deflection' mechanism dominates. The `shaking' mechanism makes the profile strongly asymmetric in both cases, through the factor $(1+\sin\gamma)$. The shape of the DNSP rate as a function of $\gamma$ can then hint at the dominant mechanism.
\item[\mySymbol] In experiments with a single dot, the DNSP will be typically done by repeating a cycle including the electron spin initialization, driving, and, perhaps, measurement. In this case, one should renormalize to the rate observed over the laboratory time by $\Gamma \to \Gamma \times (T_\mathrm{pulse} / T_\mathrm{cycle})$, reflecting that the cycle contains `dead time' with respect to the DNSP.
\item[\mySymbol] During a single cycle, Eq.~\eqref{eq:rateDNSP} is valid only up to time $T_\mathrm{pulse}$ such that $\Gamma_\mathrm{tot} T_\mathrm{pulse} \lesssim1$, since the electron spin can not change by more than a single full flip.\footnote{Maximizing the portion of the electron spin transferred to nuclei over one cycle was done in Ref.~\cite{henstra_dynamic_2014}.} 
\item[\mySymbol] Since the total spin of nuclei is difficult to measure directly, it is useful to convert the nuclear polarization into quantities directly observable through the electron. In Appendix~\ref{app:conversion} we express the effects of the DNSP given in Eq.~\eqref{eq:mainResult} as the change of the electron Larmor frequency, due to the change of the Overhauser field,
\be
\partial_t \left( g_e \mu_B B_\mathrm{Ov} \right) = \sum_i \fractionIsotope A_i I_i \partial_t p_i,
\ee
and as the change of the detuning,
\be
\partial_t \frequencyDetuning =-\frac{1}{2 \pi \hbar} \mathrm{sgn} \left( g_e \right) \sum_i \fractionIsotope |A_i| I_i \partial_t p_i.
\label{eq:dfdt}
\ee
\item[\mySymbol] In the far-off-resonance limit, corresponding to $\gamma \to \pm \pi/2$ in our notation, one of the adopted assumptions is not fulfilled, see Eq.~\eqref{eq:fourDifferences} below.\footnote{The far-off-resonance limit was considered in Ref.~\cite{jain_off-resonance_2017}.} While we believe that Eq.~\eqref{eq:mainResult} can still be used for qualitative estimates, it might break down in certain limits, one example given in Appendix~\ref{app:laird}.
\item[\mySymbol] The micromagnet was essential for several elements: the primary Rabi oscillations of the electron, the deflection of the quantization axes of the electron and the nucleus, and the dispersion of the nuclear Larmor frequencies across the dot. In the next section we argue that there are intrinsic sources for the latter two, so that they are present in comparable magnitudes in experiments without a micromagnet. The analysis here then applies also if the micromagnet, as the source of the Rabi oscillations, is replaced by the intrinsic spin-orbit interaction. In other words, it applies also for holes, as long as Eq.~\eqref{eq:wavefunction} is still applicable, see Appendix~\ref{app:hole}. If it is not, meaning the spin-orbit length is smaller than the size of the dot, we expect DNSP with nontrivial spatial textures analogous to those predicted in Ref.~\cite{rudner_electrically_2007}. 
\item[\mySymbol] Considering the nuclear spins in isolation, as we have done at the outset of the derivation of Eq.~\eqref{eq:mainResult}, is quite a cavalier approximation. We believe that it suffices for what we aim at, being a rough estimation of the DNSP rate. Another motivation to adopt it is the fact that the full problem---of an electron spin relaxing into an interacting dipole-dipole coupled nuclear system---is too difficult: While a formal expression for the rate can be found in the literature (see Eq.~(2.36) in Ref.~\cite{demco_dynamics_1975}, Eq.~(4.1) in Ref.~\cite{vega_double-quantum_1980}, or Eq.~(13) in \cite{mcarthur_rotating-frame_1969}), its evaluation is not easy, see the discussion in the introduction of Ref.~\cite{demco_dynamics_1975} and in Ref.~\cite{vega_double-quantum_1980}.
\end{itemize}

After deriving the DNSP rate within our model, we now generalize the resulting formula to grasp effects important in real-world experiments.

\section{Model limitations and extensions}

\label{sec:extensions}

The above DNSP effects rely on fulfilling the Hartmann-Hahn condition, Eq.~\eqref{eq:Hartmann-Hahn}. Specifically, the rotating wave approximation that we adopted in Sec.~\ref{sec:secondaryRabi} in describing the secondary Rabi oscillations assumes that among the four energies 
\be
\label{eq:fourDifferences}
\hbar \angularFrequencyNucleus, \,\hbar \angularFrequencyRabi,\, \hbar \angularFrequencyNucleus + \hbar \angularFrequencyRabi,\, \hbar \angularFrequencyNucleus - \hbar \angularFrequencyRabi,
\ee
the last is by far the smallest. We now look with what precision these energies (or frequencies) and their differences are defined.

Concerning the nuclear Larmor frequencies, we have considered their smearing across the quantum dot due to the micromagnet, arriving at a Gaussian density \eqref{eq:gaussianDensity} with the dispersion \eqref{eq:sigmaMM}. 
The parameters  given in the caption of Fig.~\ref{fig:data-fit} give the frequency dispersion of several tens of $2\pi\times$kHz.\footnote{Specifically, for a longitudinal gradient of 0.3 mT/$\mu$m and the dot lateral size $l=34$ nm we get 
$\Sigma_\mathrm{\mu M}(^{29}\mathrm{Si})=2\pi\times 43$ kHz, $\Sigma_\mathrm{\mu M}(^{69}\mathrm{Ga})=2\pi\times 52$ kHz, $\Sigma_\mathrm{\mu M}(^{71}\mathrm{Ga})=2\pi\times 66$ kHz, $\Sigma_\mathrm{\mu M}(^{75}\mathrm{Ga})=2\pi\times 37$ kHz.} This value should be compared to additional frequency smearing sources:\footnote{In the NMR literature, the dispersion of nuclear energies due to nuclear dipole-dipole interactions is often taken as Gaussian. For example, see Eq.~(A20) in Ref.~\cite{lurie_spin_1964}. Therefore, those numbers are directly comparable to our $\Sigma_\mathrm{\mu M}$.}
On the one hand, the bulk value for both the intrinsic nuclear linewidth deduced from the $T_2$ times and the local field (dipolar and other) from other nuclei look negligible.\footnote{Ref.~\cite{ota_decoherence_2007} found $\Sigma_\mathrm{T_2} \lesssim 2\pi \times 1$ kHz. Slightly larger values for $^{75}$As and $^{71}$Ga in lattice-matched dots are collected from other references in Ref.~\cite{chekhovich_suppression_2015}. Refs.~\cite{sundfors_exchange_1969,hester_nuclear-magnetic-resonance_1974} give the nuclear local field in GaAs as up to a few Gauss (it is anisotropic), corresponding to $\Sigma_\mathrm{dip} \sim 2\pi \times$ (a few) kHz.} On the other hand, in a nanostructure the inhomogeneous strain and electric fields amplify line widths: the quadrupole splitting $\Sigma_\mathrm{Q} \gtrsim 2\pi\times 10$ kHz \cite{yusa_controlled_2005} or the Knight field from the electron $\Sigma_\mathrm{K}$ of a similar magnitude are typical (these values are for GaAs).\footnote{For our parameters, we estimate $\Sigma_\mathrm{K} \sim J \lesssim 2\pi\times 10$ kHz. More precisely, for quantum-dot parameters $l_z=10$ nm and $l=34$ nm, the electron-frequency shift upon a single nuclear spin flip, $J_n/\hbar$, is equal to $2\pi\times 6$ kHz for $^{69}$Ga, $2\pi\times 7.5$ kHz for $^{71}$Ga, $2\pi\times 7$ kHz for $^{75}$As; and for $l_z=6$ nm and $l=20$ nm, it is $2\pi\times (-1.7)$ kHz for $^{29}$Si. In self-assembled dots, the Knight fields are much larger, and the single-nuclear-flip electron-frequency shift of 200 kHz could be detected in Ref.~\cite{jackson_quantum_2021}.} The total frequency span of a given isotope might crawl to 100 kHz.\footnote{See Fig. 3a in \cite{noorhidayati_resistively_2020} or Fig.~2 in Ref.~\cite{kawamura_electronic_2015}, showing the line profile of $^{75}\mathrm{As}$ at high magnetic fields.} All these sources can be included in our formula by simply adding the corresponding variances, redefining the parameter $\Sigma$ in Eq.~\eqref{eq:gaussianDensity} as follows
\begin{equation}
\Sigma^2 \to \Sigma_\mathrm{\mu M}^2 + \Sigma_\mathrm{T_2}^2+ \Sigma_\mathrm{dip}^2+\Sigma_\mathrm{Q}^2+\Sigma_\mathrm{K}^2.
\label{eq:frequencyDispersion}
\end{equation}
As an important consequence, one \emph{expects the discussed DNSP effects even in samples without a micromagnet}: The longitudinal magnetic field gradient is effectively replaced by the sources given on the right-hand side of Eq.~\eqref{eq:frequencyDispersion} without the first term which then equals zero. Similarly, some of these terms contribute also to the deflection of the quantization axis of nuclear spins, that is, an effective transverse gradient. The quasi-static dipole field of other nuclei parameterized by $\Sigma_\mathrm{T_2^*}^2$ is isotropic and can be thus taken as an effective contribution to the gradient $\nabla_\perp B$ in Eq.~\eqref{eq:XM}. The quadrupolar fields also contribute, though they are anisotropic so that the contributing part depends on the direction of the magnetic field and the details of the atomic electric field gradients.\footnote{\label{fnt:non-collinear}In experiments with self-assembled quantum dots, the quadrupolar fields are thought to dominate the DNSP effects \cite{urbaszek_nuclear_2013}. One important consequence of considering quadrupolar interaction explicitly (we do it in Appendix~\ref{app:quadrupolar}), is that it allows for double spin-flip transitions, $\Delta \spinNucleus_z =\pm 2$, in addition to single-flip ones, $\Delta \spinNucleus_z =\pm 1$. The multiple resonance peaks, corresponding to Raman-transition detuning equal to once \emph{and twice} the nuclear Zeeman energy, were observed in Refs.~\cite{bodey_optical_2019, gangloff_quantum_2019, gangloff_witnessing_2021}.}
 Finally, the Knight field from the electron is fast oscillating which averages out its components perpendicular to the external magnetic field. The remaining component is along the external magnetic field and does not give any deflection. In sum, for experiments without a micromagnet the effective transverse gradient entering Eq.~\eqref{eq:XM} should be assigned a value according to a conversion formula
\be
\label{eq:BperpChange}
\frac{l \nabla_\perp B}{B} \to \frac{\Sigma}{\omega_i},
\ee
with $\Sigma$ somewhat smaller than the one given by Eq.~\eqref{eq:frequencyDispersion}.

We now turn to the frequency of the electron as another source of uncertainty in Eq.~\eqref{eq:Hartmann-Hahn}. Copying the formula here again, the electron Rabi frequency is $\angularFrequencyRabi = \sqrt{(\angularFrequencyRabiResonant)^2 + (\angularFrequencyRF - \angularFrequencyElectron)^2}$. 
First, during the driving the Overhauser field will diffuse, changing the electron Larmor frequency $\angularFrequencyElectron$. However, for pulses of order microseconds, we find that the resulting shift is smaller than a few $2\pi\times$kHz and thus negligible for the discussion here.\footnote{\label{fnt:diffusion}For $T_\mathrm{pulse}=1$ $\mu$s, we estimate the diffusion-induced variance of the Overhauser field, $\Sigma_B$, of $2\pi\times 8$ kHz from the measurements of Ref.~\cite{nakajima_coherence_2020}, $2\pi\times 7$ kHz from Ref.~\cite{shulman_suppressing_2014}, or $2\pi\times 6$ kHz from Ref.~\cite{delbecq_quantum_2016} (values for GaAs).} 
More importantly, within a finite time interval $T$, no frequency can be defined with uncertainty much below $\delta \omega \sim 1/T$.\footnote{The numerical prefactor $c$ to use in the relation $\delta \omega = c \times 1/T$ is not obvious. We define it by demanding $\int_{-\delta \omega}^{\delta \omega} f(\omega) \mathrm{d}\omega =1/2$, with $f(\omega)$ being the spectral density. For $f$ equal to a Lorenzian, such as Eq.~\eqref{eq:lorenzianDensity}, one has $\delta \omega=\Sigma$ and thus $c=1$. For $f$ equal to the left-hand side of Eq.~\eqref{eq:promotionToDeltaFunction} with $t=T$, we get $\delta \omega \approx \pi/2\times 1/T$, a value that we adopt in plots.} A Rabi pulse applied for 1 $\mu$s gives $\delta \omega \sim 2\pi \times 160$ kHz. This smearing should be assigned to the equality sign in Eq.~\eqref{eq:Hartmann-Hahn}, rather than to any individual frequency, but let us interpret it as an effective electron lifetime. In general, one considers it together with the lifetime of Rabi oscillations, or the Rabi decay time $T_2^\mathrm{Rabi}$,\footnote{We use the notation of Ref.~\cite{stano_review_2022}. The Rabi decay $T_2^\mathrm{Rabi}$ is contributed by the decay and decoherence times in the rotated frame, often denoted by $T_{1\rho}$ and $T_{2\rho}$ \cite{laucht_dressed_2016}, the former introduced by Ref.~\cite{redfield_nuclear_1955} denoted therein as $T_{2e}$.} adding $(\pi/2)\times 1/T_\mathrm{pulse}$ and $1/  T_2^\mathrm{Rabi}$ in square. Nevertheless, since the latter is negligible in our scenario, we define $\Sigma_\mathrm{p}= (\pi/2) \times 1/T_\mathrm{pulse}$.\footnote{Previous works on DNSP arising from ESR in quantum dots \cite{danon_nuclear_2008,danon_multiple_2009,rudner_electrically_2007} considered that $1/T_2^\mathrm{Rabi}$ as just described dominates all other time-decay or frequency-smearing scales. These works do not even consider the nuclear hyperfine energy. This approximation was probably motivated by early experiments \cite{koppens_driven_2006, nowack_coherent_2007} where only a few Rabi oscillations were discernible. More recently, Rabi oscillations of single spins of much higher quality were achieved: the decay time $T_2^\mathrm{Rabi}$ was larger than the Rabi oscillation period by the factor 42.5 in Ref.~\cite{nakajima_coherence_2020} (GaAs), 70 in Ref.~\cite{takeda_fault-tolerant_2016} (natural Si) and 444 in Ref.~\cite{yoneda_quantum-dot_2017} (isotopically purified Si). In other words, for current experiments, it might be reasonable to assume $T_\mathrm{pulse} \ll T_2^\mathrm{Rabi}$.}
 An important difference to the sources in Eq.~\eqref{eq:frequencyDispersion} discussed in the previous paragraph is that this type of smearing, essentially originating from the Heisenberg uncertainty relation, leads to a Lorenzian, rather than a Gaussian, spectral density
\begin{equation}
\label{eq:lorenzianDensity}
F_\Sigma(\omega) = \frac{1}{\pi} \frac{\Sigma}{(\omega-\omega_R)^2 + \Sigma^2}.
\end{equation}
This smearing could be included in the main result, Eq.~\eqref{eq:rateDNSP}, by replacing the spectral density in Eq.~\eqref{eq:gaussianDensity} by the convolution
\begin{equation}
\label{eq:convolutionGF}
G_\Sigma (\omega_R) \to \left( G_\Sigma \star F_{\Sigma_\mathrm{p}} \right) (\omega_R).
\end{equation}
However, we will not use Eq.~\eqref{eq:convolutionGF}.
Since the reasoning that lead to both Eq.~\eqref{eq:gaussianDensity} and Eq.~\eqref{eq:lorenzianDensity} was only qualitative, dwelling on an exact expression in Eq.~\eqref{eq:convolutionGF} is not meaningful. Instead, we simply add the finite-lifetime smearing $\Sigma_\mathrm{p}$ into the list in Eq.~\eqref{eq:frequencyDispersion} and use that as the width of the spectral density function entering Eq.~\eqref{eq:mainResult} with either the Gaussian or the Lorenzian profile.

To complete the list of smearing mechanisms, note that due to the nuclear and electrical noise, in an experiment the electron detuning frequency varies with time and thus can be known and controlled only approximately. In experiments employing estimation and feedback, similar to the one producing the data in Fig.~\ref{fig:data-fit}, the resulting uncertainty was $2\pi\times$288 kHz in Ref.~\cite{nakajima_coherence_2020}, several hundreds of $2\pi\times$kHz in Ref.~\cite{delbecq_quantum_2016} and several times $2\pi\times$78 kHz (the frequency bin) in Ref.~\cite{shulman_suppressing_2014}. This uncertainty is yet another source of averaging: The experimentally measured polarization rate corresponds to
\begin{equation}
\langle \partial_t p_i \rangle (\angularFrequencyDetuning) = \int_{-\infty}^\infty \mathrm{d} \omega_\mathrm{err} \, G_{\Sigma_\mathrm{err}}(\omega_\mathrm{err}) \, \partial_t p_i (\angularFrequencyDetuning+\omega_\mathrm{err}) ,
\end{equation}
where $\Sigma_\mathrm{err}$ is the precision with which the detuning angular frequency can be fixed during the collection of data assigned to a single point on the curve such as plotted in Fig.~\ref{fig:data-fit}. This averaging is different from the previous two, since now it is not only the spectral density that is smeared, but also the angle $\gamma$ dependency that is averaged. Therefore, it would suppress the anti-symmetric-in-$\gamma$ parts of the polarization rates, which can be identified easily by looking at Eqs.~\eqref{eq:XM}--\eqref{eq:pe}.

To conclude, there are three different types of averaging that need to be done with Eq.~\eqref{eq:rateDNSP}: a Gaussian and a Lorentzian smearing of the spectral function, and a Gaussian averaging of the whole formula. Roughly, we replace them by adding all the smearing sources to $\Sigma$ used in \eqref{eq:gaussianDensity}.

With the polarization rate derived and analyzed in detail, we next move to examining system dynamics in the presence of DNSP pumping.

\section{Polarization-rate profile, system dynamics, and feedback}

\label{sec:feedback}

In this section, we look at three topics. First, we illustrate the polarization-rate magnitude expected in a typical quantum dot, and discuss the rate inversion-symmetry with respect to the zero detuning $\frequencyDetuning=0$. Second, we examine polarization-rate feedback induced by changes in the detuning aiming at a substantial nuclear polarization. Third, we analyze the effects of the feedback on suppressing or enhancing detuning fluctuations, which influence qubit gate fidelities. 

\subsection{Polarization rate profile}

\begin{figure}
\begin{center}
\includegraphics[width=\linewidth]{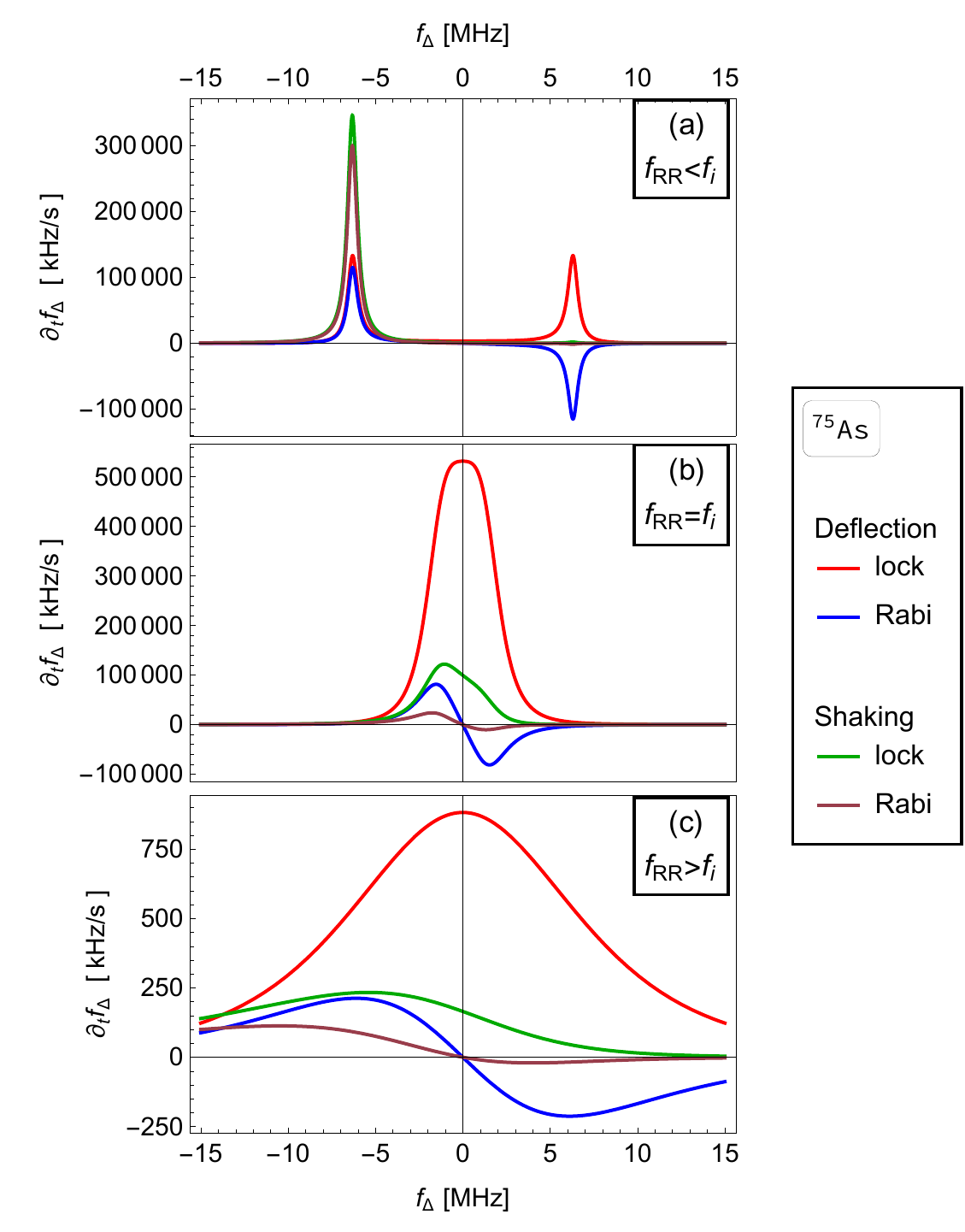}
\end{center}
\caption{The polarization rate as a function of the electron detuning frequency for $^{75}$As. In the upper panel $\frequencyRabiResonant = f_i/2$, in the middle panel $\frequencyRabiResonant = f_i$, and in the lower panel $\frequencyRabiResonant =2 f_i$. 
Further parameters are as in Fig.~\ref{fig:data-fit} except for $T_\mathrm{pulse}=10$ $\mu$s.}
\label{fig:As}
\end{figure}

We illustrate the behavior of the polarization rate derived in Eq.~\eqref{eq:mainResult} by plotting it for the arsenic isotope as a function of the detuning in Fig.~\ref{fig:As}. Analogous plots for other nuclei of GaAs, and for a Si dot where the rate is orders of magnitude smaller, are in Appendix~\ref{app:analogs}. Figure \ref{fig:As}a shows the rate in the regime where the electron Rabi frequency at resonance is smaller than the nuclear Larmor frequency. The rate has a resonant peak at a finite detuning, where the electron Rabi and nuclear Larmor frequencies become equal. At this resonance the rate can reach large values, depending on the resonance width, which has been discussed in Sec.~\ref{sec:extensions}. The deflection mechanism corresponds to a rate with a definite left-right symmetry in the figure, symmetric for a lock-in initial state and antisymmetric for an initial state along the magnetic field.\footnote{In other words, the EDSR-scenario curves cross zero at zero detuning, due to the factor $p_e=\pm \sin \gamma$. Polarization rates with this profile are called 'cooling functions' in Refs.~\cite{gangloff_quantum_2019,gangloff_witnessing_2021}. In those experiments, the polarization is explained as due to asymmetry in the density of final states \cite{yang_collective_2012,xu_optically_2009}.}
The shaking mechanism corresponds to a strongly asymmetric rate, with appreciable values at negative detunings only.

Figure \ref{fig:As}c shows the case with the Rabi frequency at zero detuning larger than the nuclear Larmor frequency. In this case, the condition of the Hartmann-Hahn resonance can not be reached for any detuning. While the symmetry properties of the rate components discussed in the previous paragraph still hold, there is no resonance peak and the rates are much smaller overall. This difference, between the resonant and nonresonant regime, is the larger the larger is the ratio $\frequencyRabiResonant / f_i$. Finally, at large detuning the rates fall off as $1/\frequencyDetuning^2$, which is the same in the upper panel, though hard to see there because of the resonant peak. 

Figure \ref{fig:As}b shows the crossover case $\frequencyRabiResonant = f_i$. Here, the two resonance peaks visible in the upper panel merge into one. Compared to those two resonances, the merged peak is broader and (for spin locking) has a somewhat anomalous shape (it has a flat top). This property can be understood by noting that the derivative $\partial \frequencyRabi /\partial \frequencyDetuning$ becomes zero at zero detuning.

\subsection{Feedback}

Equation \eqref{eq:dfdt} hints at feedback effects. The detuning frequency $\frequencyDetuning$ changes if the nuclear polarization changes, since the electron feels it as the Overhauser field. However, the polarization rates themselves strongly depend on the detuning. Such mutual dependence of the nuclear polarization rate and its effects, the accumulated nuclear polarization, has been studied at length (see the second paragraph of the introduction and the references therein). 

Motivated by those works, we now look at the feedback effects in our system. We start by pointing out one crucial difference. Here, the DNSP polarization is a resonance phenomenon, so that the dependence of the polarization rate on the electron detuning might become (close to resonance) much more sensitive than the dependence in the Pauli spin blockade setups \cite{rudner_nuclear_2011}. While this fact will make building up large polarizations more difficult, it might allow for more efficient Overhauser field stabilization and the associated dephasing suppression.

To appreciate this sensitivity, we copy here Eq.~\eqref{eq:dfdp-app} derived in Appendix~\ref{app:conversion} [it also follows from Eq.~\eqref{eq:dfdt}]
\be
\frac{\partial \frequencyDetuning}{\partial p_i } = -\frac{\mathrm{sgn}(g_e)}{2 \pi \hbar} \fractionIsotope |A_i| I_i.
\label{eq:dfdp}
\ee
This equation relates changes in the nuclear polarization $p_i$ to changes in the electron detuning \emph{at a fixed value of the driving frequency}. Evaluating the constants on the right-hand side, we get 25 MHz in natural silicon (it would be sixty times less in isotopically-purified 800 ppm silicon), and from about 7 to 17 GHz for the three isotopes in GaAs. Therefore, especially in the latter material, a tiny change in the nuclear polarization---say a few of 0.01\%---can bring the system into and out of the resonance, turning on and off the DNSP rate.

\begin{figure}
\begin{center}
\includegraphics[width=0.9\linewidth]{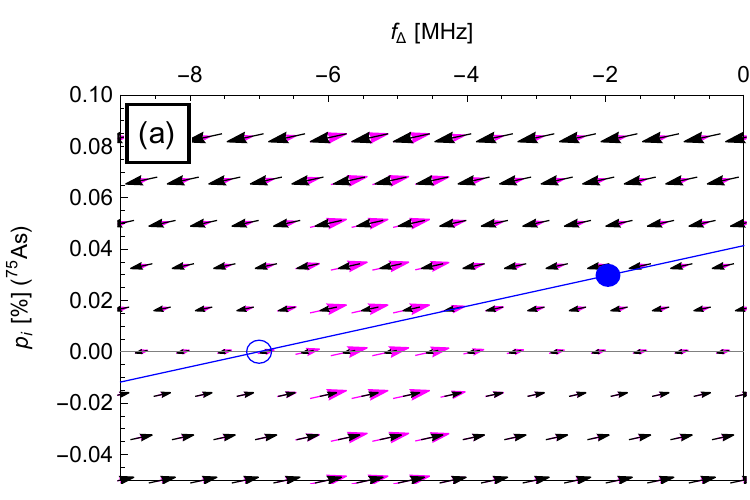}
\includegraphics[width=0.88\linewidth]{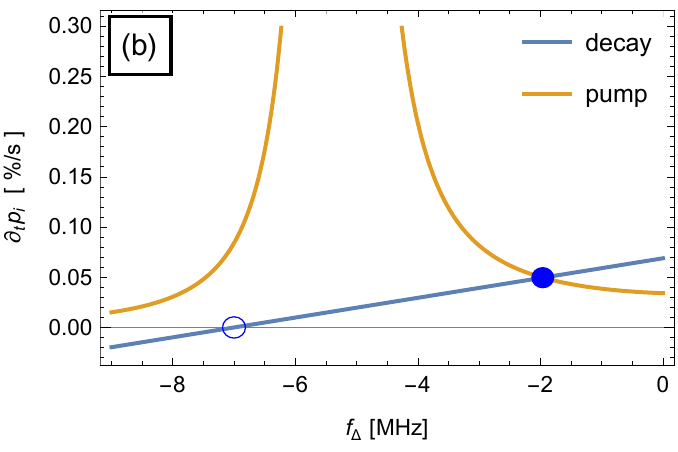}
\end{center}
\caption{\textbf{System dynamics under DNSP and decay.} The parameters are the same as in Fig.~\ref{fig:As} except for $\frequencyRabiResonant=5$ MHz. (a) The two axes give the system state coordinates: the electron detuning on the horizontal axis and the nuclear polarization on the vertical axis. The vectors show the direction and rate at which the system moves from a given configuration. The colored arrows depict the DNSP rate for $i= ^{75}$As, the black arrows denote the spin decay into $p_i=0$. At a fixed driving frequency the system can move along a line since all vectors throughout the plot are parallel. The position of the line is fixed by the detuning at zero nuclear polarization (empty circle), here fixed to $-7$ MHz. (b) The polarization ['pump'; the right-hand side of Eq.~\eqref{eq:rateDNSP}] and decay [$R p_i$, the right-hand side of Eq.~\eqref{eq:decayRateR}] rates at fixed driving frequency. The equilibrium is where the two rates are equal, denoted by the filled circle. \\
In (a) the arrows are scaled for visibility: While a larger arrow means a larger rate, the proportionality is not linear for a given color and not to scale between different colors. The arrows' map is only illustrative. The rate magnitudes are quantitative in (b). 
\label{fig:quiverPlot}
}
\end{figure}

To shed light on the possible system dynamics, we plot the DNSP rates in Fig.~\ref{fig:quiverPlot} in a two-dimensional plot. We assume the 'spin locking' scenario, see Eq.~\eqref{eq:pe}, where the rates are somewhat larger than for the 'EDSR' choice.\footnote{A feedback exploiting the EDSR scenario was implemented in Ref.~\cite{jackson_optimal_2022}.} The horizontal axis is the detuning, the vertical the nuclear polarization. The colored arrows show the polarization rate: the arrow length scales with the rate magnitude and the arrow direction shows which way the system evolves at a fixed driving frequency. The black arrows represent nuclear spin-polarization decay, due to diffusion or other means, according to
\be
\label{eq:decayRateR}
\partial_t p_i =-R p_i.
\ee
The decay constant $R$ depends on the material nuclear spin diffusion constant, the dot geometry, and possibly on the isotope. Since these dependencies might be complicated,\footnote{For example, Ref.~\cite{malinowski_spectrum_2017} converts the observed Overhauser field dynamics into the effective material diffusion constant and finds that its value changes strongly with the magnetic field.} it is more practical to extract the decay scale $R$ from experimental data rather than to calculate it from first principles. Typical decay times of nuclear polarization in dots is from seconds (see Ref.~\cite{malinowski_spectrum_2017} or the estimates of parameter $\kappa$ in Appendix~\ref{app:randomWalk}) to minutes (see Fig.~3 in Ref.~\cite{nakajima_coherence_2020} or Fig.~3e in Ref.~\cite{jang_wigner-molecularization-enabled_2022}).

One can understand the system behavior from Fig.~\ref{fig:quiverPlot}a. As a simple example, it shows the rate for $^{75}$As isotope in a GaAs dot with the driving frequency fixed to a certain value corresponding to the detuning $-7$ MHz at zero nuclear polarization. This state is denoted by the empty circle in Fig.~\ref{fig:quiverPlot}. With the driving frequency fixed, the system can move only along the blue line. It will reach a steady state at a finite positive polarization $p=p_{eq}$ where the polarization and decay rates are equal (they are shown in Fig.~\ref{fig:quiverPlot}b). The system will stay at such finite polarization as long as random (thermal) fluctuations do not take it out of the window where the DNSP rate is sizable. The larger the value of the equilibrium polarization $p_{eq}$, the stronger the forces on the system at the equilibrium and the smaller the fluctuations around the steady state \cite{danon_nuclear_2008,tenberg_narrowing_2015}.\footnote{The decrease of fluctuations when the forces become larger can be also understood from the model in Appendix~\ref{app:randomWalk}: Eq.~\eqref{eq:DOmega} states that the fluctuations $\sigma_\Omega^2$ are proportional to the inverse of the decay rate $2/\kappa$.} On the other hand, also the more volatile the steady state becomes and the more easily it can be kicked off by thermal fluctuation into $p=0$. In other words, if $p_{eq}$ is large enough, the system will be bistable. What is large enough is decided by the width of the resonance, in turn given also by the inverse of the micromagnet gradient $B_{||}$ and additional sources according to the discussion around Eq.~\eqref{eq:frequencyDispersion}.

One can consider more complicated evolutions when the driving frequency is changed. A change in the driving frequency translates into a horizontal shift of the blue line. When the system state is represented by the filled circle in the figure, a sudden change of the driving will move it together with the blue line horizontally, that is, keeping the current value of the polarization. In changes that are more adiabatic, the system state will tend to follow the local equilibrium position on the blue line. A simple scenario would be a slow increase of the rf-frequency, starting at a negative detuning $\frequencyDetuning = \frequencyRF - \frequencyLarmorElectron$. The polarization would steadily increase until the equilibrium polarization would become too large to be sustained.  The required speed of change of the driving frequency can be read off from Fig.~\ref{fig:quiverPlot}b, or directly from a plot like Fig.~\ref{fig:As}: the optimal speed to built a large polarization is a value somewhat smaller than the polarization rate at the peak, which is a few hundreds of MHz/s for these parameters.

\subsection{Restoring force}

\label{sec:restoringForce}

To elaborate on the previous section, we next consider the electron spin coherently driven by EDSR with the goal of performing a qubit gate. One typical situation is that the electron is driven at zero detuning and starts polarized along the external field. It differs from the previous by having now $p_e=\pm \sin \gamma$. We are interested in how the arising DNSP polarization affects gate precision. Specifically, we analyze the DNSP influence on the stability of the desired condition $\frequencyDetuning=0$. Using Eqs.~\eqref{eq:mainResult} and \eqref{eq:dfdt}, we get
\be
\begin{split}
\label{eq:backforce}
\partial_t \frequencyDetuning^* & = \mathrm{sgn}(p_e^*) \mathrm{sgn}(g_e)\frac{\frequencyDetuning^*}{\frequencyRabiResonant} \left(X_\mathrm{df}^{*2} + X_\mathrm{sh}^{*2} \right) \frac{1}{6\hbar^3} \\
&\qquad \times
\sum_i \fractionIsotope |A_i| I_i(I_i+1)  G_\Sigma\left(\angularFrequencyRabiResonant - \omega_i \right).
\end{split}
\ee
To arrive at these formulas, we have used Eq.~\eqref{eq:peForEDSR}, expanded Eq.~\eqref{eq:rateDNSP} in the limit around $\gamma\approx 0$, and, to simplify, dropped the polarization $p_i$ from the right-hand side and used $\alpha_I$ for the small $p_n$ limit. The star as the superscript denotes a relation to the limit $\gamma \to 0$. Specifically, for the matrix elements $X_\mathrm{sh}$ and $X_\mathrm{df}$ the star means that they are evaluated using Eqs.~\eqref{eq:XM} and \eqref{eq:XT} with $\gamma=0$. Also, we have added the initial state specification as $\mathrm{sgn}(p_e^*)$ with the value $+1$ for the ground state and $-1$ for the excited state. Finally, we also note that except of $g_e$, $p_e^*$, and $\frequencyDetuning$, quantities in the expression are positive. 

Equation \eqref{eq:backforce} describes a simple feedback, since the rate of change of the detuning is proportional to the detuning value. Whether the feedback is negative (fluctuations suppressed) or positive (fluctuations amplified) is decided by the overall sign, the product of signs of the electron $g$ factor $g_e$ and the initial state $p_e^*$. This latter product can be contracted to 'electron spin initially along $\bFieldElectron$' being the sign $-1$ (negative feedback) and 'electron spin initially opposite to $\bFieldElectron$' being the sign +1 (positive feedback).\footnote{The fact that the feedback switches from positive (`resonance seeking') to negative (`resonance avoiding ) upon inverting the electron spin was pointed out in Ref.~\cite{hogele_dynamic_2012}.} Let us first discuss the first alternative.

A negative feedback means that driving the electron spin \emph{stabilizes} the desired condition $\frequencyDetuning=0$. To quantify this effect, we write Eq.~\eqref{eq:backforce} in the form
\be
\label{eq:feedback}
\partial_t \frequencyDetuning^* = -\Gamma^* \frequencyDetuning^*,
\ee
introducing $\Gamma^*$ as the feedback strength with the units of inverse time. To assess how efficient the stabilization is, we judge it against the intrinsic thermal fluctuations of the nuclei. However, the comparison is not straightforward, since these thermal fluctuations proceed as a diffusion of the Overhauser field, characterized by a diffusion constant, which is not a rate. To bridge this gap, in Appendix~\ref{app:randomWalk} we describe this diffusion by a bounded random walk model, which contains two parameters: the diffusion constant $D_\Omega$ and a time $\kappa$ related to the restoring force that keeps the Overhauser-field fluctuations bounded.

The behavior of the system can be then understood as follows: Let us assume that the detuning is set to the desired value $\frequencyDetuning=0$. At this value, the polarization rate is zero. The detuning will diffuse away from the desired condition according to the diffusion constant $D_\Omega$. This short-time diffusion speed is not affected by the DNSP and the feedback. Without any feedback, the Overhauser field will reach the long-time variance $\sigma_\Omega^2 = D_\Omega \kappa/2$. In Appendix~\ref{app:randomWalk} we show that the restoring force in this process can be represented in a form identical to Eq.~\eqref{eq:feedback} upon identifying the constant $\Gamma$ with $1/\kappa$. Thus, one can assign an intrinsic restoring force $\Gamma_0 \equiv 1/\kappa_\mathrm{th}$ to the thermal diffusion. The DNSP feedback increases the restoring force by adding $\Gamma^*\equiv 1/\kappa_\mathrm{fb}$ to the intrinsic component. The fluctuations are then described by the variance
\be
\sigma_\Omega^{*2} = D_\Omega \left( \frac{1}{\kappa_\mathrm{th}} + \frac{1}{\kappa_\mathrm{fb}} \right)^{-1}.
\ee
To quantify the efficiency, one should compare the intrinsic rate $\Gamma_0$ to the one due to feedback $\Gamma^*$. If $\Gamma^* \ll \Gamma_0$, the feedback is negligible. If $\Gamma^* \gg \Gamma_0$, the feedback substantially decreases the magnitude of the fluctuations, cutting the resulting variance by factor $\Gamma^* / \Gamma_0$.

\begin{figure}
\begin{center}
\includegraphics[width=0.99\linewidth]{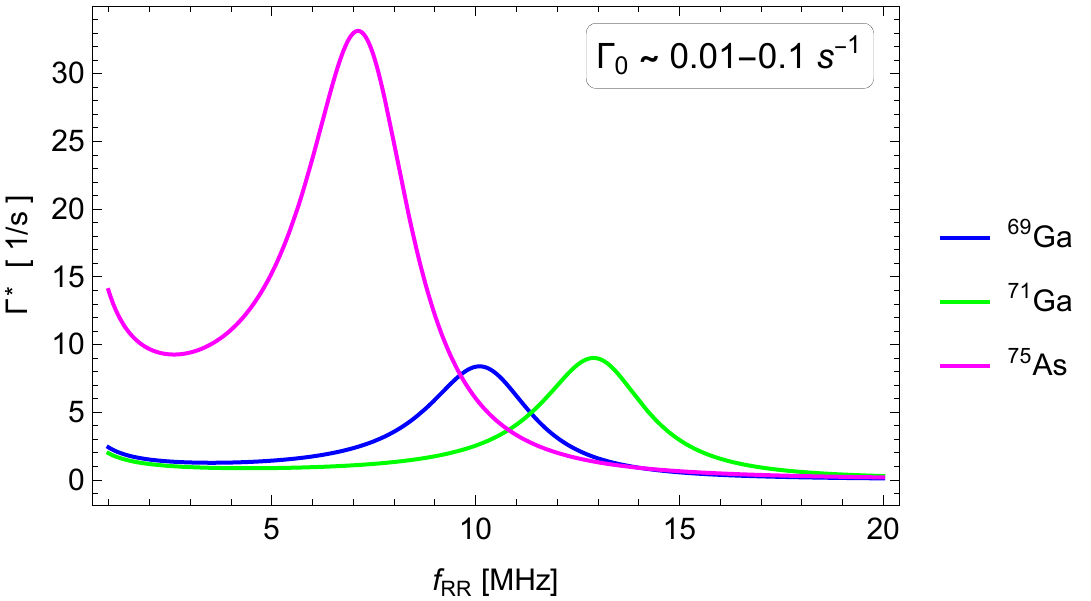}
\end{center}

\caption{\textbf{Stabilization by feedback in GaAs}. The EDSR-driven electron starts in its ground state and precesses around an in-plane axis of the Bloch sphere, meaning $\frequencyDetuning = 0$. The plot shows the restoring rate $\Gamma^*$ as a function of the Rabi frequency at resonance for the three isotopes. We took $T_\mathrm{burst} = T_\mathrm{cycle} = 1$ $\mu$s, zero additional broadening, and the remaining parameters as in Fig.~\ref{fig:data-fit}. The value in the box is the expected range for the intrinsic restoring force $\Gamma_0= \kappa_\mathrm{th}^{-1}$. The analogous figure for an electron qubit in Si is in Appendix~\ref{app:analogs6} and for a hole qubit in SiGe in Appendix~\ref{app:hole}. 
\label{fig:feedback-1-GaAs}
}
\end{figure}

We plot the quantity $\Gamma^*$ in Fig.~\ref{fig:feedback-1-GaAs}. It shows that the DNSP-induced feedback might be indeed substantial, with $\Gamma^*$ larger than $\Gamma_0$ by up to two or three orders of magnitude for our parameters at the resonance with the arsenic isotope. The effect on the gate fidelity is more complicated, since the feedback depends on the initial state. While for some input states the feedback improves the fidelity by stabilizing the detuning, the effects are opposite for other input states. Concerning gate fidelities, it seems advisable to keep the system away from the Hartmann-Hahn resonance.

\subsection{Closing remarks}

We note that the simplified picture presented in the above by discussing a single isotope is complicated in GaAs by having three different isotopes with different resonance frequencies. Still, the DNSP rates depend, through the actual value of the detuning, only on the sum of the corresponding Overhauser fields. The simplest-looking scheme to stabilize the total Overhauser field is to use the Hartmann-Hahn resonance of a single isotope with the most efficient polarization, being $^{75}$As in our estimates.

Let us also reiterate a point crucial for both observing and exploiting the DNSP rates discussed in this paper. As already stressed several times, these rates are resonance phenomena, sensitive to the electron detuning, in turn to its Larmor frequency. A change in the frequency by a few MHz can substantially change the polarization rates. Therefore, adjusting the driving frequency to the instantaneous value of the electron Larmor frequency is essential. It can be possibly done by periodic estimation of this frequency \cite{shulman_suppressing_2014,nakajima_coherence_2020}. Another possibility is to use chirps of the driving frequency\cite{shafiei_resolving_2013,henstra_dynamic_2014}. 

Let us conclude by saying that there is room for more investigations of feedback effects based on Hartmann-Hahn DNSP in gated quantum dots.

\section{Conclusions}

In this paper, we have investigated dynamical nuclear spin polarization arising in a quantum dot with a single electron whose spin is electrically driven to perform Rabi oscillations. We considered the coherent regime where many Rabi oscillations happen before the electron leaves the dot or its spin decoheres. In this regime, the electron spin can polarize nuclear spins in the quantum dot volume through an analog to the Hartmann-Hahn effect known from NMR \cite{hartmann_nuclear_1962}. This is a resonance phenomenon, occurring when the electron Rabi frequency becomes equal to the nuclear Larmor frequency.

We have derived the corresponding nuclear-spin polarization rate under general conditions in Sec.~\ref{sec:rate} so that the main result, Eq.~\eqref{eq:mainResult}, covers both GaAs and Si dots, and, with slight adjustments,\footnote{We give the main result, Eq.~\eqref{eq:mainResult}, assuming isotropic hyperfine tensor, $\propto \spinElectronVector \cdot \spinNucleusVector$. It remains unchanged for `secular' hyperfine tensor, $\spinElectron_z \spinNucleus_z$. If additional, `non-secular', terms are present, often the case for holes, they will generate additional terms in Eq.~\eqref{eq:X}, which need to be reflected in Eq.~\eqref{eq:mainResult} using Table~\ref{tab:ManisotropicJ}. Also, since micromagnets are not needed for hole spin qubits \cite{bulaev_electric_2007,froning_ultrafast_2021,bosco_phase_2023}, the micromagnet-related parameters entering Eq.~\eqref{eq:mainResult} need to be reinterpreted as discussed in Sec.~\ref{sec:extensions}, see especially Eqs.~\eqref{eq:frequencyDispersion} and \eqref{eq:BperpChange}.} even Ge or Si hole dots. 

When converted to changes in the electron detuning from the Rabi resonance, the nuclear-spin polarization rates in GaAs reach tens to hundreds of MHz/s. The theory fares well with a preliminary measurement in a GaAs sample presented in Fig.~\ref{fig:data-fit}. In Si, the rates are orders of magnitude smaller. While we do not present data for Si, our theory predicts rates of order tens of kHz/s.

We have identified two essential differences to the standard Hartmann-Hahn scenario: (1) The micromagnet magnetic-field gradient slightly deflects the spin-quantization axes of the electron and nuclei. (2) The electric driving slightly wiggles the electron with respect to the atomic lattice. These two effects correspond to two different mechanisms of polarization. In Sec.~\ref{sec:extensions}, we have reasoned that the polarization will be present even in samples without a micromagnet, and we have provided estimates with which a polarization rate can be assigned to this scenario.

Finally, we have analyzed the feedback in the system. It stems from the fact that the polarization rate is sensitive to the electron detuning from the Rabi resonance, which in turn is sensitive to the accumulated nuclear polarization through the Overhauser field. In Sec.~\ref{sec:feedback}, we have looked at the possibility of reaching a sizable nuclear polarization and the consequences of the Hartmann-Hahn resonance on the fidelity of a gate implemented as coherent Rabi precession. Concerning the first, the achievable nuclear polarization is ultimately set by how sharp the resonance can be made, in turn dependent on the electron spin coherence and the micromagnet gradient. If used as active feedback, we estimate that exploiting the resonance can decrease the fluctuations of the Overhauser field by two or more orders of magnitude (in GaAs). Concerning the gate fidelities, we have found that it is improved for some input states and worsened for others. We do not evaluate the fidelities, and remain at the advice of avoiding the resonance when implementing quantum gates on an electron or hole spin qubit.

\vspace{-0.3cm}

\acknowledgments

\vspace{-0.3cm}

PS would like to thank Minoru Kawamura for a useful discussion. We thank for the financial support from CREST JST grant No.~JPMJCR1675, JST Moonshot R\&D grant No.~JPMJMS226B-1, JST PRESTO grant No.~JPMJPR2017, JSPS KAKENHI grant No.~18H01819 and from the Swiss National Science Foundation and NCCR SPIN grant No.~51NF40-180604. 

\vspace{-0.3cm}

\appendix

\section{Density of nuclear Larmor frequencies}
\label{app:nuclearDensity}

\begin{figure}
\begin{center}
\includegraphics[width=0.8\linewidth]{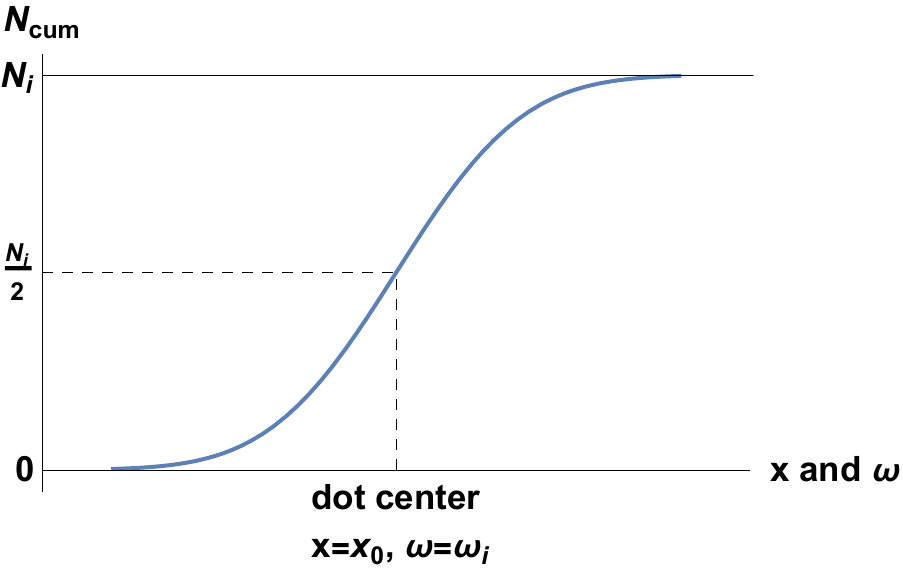}
\end{center}

\caption{\textbf{Cumulative distribution of the effective number of nuclei}. The horizontal axis shows both the real space coordinate $x$, here along the gradient of the longitudinal component of the magnetic field, and the nuclear Larmor frequency. The frequency increases monotonically with $x$ and in the dot center it equals $\omega_i$.
\label{fig:cumulativeG}
}
\end{figure}

In this section, we introduce the distribution $g$ used in Eq.~\eqref{eq:GammaTotalAsSum} to replace the summation over discrete nuclei. The quantity being summed contains a factor $v_0^2 |\Psi_n|^4$, arising from the square of the right-hand side of Eq.~\eqref{eq:hyperfine}. Here, we have denoted $\Psi_n \equiv \Psi(\mathbf{r}_n,z_n)$. Therefore, let us consider
\be
\sum_{n \in \mathrm{set}} v_0^2|\Psi_n|^4 \times \mathrm{constants},
\ee
where the `set' defines which nuclei are included. In our case, it is nuclei of isotope $i$ with a given Larmor frequency. Also, `constants' are further terms that can depend on the isotope, but not on the nucleus spatial position. Since such constants only propagate through all the formulas below, we omit them. We rewrite the sum as 
\be
\frac{1}{N_\mathrm{tot}^2} \sum_{n \in \mathrm{set}}  N_\mathrm{tot}^2 v_0^2|\Psi_n|^4.
\ee
taking out a dimensionless factor $N_\mathrm{tot}^{-2}$. In line with the existing literature, we move this factor into the matrix elements $X$, see Eqs.~\eqref{eq:XM}-\eqref{eq:XT}. These matrix elements then contain the `average' hyperfine strength $A_i/N_\mathrm{tot}$. The dividing factor, written suggestively as  $N_\mathrm{tot}$, is interpreted as the total (counting all isotopes) effective number of atomic nuclei within the dot, defining it by
\be
N_\mathrm{tot} = \frac{1}{v_0 \int \mathrm{d}\mathbf{r} \, \mathrm{d}z |\Psi(\mathbf{r},z)|^4}.
\label{eq:Ntot}
\ee
With this rescaling, the sum that we are interested in is
\be
\sum_{n \in \mathrm{set}}  N_\mathrm{tot}^2 v_0^2|\Psi_n|^4.
\ee
Since the nuclear Larmor frequency is a smooth function of the nuclear position, we replace the discrete summation by integration in space with the three-dimensional volume element $\mathrm{d}V$. As we only include the isotope $i$, the volume density of nuclei is $\fractionIsotope/v_0$. We get 
\be
\int_D \mathrm{d}V \frac{\fractionIsotope}{v_0} N_\mathrm{tot}^2 v_0^2|\Psi|^4,
\ee
where the restriction $n\in \mathrm{set}$ has been expressed as a volume $D$. We define $N_i \equiv \fractionIsotope N_\mathrm{tot}$ as the effective number of isotope-$i$ nuclei, and use Eq.~\eqref{eq:Ntot} to finally get
\be
\mathrm{d}N_i = N_i \frac{ \mathrm{d}V |\Psi|^4}{\int \mathrm{d}V |\Psi|^4},
\ee
as the effective number of nuclei of isotope $i$ within a volume element $\mathrm{d}V$. In this expression, the denominator normalizes the density $\mathrm{d}V N_i |\Psi|^4$ into a dimensionless quantity. If integrated over all space, it gives the effective number of isotope-$i$ nuclei in the dot. 

We now consider the desired restriction on the nuclei included in the sum, being a given value of their Larmor frequency. The latter is proportional to the magnitude of the magnetic field at the position of the corresponding nucleus, $B_n$. In the lowest order of the magnetic-field gradients and neglecting the shifts along the $z$ coordinate, this magnitude varies linearly over the dot in-plane coordinates,
\be
B (x_n) \approx B (x_0) + (x_n-x_0) \nabla_{||} B, 
\ee
where we choose the in-plane coordinates such that $x$ is along the gradient of the magnetic field longitudinal component (the component along the direction of the magnetic field at the dot center; see also Appendix~\ref{app:derivationX}) and $\nabla_{||} B$ is the magnitude of this gradient. 

The restriction on the nuclear Larmor frequency is then a restriction on the $x$-coordinate, and we can integrate out the remaining two coordinates $y$ and $z$,
\be
\mathrm{d}N_i = N_i \frac{\sqrt{2}}{l \sqrt{\pi}} \exp\left( -\frac{2(x-x_0)^2}{l^2} \right) \mathrm{d}x,
\ee
where we have used the Gaussian form for the in-plane wave function, Eq.~\eqref{eq:wavefunction}.

The desired density can be now obtained from the cumulative distribution (see Fig.~\ref{fig:cumulativeG} for an illustration)
\be
\begin{split}
\int_{-\infty}^{\omega} g(\omega) \mathrm{d}\omega&= \int_{-\infty}^{x(\omega)} \frac{\mathrm{d}N_i}{N_i} \\
&=\int_{-\infty}^{x(\omega)}  \frac{\sqrt{2}}{l \sqrt{\pi}} \exp\left( -\frac{2(x^\prime-x_0)^2}{l^2} \right) \mathrm{d}x^\prime,
\end{split}
\label{eq:cumulative}
\ee 
where $x(\omega)$ is the coordinate at which the Larmor frequency is $\omega$. It can be obtained from the relation
\be
\frac{\omega - \omega_i}{\omega_i} = \Big(x(\omega)-x(\omega_i)\Big) \frac{\nabla_{||} B}{B},
\label{eq:xw}
\ee
where $\omega_i$ is the (isotope-dependent) nuclear Larmor frequency at the dot center, $x(\omega_i)= x_0$. 

Differentiating Eq.~\eqref{eq:cumulative} with respect to $\omega$, and using Eq.~\eqref{eq:xw}, we get
\be
\label{eq:gi}
g_i(\omega) = \frac{1}{\omega_i} \frac{B}{l\nabla_{||}B} \frac{\sqrt{2}}{\sqrt{\pi}}\exp\left[ -2\left(\frac{\omega-\omega_i}{\omega_i} \frac{B}{l\nabla_{||}B}\right)^2 \right].
\ee 
The first term sets the scale, the rest is a dimensionless peak profile centered at $\omega = \omega_i$. It encodes the resonance character of the problem: since the nuclear $g$ factors differ for different isotopes, they become resonant at different values of the electron Rabi frequency $\omega_R$. The width of resonance is a fraction of the nuclear Larmor frequency proportional to the gradient of the magnetic-field longitudinal component.

\section{DNSP rate expressed as the detuning change}
\label{app:conversion}

Here, we give the electron Larmor frequency including the contribution from the nuclear polarization. While the formulas might look too straightforward even for an appendix, they might be useful when considering materials with different signs of the $g$ factors.

The electron spin couples to the external magnetic field and the effective field arising from polarized nuclei, called also the Overhauser field,
\be
H_e^Z = g_e \mu_B \mathbf{B} \cdot \mathbf{s} + \sum_i \fractionIsotope A_i I_i p_i \axisEnergyNucleus \cdot \mathbf{s}.
\ee
We used the definition of the polarization $p_i$ to be along the unit vector $\axisEnergyNucleus = \mathrm{sgn}(g_n) \mathbf{B}/B$. With this, and using also $A_i =\mathrm{sgn}(g_i)|A_i|$, we can write the above Hamiltonian as
\be
H_e^Z = \mathrm{sgn}(g_e) |g_e| \mu_B \mathbf{B} \cdot \mathbf{s} \left(1 + \sum_i \fractionIsotope |A_i| I_i p_i \frac{\mathrm{sgn}(g_e)}{|g_e| \mu_B B} \right).
\ee
The electron Larmor frequency is the magnitude of the vector multiplying the electron spin operator,
\be
\hbar \angularFrequencyElectron =  |g_e| \mu_B B  \left|1 + \sum_i \fractionIsotope |A_i| I_i p_i \frac{\mathrm{sgn}(g_e)}{|g_e| \mu_B B} \right|.
\ee
Most often, the nuclear polarization is not so large as to make the Overhauser field bigger than the external field. Then, the second term inside the absolute value is smaller in magnitude than the first, making their sum positive and the absolute value sign unnecessary,
\be
\hbar \angularFrequencyElectron =  |g_e| \mu_B B  + \mathrm{sgn}(g_e) \sum_i \fractionIsotope |A_i| I_i p_i .
\ee
We can now covert the DNSP polarization rate into the rate of change of the electron Larmor frequency and the detuning,
\be
\partial_t \frequencyDetuning = -\partial_t \frequencyLarmorElectron = - \frac{\mathrm{sgn}(g_e)}{2\pi \hbar} \sum_i  \fractionIsotope |A_i| I_i \partial_t p_i,
\ee 
the first equation following from our definition $\frequencyDetuning = \frequencyRF - \frequencyLarmorElectron$. Finally, we note the relation between the change of the electron detuning with respect to the change in the nuclear polarization,
\be
\frac{\partial \frequencyDetuning}{\partial p_i } = -\frac{\mathrm{sgn}(g_e)}{2 \pi \hbar} \fractionIsotope |A_i| I_i.
\label{eq:dfdp-app}
\ee
It becomes useful when considering possible feedback in the system.

\section{Derivation of Eq.~(\ref{eq:pzI})}
\label{app:pzI}

\begin{figure}
\begin{center}
\includegraphics[width=\linewidth]{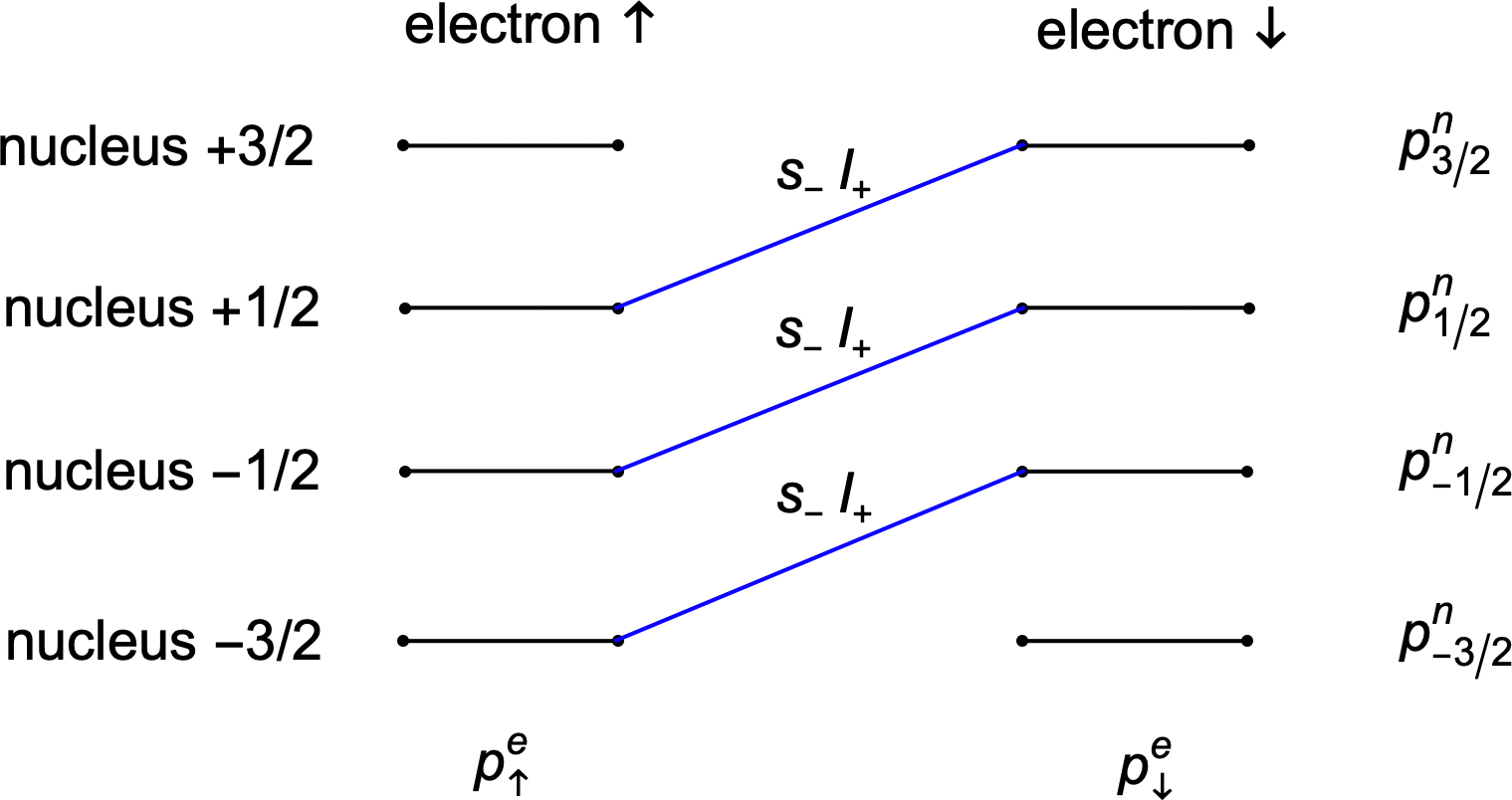}
\end{center}

\caption{\textbf{Spin polarization transitions}. The diagram shows the states of a system composed of an electron spin and a nuclear spin, the latter illustrated for $I=3/2$. The electron spin can be either up or down, with corresponding probabilities $p^e_\uparrow$ and $p^e_\downarrow$. The nuclear spin can be in one of the four states, with corresponding probabilities $p^n_j$. The matrix element $\mathcal{I}_+$ connects states as denoted by the blue lines. A transition increasing the nuclear polarization corresponds to going along one of the blue lines upwards, from left to right. 
\label{fig:pzI}
}
\end{figure}

Here, we derive Eq.~\eqref{eq:pzI}. The line over the left-hand side of that equation denotes the average over the probability distributions, or density matrices, of the electron and the nuclear spin. Figure \ref{fig:pzI} helps to visualize the transitions, and shows why the matrix element $\mathcal{I}_+$ should be averaged together with (and not independently to) the polarization $p_z$. 

To perform the calculation, one needs to quantify the probabilities of the basis states $|sj\rangle$. As explained in the main text, we consider them separable into the corresponding probabilities for the electron spin $s$ and the nuclear spin $j$, $p_{sj} = p^e_s p^n_j$. With the electron spin having only two states available, their probabilities can be expressed through a single number, let us denote it by $p_e$, because of the normalization $p^e_\uparrow + p^e_\downarrow =1$. Namely, 
\begin{subequations}
\label{eq:electronPolarization}
\begin{align}
p^e_\uparrow = \frac{1}{2}\left( 1 + p_e \right),\\
p^e_\downarrow = \frac{1}{2}\left( 1 - p_e \right).
\end{align}
\end{subequations}
These relations then define $p_e$ as the initial electron-spin-polarization along the axis $\axisEnergyElectronO$, and lead to Eq.~\eqref{eq:pe}. 

On the other hand, there might be more than two nuclear spin states in general. Still, we define the nuclear polarization by
\be
p_n \equiv \frac{\langle I_z \rangle}{I} =\frac{1}{I} \sum_{j=-I}^I j p^n_j.
\label{eq:pn}
\ee
This single number, together with the normalization $\sum_j p^n_j =1$, is not enough to specify the probabilities $p_j^n$ uniquely.\footnote{For example,  the occupations of the four sublevels of spin 3/2 got far from thermal distribution under the feedback employed in Ref.~\cite{gangloff_witnessing_2021}.} 
Nevertheless, starting with 
\be
\begin{split}
\overline{ p_z(0) |\mathcal{I}_+|^2} = \sum_j & p_{\uparrow j} |\langle \downarrow, j+1| s_- I_+ | j \uparrow \rangle|^2 \\
&- p_{\downarrow j} |\langle \uparrow, j-1| s_+ I_- | j \downarrow \rangle|^2,
\end{split}
\label{eq:pz01}
\ee
it is a few-line algebra to get
\be
\overline{ p_z(0) |\mathcal{I}_+|^2} = -\langle I_z \rangle + p_e \left[ I(I+1) - \langle I_z^2 \rangle \right].
\label{eq:pz02}
\ee
Equation~\eqref{eq:pz01} expresses the rate (proportionality factor) for building the nuclear polarization as the difference of the rates for transitions increasing the value of spin $j$ and the rate decreasing it, see Fig.~\ref{fig:pzI}. 
Using the definition of $p_n$ we can then write
\begin{subequations}
\label{eq:pz03}
\begin{align}
\overline{ p_z(0) |\mathcal{I}_+|^2} &=  I \times \left( p_e \alpha_I (p_n)  - p_n\right),\\
\label{eq:pz03b}
\alpha_I(p_n)&\equiv (I+1) -\langle I_z^2 \rangle /I.
\end{align}
\end{subequations}
For small nuclear polarization, one has $\langle I_z^2 \rangle = I(I+1)/3 +O(p_n^2)$.  For the opposite limit, $p_n \to 1$, we got $\langle I_z^2 \rangle  = I^2 - O[(1-p_n)]$. Therefore
\begin{subequations}
\label{eq:alpha}
\begin{align}
\alpha_I &= \frac{2}{3}(I+1), \quad &\mathrm{for} \, \, p_n\to 0,\\
\alpha_I &= 1, \quad &\mathrm{for} \,\, p_n\to 1.
\end{align}
\end{subequations}
At intermediate polarization, $\alpha$ will be somewhere between these two limiting values. For spin one-half the two limiting values are the same and $\alpha=1$ for any $p_n$.

\section{Derivation of Eq.~(\ref{eq:X2})}
\label{app:derivationX}

Our goal is to calculate the transition matrix element 
\be
Y = \langle s^\prime, j^\prime | \delta \mathbf{I} \cdot {J_n^\prime}^{(-2s)} \cdot \mathbf{s} | s, j  \rangle.
\ee
Here, the spin indexes are related by Eq.~\eqref{eq:spinIndexRelation}, the corresponding quantization axes are $\axisEnergyElectronO$ and $\axisEnergyNucleus$, see Eq.~\eqref{eq:resonantPair}, and the time-dependent tensor ${J_n^\prime}$ is defined in Eqs.~\eqref{eq:U}, \eqref{eq:Jtilde}, and \eqref{eq:Jprime}. Using these relations, and choosing $s=1/2$, we write the matrix element in a more concrete form
\be
Y = \langle \downarrow, j+1 | \delta \mathbf{I} \cdot {J_n^\prime}^{(-1)} \cdot \mathbf{s}| \uparrow, j  \rangle.
\label{eq:Y1}
\ee
We now do two straightforward transformations. First, we express the spin-operator vectors in coordinates aligned with their quantization axes. For example, the electron spin $\uparrow$ is an eigenstate of the operator
\be
\sigma_{\axisEnergyElectronO} \equiv \boldsymbol{\sigma} \cdot \axisEnergyElectronO =  \boldsymbol{\sigma} \cdot R_{\axisEnergyElectron \to \axisEnergyElectronO}[\axisEnergyElectron] = \left( R^{-1}_{\axisEnergyElectron \to \axisEnergyElectronO}[\boldsymbol{\sigma}] \right) \cdot \axisEnergyElectron,
\ee
where we denote $R_{\mathbf{n} \to \mathbf{m}}$ a rotation operator taking unit vector $\mathbf{n}$ to unit vector $\mathbf{m}$. Second, we introduce `ladder' operators for spins; for example, 
\be
\left( \begin{tabular}{c} $\sigma_x$\\ $\sigma_y$\\ $\sigma_z$ \end{tabular} \right)
=
\left( \begin{tabular}{ccc} $1/2$ & $1/2$ & 0\\ $-\mathrm{i}/2$ & $\mathrm{i}/2$&0\\ 0&0&1 \end{tabular} \right)
\cdot
\left( \begin{tabular}{c} $\sigma_+$\\ $\sigma_-$\\ $\sigma_z$ \end{tabular} \right).
\label{eq:L}
\ee 
With these, the operator of interest can be written
\be
\delta \mathbf{I} \cdot {J_n^\prime}^{(-1)} \cdot \mathbf{s} = \delta \mathbf{I}_{\axisEnergyNucleus,L} \cdot 
\left( L^\mathrm{T}  R^\mathrm{T}_{\axisEnergyElectron \to \axisEnergyNucleus} {J_n^\prime}^{(-1)} 
R_{\axisEnergyElectron \to \axisEnergyElectronO} L \right)   \cdot 
\mathbf{\sigma}_{\axisEnergyElectronO,L},
\ee
where $L$ is the three by three matrix in Eq.~\eqref{eq:L} and the subscript $(\mathbf{n},L)$ on the spin-operator vector states that the vector components are in the ladder operators basis in the coordinate frame with its third axis along $\mathbf{n}$. The advantage of such transformation is that in this basis we can treat the spin quantum numbers $s, j$ as representing the `usual' basis with the spin quantization axis along `$z$'. Also, as the only possibly nonzero component of the polarization $\langle \mathbf{I} \rangle$ is z, we can drop the polarization, $\delta \mathbf{I} \to \mathbf{I}$, and get the matrix element in Eq.~\eqref{eq:Y1} as
\be
\begin{split}
Y&=\langle \downarrow, j+1 | I_+ [J_n(t) M(t)]^{(-1)}_{+ -} s_- | \uparrow, j  \rangle \equiv  X \mathcal{I}_+,
\end{split}
\ee
where we used Eq.~\eqref{eq:Jtilde} and Eq.~\eqref{eq:J} with the time dependence according to Eq.~\eqref{eq:dotShift}, to express the element through the following short hands:
\begin{align}
\mathcal{I}_+ &= \langle \downarrow, j+1 | I_+ s_- |\uparrow, j\rangle,\\
X & = [J_n(t) M(t)]^{(-1)}_{+ -},\\
J_n(t)&=J_n \left( 1-2\frac{\dotShift(t) \cdot (\mathbf{r}-\mathbf{r}_0) }{l^2} \cos(\omega_\rf t -\phi_\rf) \right),\\
\label{eq:Mt}
M(t) &= L^\mathrm{T}  R^\mathrm{T}_{\axisEnergyElectron \to \axisEnergyNucleus} 
R_{\axisEnergyNucleus, \omega_\rf t} R_{\axisEnergyElectron, \omega_\rf t}^{-1}
R_{\axisEnergyElectron \to \axisEnergyElectronO} L.
\end{align}
In the last line, we used an alternative notation for rotations, putting $R_{\mathbf{n},\alpha}$ for the matrix implementing rotation around unit vector $\mathbf{n}$ by angle $\alpha$.

\begin{table*}
\begin{tabular}{@{}c@{\quad}c@{\quad}c@{\quad}c@{\quad}c@{}}
\toprule
Fourier& \multicolumn{4}{c}{matrix elements in ladder basis }\\
 index $k$ & $M_{++}^{(k)}$ & $M_{+-}^{(k)}$ & $M_{-+}^{(k)}$ & $M_{--}^{(k)}$ \\
\midrule
$\,\,\,\,\,0$ 	& $\frac{e^{\mathrm{i} \delta^\prime}(\sin\gamma-1)(1+\cos\delta)}{8}$
		& $\frac{e^{\mathrm{i} \delta^\prime}(\sin\gamma+1)(1+\cos\delta)}{8}$
		& $\frac{e^{-\mathrm{i} \delta^\prime}(\sin\gamma+1)(1+\cos\delta)}{8}$
		& $\frac{e^{-\mathrm{i} \delta^\prime}(\sin\gamma-1)(1+\cos\delta)}{8}$\\\\
$-1$ 	& $\frac{\cos\gamma \sin\delta}{4}$
		& $\frac{\cos\gamma \sin \delta}{4}$
		& 0
		& 0\\\\
$+1$ 	& 0
		& 0
		& $\frac{\cos\gamma \sin\delta}{4}$
		& $\frac{\cos\gamma \sin\delta}{4}$\\\\
$-2$ 	& $\frac{e^{-\mathrm{i} \delta^\prime}(\sin\gamma+1)(\cos\delta-1)}{8}$
		& $\frac{e^{-\mathrm{i} \delta^\prime}(\sin\gamma-1)(\cos\delta-1)}{8}$
	        & 0
		& 0\\\\
$+2$	& 0
		& 0
	        & $\frac{e^{\mathrm{i} \delta^\prime}(\sin\gamma-1)(\cos\delta-1)}{8}$
		& $\frac{e^{\mathrm{i} \delta^\prime}(\sin\gamma+1)(\cos\delta-1)}{8}$\\\\
\bottomrule
\end{tabular}\\
\caption{
\label{tab:M}
Matrix elements of $M(t)$ defined in Eq.~\eqref{eq:Mt}. We do not give the elements $M_{z\cdot}$ and $M_{\cdot z}$ as they do not couple resonant states. The elements have symmetry $M^{(k)}_{f f^\prime} = (M^{(-k)}_{\overline{f},\overline{f}^\prime})^*$, which we interpret as the amplitude for a spin-spin transition at energy quantum $k$ being complex conjugate of a reverse transition at opposite energy. Here, the index inversion is defined by $\overline{+} = -$, $\overline{-} = +$, and $\overline{z} = z$.
}
\end{table*}

Since $J_n(t)$ contains only $-1$, $0$, and $+1$ Fourier components, to get the component $\pm1$ of the product $J_n(t)M(t)$, we need the Fourier components of $M(t)$ up to $\pm 2$. They are given in Table~\ref{tab:M} for the parts of interest. 
The desired matrix element is
\be
\label{eq:X}
\begin{split}
X &\equiv [J_n(t) M(t)]^{(-1)}_{+ -} \\
&= J_n^{(-1)}M^{(0)}_{+-}+  J_n^{(0)}M^{(-1)}_{+-} +  J_n^{(1)}M^{(-2)}_{+-},
\end{split}
\ee
and the three terms are, respectively,
\begin{subequations}
\label{eq:terms}
\begin{align}
-&J_n\frac{1+\cos\delta}{8} e^{\mathrm{i} (\angleRF +\delta^\prime)}(1+\sin \gamma)\frac{\dotShift(t) \cdot (\mathbf{r}-\mathbf{r}_0) }{l^2},\label{eq:term1}\\
&J_n \frac{\sin \delta}{4} \cos\gamma,\label{eq:term2}\\
-&J_n\frac{1-\cos\delta}{8} e^{-\mathrm{i} (\angleRF +\delta^\prime)}(1-\sin \gamma)\frac{\dotShift(t) \cdot (\mathbf{r}-\mathbf{r}_0) }{l^2},\label{eq:term3}
\end{align}
\end{subequations}
where $\delta^\prime$ and $\delta$ are the two Euler angles of the rotation $R_{\axisEnergyElectron \to \axisEnergyNucleus} =  R_{\axisEnergyElectron,\delta^\prime} \circ R_{\mathbf{y}_e,\delta}$, see Fig.~\ref{fig:angles}. 
\newcommand{\bFieldDot}{\langle \bField \rangle} 

For realistic micromagnet gradients and quantum dot sizes, the change of the magnetic field across the quantum dot is small compared to the magnetic field magnitude. The angle $\delta$ is then close to either 0, when $\mathrm{sgn}(g_e g_i)=-1$ (the case of both Si and GaAs conduction band), or $\pi$, when $\mathrm{sgn}(g_e g_i)=+1$. Out of the two terms in Eqs.~\eqref{eq:term1} and \eqref{eq:term3}, these two scenarios imply that the second or the first can be neglected, respectively. The matrix elements in Eq.~\eqref{eq:terms} show that these two scenarios map to each other upon inverting the sign of $\gamma$. Therefore, the relative sign of the electron and nuclear $g$ factors implies no essential difference for the arising DNSP rate magnitude. 

\begin{table*}
\begin{tabular}{@{}c@{\qquad}l@{}}
\toprule
& \multicolumn{1}{c}{Matrix element value}\\
\midrule
$M_{+-}^{(0)}$&$\frac{1+\sin\gamma}{8}\Big( ( J_{xx} \cos\delta+J_{yy}+i  J_{xy}\cos\delta-iJ_{yx}) \cos\delta^\prime- (J_{zx}+iJ_{zy})\sin\delta+(iJ_{xx}-J_{yx}+  J_{yx}\cos\delta+i J_{yy}\cos\delta) \sin\delta^\prime \Big)$\\\\
$M_{+-}^{(-1)}$&$-\frac{\cos\gamma}{4}\Big( ( J_{xz}\cos\delta-iJ_{yz})\cos\delta^\prime-J_{zz}\sin\delta + ( J_{yz}\cos\delta+iJ_{xz})\sin\delta^\prime \Big)$\\\\
$M_{+-}^{(+1)}$& $0$\\\\ 
$M_{+-}^{(-2)}$&$\frac{-1+\sin\gamma}{8}\Big( ( J_{xx}\cos\delta-J_{yy}-i  J_{xy}\cos\delta-iJ_{yx}) \cos\delta^\prime - (J_{zx}-iJ_{zy})\sin\delta+ (iJ_{xx}+J_{xy}+ J_{yx}\cos\delta-i J_{yy}\cos\delta) \sin\delta^\prime \Big)$\\\\
 $M_{+-}^{(+2)}$& $0$\\
\bottomrule
\end{tabular}
\caption{
\label{tab:ManisotropicJ}
Matrix elements of $M(t)$ defined in Eq.~\eqref{eq:Mt} for a general hyperfine interaction, given by $\spinNucleus_i \spinElectron_j J_{ij}$. We give only the elements $M_{+-}$. The symmetry $M^{(k)}_{f f^\prime} = (M^{(-k)}_{\overline{f},\overline{f}^\prime})^*$ still holds.
}
\end{table*}

On the other hand, the micromagnet gradient makes the angles $\delta, \delta^\prime$ dependent on the position within the dot, complicating the analysis. We consider a simplified scenario. The restriction is insignificant for the results presented in this paper, but simplifies the notation and calculations. Namely, the gradient of the magnetic field at the dot position is given by the tensor $\nabla_i \langle B_j \rangle$.\footnote{The derivative is with respect to the dot center $\mathbf{r}_0$.} For in-plane displacements, the six derivatives, $\nabla_{x_0} \bFieldDot$ and $\nabla_{y_0} \bFieldDot$, enter the problem. We split them to the gradient of the field along its direction (also denoted as the field longitudinal component), $\boldsymbol{\nabla} (\bFieldDot \cdot \axisEnergyElectron)$, and the two gradients of the two remaining transverse components. The former is important for the resonance width, see Eq.~\eqref{eq:gi} in Appendix~\ref{app:nuclearDensity}. The latter can be represented by a two by two matrix, schematically denoted by $\boldsymbol{\nabla} (\bFieldDot \times \axisEnergyElectron)$. Our simplified scenario corresponds to neglecting the smaller-in-magnitude of the two singular values $w_1$ and $w_2$ of this matrix. Assuming $w_1$ is the larger one, the component of the magnetic field perpendicular to $\axisEnergyElectron$ is given by $(\mathbf{r}-\mathbf{r}_0) \cdot \mathbf{u}_1 w_1 \mathbf{v}_1$, where $\mathbf{u}_i$ and $\mathbf{v}_i$ are the unit vectors from the singular value decomposition.\footnote{We use the notation of  Chapter 2.9 of Ref.~\cite{press_numerical_2007}. See therein for details on the singular value decomposition.} More important than their values, we note that in this case the angle $\delta^\prime$ is fixed, given by the direction of the vector $\mathbf{v}_1$, while $\delta$ is position dependent, given by\footnote{The plus sign applies if the axes $\axisEnergyElectron$ and $\axisEnergyNucleus$ are close to parallel ($\delta \approx 0$) and minus sign if they are close to antiparallel $(\delta\approx \pi)$. If it is the minus sign here, it inverts the relative sign of the interference term in Eq.~\eqref{eq:xi}. To ease the notation, we include this possible minus sign by redefining $\delta^\prime$, adding $\pi$ to it.}
\be \begin{split}
\label{eq:delta}
\pm \sin \delta \approx \tan \delta &= \frac{((\mathbf{r}-\mathbf{r}_0)\cdot \boldsymbol{\nabla}) (\bFieldDot \times \axisEnergyElectron) \cdot \mathbf{v}_1}{\bFieldDot}\\
& \approx \frac{(\mathbf{r}-\mathbf{r}_0)\cdot \mathbf{u}_1 w_1}{\bFieldDot}.
\end{split}\ee
In the main text, we use a shorthand notation $\nabla_\perp B \equiv w_1 = |\boldsymbol{\nabla} (\bFieldDot \times \axisEnergyElectron) \cdot \mathbf{v}_1|$ as the gradient size of the transverse component of the magnetic field, and $\nabla_{||} B \equiv |\boldsymbol{\nabla} (\bFieldDot \cdot \axisEnergyElectron)|$ as the gradient size of the longitudinal component. Also, we denote $\phi$ as the angle of vectors $\dotShift$ and $\mathbf{v}_1$.

With the geometry clarified, let us go back to Eq.~\eqref{eq:X}. It is a sum of three terms. To be specific, let us take the $\delta\approx0$ scenario. Each term contains a small factor: in the first, it is the dot shift compared to its size $O(\dotShift/l)$, in the second the deflection angle $O(\delta)$, and in the third there are both $O(\dotShift/l)$ and $O(\delta^2)$. As already noted, the third term can be neglected with respect to the first. (If $\delta\approx \pi$, the roles of the first and third terms would be swapped). The transition amplitude is thus a sum of two terms. The complex factor $\exp[\mathrm{i}(\angleRF+\delta^\prime)]$ makes the two terms interfere in the matrix element squared magnitude $|X|^2$. Once again, we are interested in the average of this expression over the dot. Equation \eqref{eq:X2} follows after a short algebra using Eq.~\eqref{eq:X}, Eq.~\eqref{eq:delta}, and the following averages,
\begin{align*}
&\langle \mathbf{r} - \mathbf{r}_0 \rangle = \boldsymbol{0},\\
&\langle (\mathbf{a} \cdot (\mathbf{r} - \mathbf{r}_0)) (\mathbf{b} \cdot (\mathbf{r} - \mathbf{r}_0))  \rangle = (\mathbf{a} \cdot \mathbf{b}) l^2.
\end{align*}
We conclude with a comment to the interference strength $\xi$ given in Eq.~\eqref{eq:xi}: It is a product of two cosines. If neither of the two arguments is known, one could replace them by their average using $\xi \to \sqrt{\langle \xi^2 \rangle} =1/2$, where the average $\langle \cdot \rangle$ is the integral over the unknown angles with a uniform prior probability distribution. In other words, the interference is somewhat suppressed by the misalignment of the essentially random directions of vectors $\dotShift$, $\mathbf{v}_1$, and the angle $\angleRF$. Within our precision here, we simply neglect the interference.

\section{Anisotropic hyperfine interaction}

\label{sec:anisotropicJ}

We now extend the previous appendix by considering a more general form of the hyperfine Hamiltonian $H_\mathrm{hf}$ in Eq.~\eqref{eq:hyperfine}, with the spin-spin interaction not necessarily isotropic. It would result in the hyperfine coupling in Eq.~\eqref{eq:J} becoming a second-rank tensor, with Cartesian components denoted here as $J_{xx}$, $J_{xy}$, and so on. The relevant matrix elements for a general hyperfine tensor, calculated from Eq.~\eqref{eq:Mt}, are given in Table~\ref{tab:ManisotropicJ}.

We are motivated by the possible application of our formulas to hole qubits (see Appendix~\ref{app:hole}). Before that, we look at what tensor matrix elements are required for a non-zero DNSP, making the connection to the existing literature. From Table~\ref{tab:ManisotropicJ}, we find that without the axes deflection, neither the isotropic exchange ($J_{xx}=J_{yy}=J_{zz}$ as the only nonzero matrix elements) nor the `secular exchange' ($J_{zz}$ the only non-zero matrix element; the name is used in the NMR literature), induce transitions. Whereas the former applies for dipole-dipole interactions in liquid solutions \cite{weis_solid_2000}, the latter form of spin-spin coupling originating in dipole-dipole interaction is typically considered in the solid state \cite{lurie_spin_1964,mcarthur_rotating-frame_1969}.\footnote{However, there are also isotropic interactions: In its derivation of the polarization rate, Ref.~\cite{hartmann_nuclear_1962} considered the electron-mediated nuclear-nuclear exchange as such an isotropic interaction.}

We thus have the following analogy to the NMR and the Hartmann-Hahn effect: While there both spins are driven, here it is only one of them (the electron). Since driving a spin effectively deflects its energy quantization axis from the direction of the magnetic field, driving also the second nuclear spins in NMR is analogous to having here a nonzero deflection angle $\delta$ due to the micromagnet.\footnote{Section~IV.C of Ref.~\cite{hartmann_nuclear_1962} contains a discussion of the case where the second spin is not driven (as here) and invokes an exchange tensor with components such as $J_{zx}$ and $J_{zy}$ needed to produce finite nonzero matrix element for the polarization transition. The same anisotropic terms were considered also in  Refs.~\cite{weis_solid_2000,henstra_theory_2008}.}

In the literature on self-assembled quantum dots, the `nonsecular' (the NMR name) hyperfine interaction terms, such as $J_{xz}$, are called `noncollinear'. While in Ref.~\cite{xu_optically_2009,yang_collective_2012} such terms are assigned to the effects of the light-hole--heavy-hole mixing on the hole hyperfine tensor (see Appendix~\ref{app:hole}), in the majority of the works in that field the `noncollinearity' is understood as due to the quadrupolar fields \cite{urbaszek_nuclear_2013} (they were considered as the DNSP source in Ref.~\cite{huang_theoretical_2010} and coined as `noncollinear' hyperfine tensor in Ref.~\cite{hogele_dynamic_2012}, including the quadrupolar effects perturbatively). If due to quadrupolar fields, the `noncollinearity' then qualitatively corresponds, in our work, to the combined effect of an isotopic electron-nuclear hyperfine interaction and the deflection of the quantization axes.

\section{Additional plots.}

We present additional plots analogous to Figs.~\ref{fig:As} and \ref{fig:feedback-1-GaAs} of the main text.

\label{app:analogs}

\subsection{Plots analogous to Fig.~\ref{fig:As}.}

\label{app:analogs4}

\begin{figure}
\begin{center}
\includegraphics[width=\linewidth]{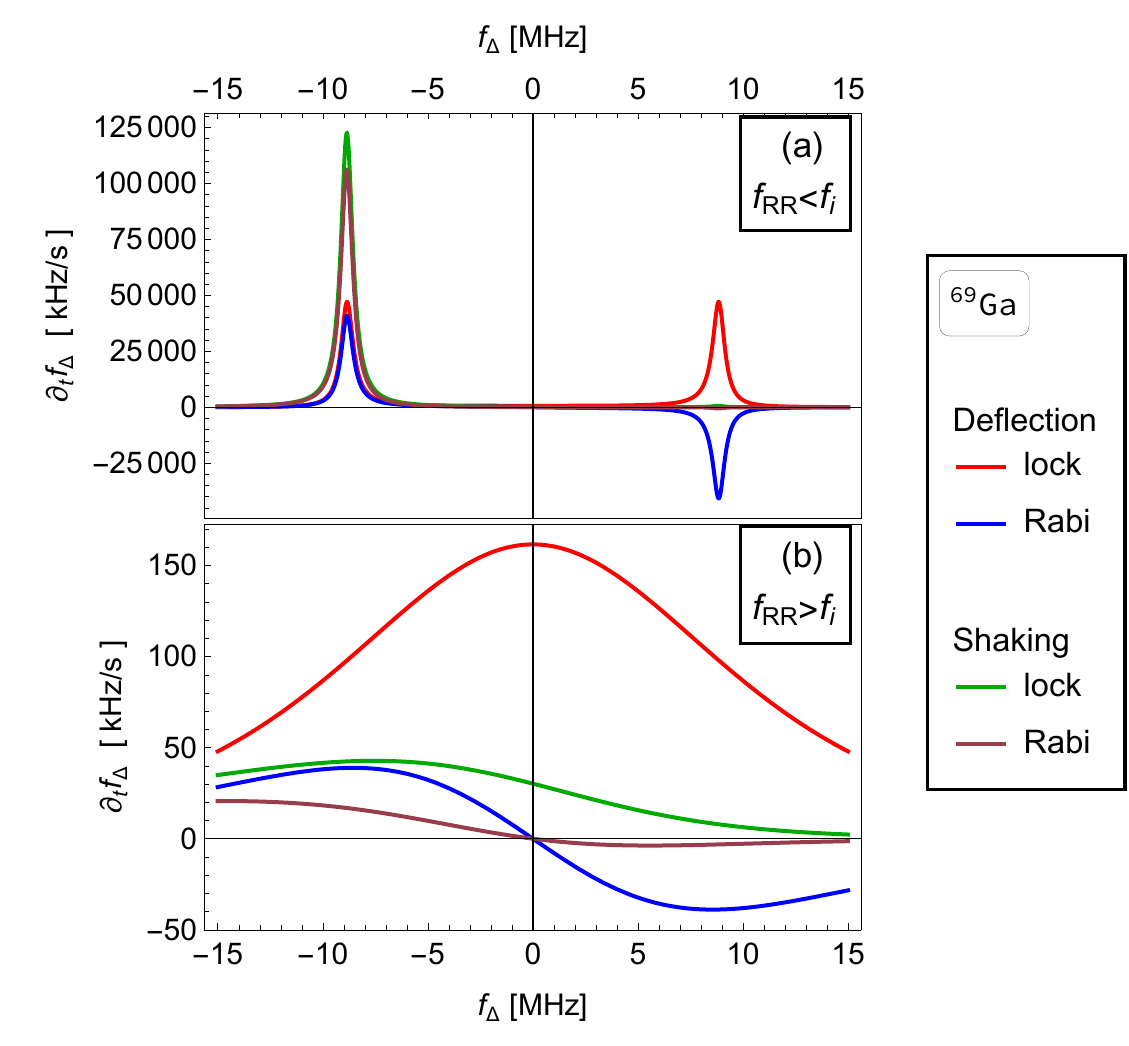}
\includegraphics[width=\linewidth]{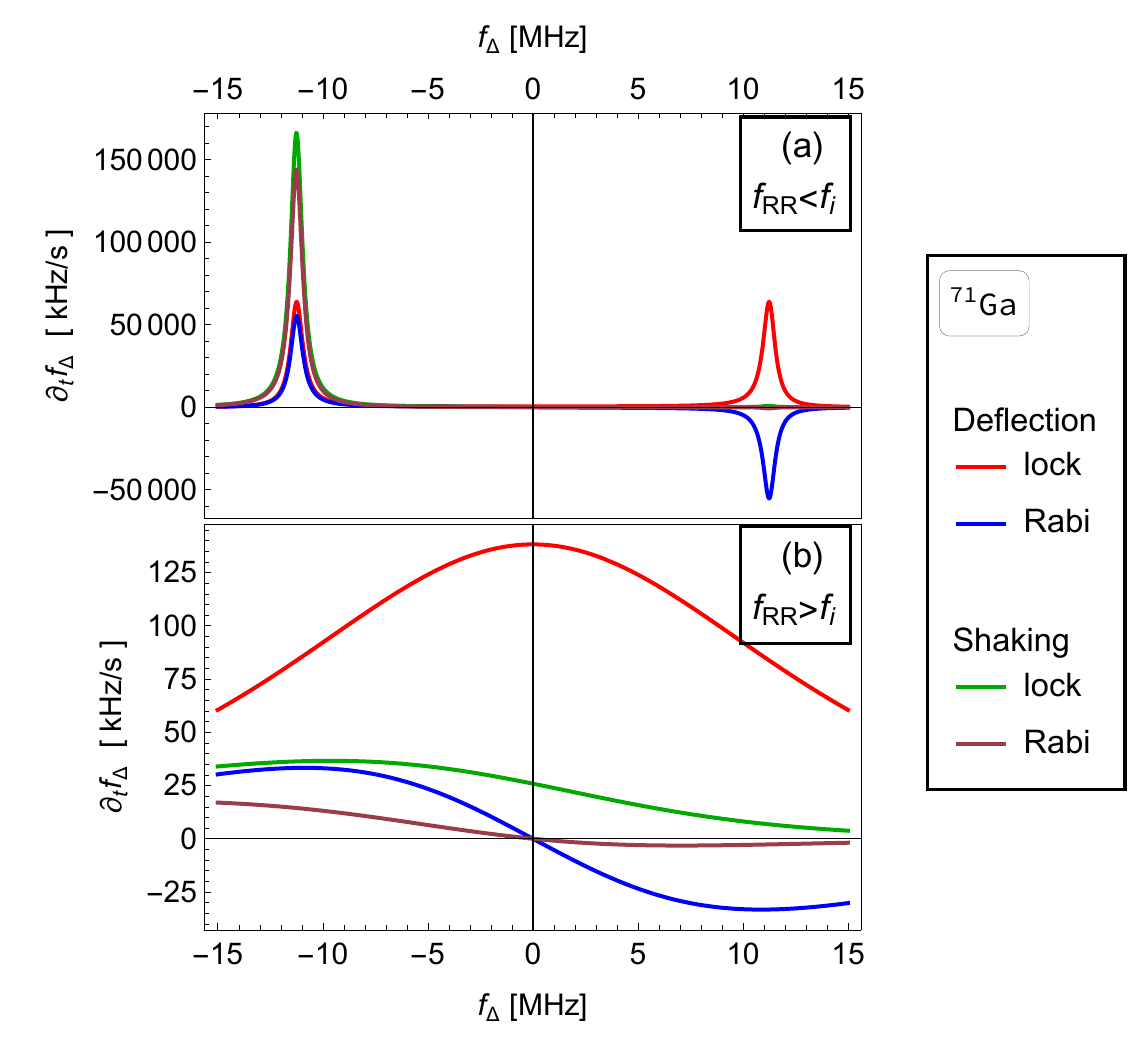}
\end{center}
\caption{Polarization rate as a function of the electron detuning frequency for gallium isotopes. The plot is analogous to Fig.~\ref{fig:As}, and all parameters are the same as there.}
\label{fig:Ga}
\end{figure}

\begin{figure}
\begin{center}
\includegraphics[width=\linewidth]{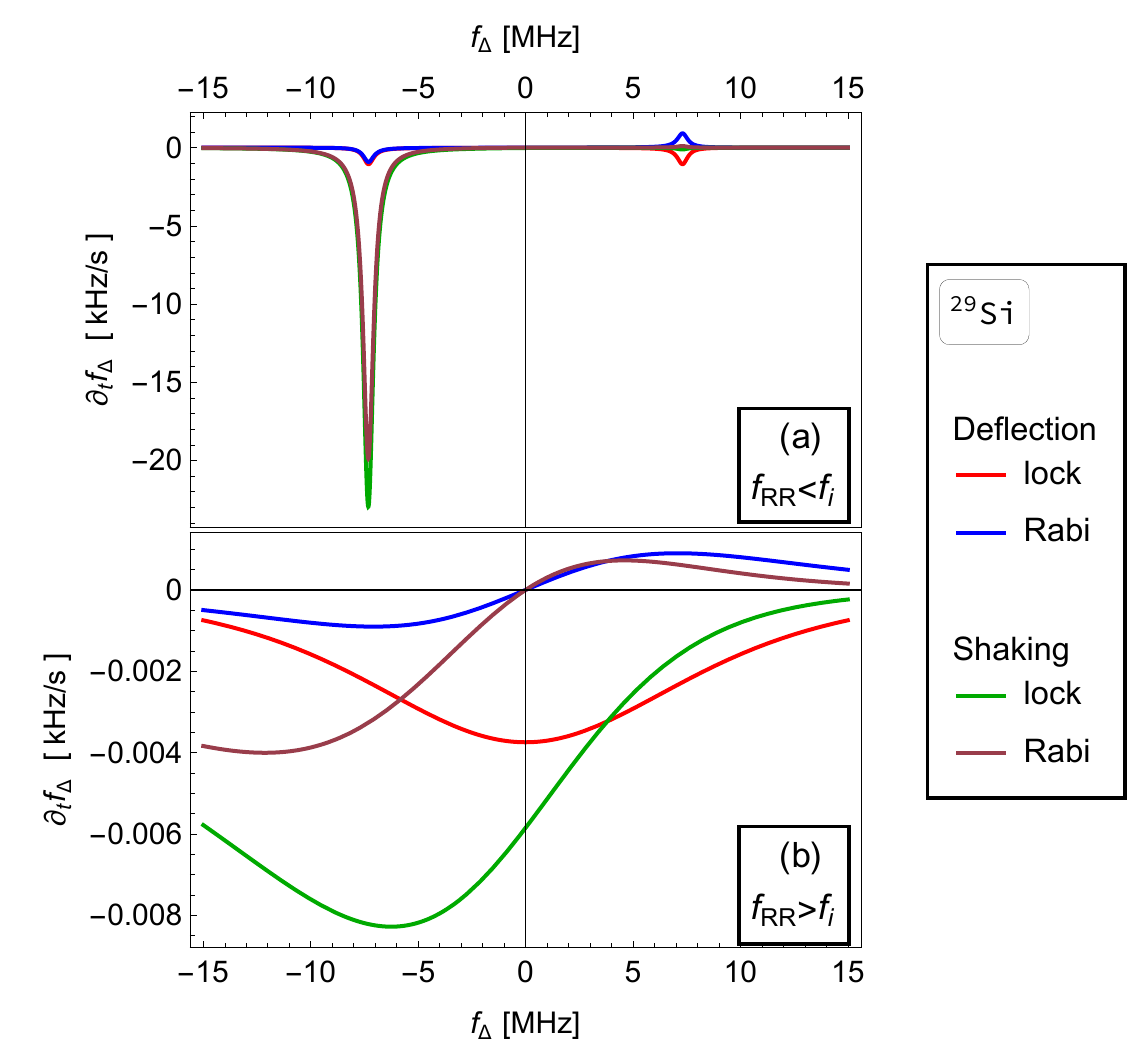}
\end{center}
\caption{Polarization rate as a function of the electron detuning frequency for $^{29}$Si in a silicon quantum dot. The plot is analogous to Fig.~\ref{fig:As} and the parameters are as there except for: pulse time $T_\mathrm{pulse}=100$ $\mu$s, cycle time $T_\mathrm{cycle} = 200$ $\mu$s, dot in-plane size $l=20$ nm, additional smearing $2\pi \times 25$ kHz.}
\label{fig:Si}
\end{figure}

Figure \ref{fig:Ga} shows DNSP rates for the two gallium isotopes of GaAs, using the same parameters as in Fig.~\ref{fig:As}. Figure \ref{fig:Si} is a similar plot for the $^{29}$Si isotope of a Si dot. There, some parameters are slightly changed, reflecting that silicon dots are typically smaller and have better coherence because of nuclear-induced dephasing being smaller than in GaAs. For these parameters, the DNSP rates in Si are about four orders of magnitude smaller than in GaAs.

\subsection{Plots analogous to Fig.~\ref{fig:feedback-1-GaAs}.}

\label{app:analogs6}

\begin{figure}
\begin{center}
\includegraphics[width=0.99\linewidth]{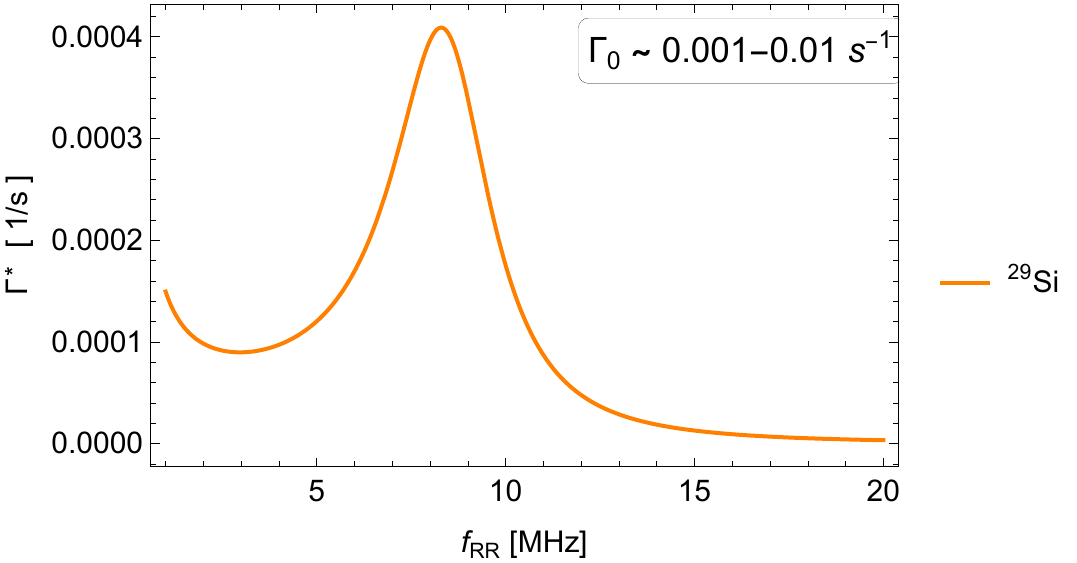}
\end{center}
\caption{\textbf{Stabilization by feedback in an electron qubit in Si}. The plot is analogous to Fig.~\ref{fig:feedback-1-GaAs} (see its caption for a plot description) and adopts the same parameters except for the material parameters of Si and a smaller dot, $l=20$ nm.}
\label{fig:feedback-1-Si}
\end{figure}

To illustrate the magnitude of the DNSP polarization rate in silicon, we plot the restoring rate $\Gamma^*$ in Fig.~\ref{fig:feedback-1-Si}.
Comparing to Fig.~\ref{fig:feedback-1-GaAs}, we observe that in Si the polarization is several orders of magnitude smaller than in GaAs. It is also much smaller than the intrinsic restoring force $\Gamma_0$ due to thermal diffusion. In this respect, the EDSR-induced dynamics of nuclear spins is minor.

\section{Laird mechanism}
\label{app:laird}

We now consider a different resonance, not of the Hartmann-Hahn type. We include it for completeness, and because both the calculation and the corresponding experiment are related to the ones we have considered. Namely, we now assume that for some reason, the electrical driving is not effective in driving the electron spin. While the dot is electrically driven as before, there is no EDSR (disregarding nuclei). The most straightforward scenario would be a dot without a micromagnet. 

Still, the electron spin can be transferred to the nuclei, so that DNSP arises. However, it happens at a different resonance condition, namely when the driving frequency equals the difference of the electron and nuclear Larmor frequencies.\footnote{Such slight detunings which lead to spin-selective electron-nuclear flip-flops were considered in Ref.~\cite{tenberg_narrowing_2015}.} As we explain here, the DNSP arises as a backaction of the torque that a random transverse component of the Overhauser field exerts on the electron.\footnote{Several previous works \cite{petrov_coupled_2009,merkulov_long-term_2010,glazov_electron_2012} analyzed the DNSP arising from a periodically reset electron spin, treating the electron and Overhauser fields as classical vectors precessing around each other. That a semiclassical model contains all the relevant physics is confirmed by the fact that it also explains the observed nuclei-induced electron-spin dephasing \cite{neder_semiclassical_2011}.}
 When the resonance condition is fulfilled, this torque results in the electron EDSR and the nuclear polarization. The effect was experimentally demonstrated by Laird \textit{et al.}~\cite{laird_hyperfine-mediated_2007}.

To estimate the strength of this mechanism, and compare it to the ones from the main text, we now make an analogous derivation of the DNSP rate for this scenario. Since the micromagnet is not relevant now, we drop it from the problem. The Hamiltonian in Eq.~\eqref{eq:electronNuclearPairH} simplifies considerably,
\be
H = 
-\hbar\angularFrequencyNucleus \spinNucleusVector_n \cdot \axisEnergyNucleus
-\hbar\angularFrequencyElectron \spinElectronVector \cdot \axisEnergyElectron
+J_n(t) \delta \spinNucleusVector_n \cdot \spinElectronVector.
\label{eq:electronNuclearPairH-Laird}
\ee
The two vectors $\axisEnergyNucleus$ and $\axisEnergyElectron$ are now parallel (or antiparallel). We go into a frame rotating with the nuclear spin,
\be
H^\prime = 
-\left[ \hbar\angularFrequencyElectron+\mathrm{sgn}(g_e g_n)\hbar\angularFrequencyNucleus \right] s_z 
+ J_n(t) \delta \spinNucleusVector_n \cdot \spinElectronVector.
\label{eq:Htilde-Laird}
\ee
In the last term, we keep only the transverse (in spin) and varying (in time) component,
\be \begin{split}
H^\prime &= 
-\left[ \hbar\angularFrequencyElectron+\mathrm{sgn}(g_e g_n) \hbar\angularFrequencyNucleus \right] s_z + [J_n(t)-J_n(0)] \spinNucleusVector_{n\perp} \cdot \spinElectronVector.
\end{split}\ee
We sum over all nuclei and consider the electron dynamics described by
\begin{subequations}
\begin{align}
&H_e = -(\hbar \angularFrequencyRF- \hbar \angularFrequencyDetuning^L) s_z \nonumber\\&\qquad\qquad+  
2\cos(\angularFrequencyRF t -\phi_\rf) \hbar \boldsymbol{\omega}_{RR}^{L} \cdot \spinElectronVector,\\
&\hbar \angularFrequencyDetuning^L = \hbar \angularFrequencyRF-\left[ \hbar\angularFrequencyElectron+\mathrm{sgn}(g_e g_n)  \hbar\angularFrequencyNucleus \right], \\
&\hbar \boldsymbol{\omega}_{RR}^{L} = \frac{1}{2} \sum_n v_0 A_n (\mathbf{d} \cdot \boldsymbol{\nabla}) |\Psi_n|^2 \spinNucleusVector_{n,_\perp},
\end{align}
\end{subequations}
Thus, we obtained a Rabi Hamiltonian $H_e$ with the detuning $\hbar \angularFrequencyDetuning^L $ and the driving field $\hbar \boldsymbol{\omega}_{RR}^{L}$. In this section, we introduce several quantities analogous to the ones in the main text, denoting them by the superscript $L$ for `Laird'. 

We estimate the typical value of the transverse field, averaging over the dot
\be \label{eq:LRR}
\begin{split}
&\langle \left( \hbar \boldsymbol{\omega}_{RR}^{L}  \right)^2 \rangle \\
&\quad = \frac{1}{4}
\langle 
\sum_{n,m} v_0^2 A_i^2 4 (\mathbf{d} \cdot \mathbf{r}_{n0})(\mathbf{d} \cdot \mathbf{r}_{m0})l^{-4} |\Psi_n|^4
\spinNucleusVector_{n,_\perp} \cdot \spinNucleusVector_{m,_\perp} \rangle\\
&\quad = 2\frac{I(I+1)}{3}(d/l)^2 \fractionIsotope A_i^2 v_0 / V_D,
\end{split}\ee
with the short hand $\mathbf{r}_{n0}=\mathbf{r}_{m}-\mathbf{r}_{0}$. In the averaging, we assume that the polarization $p_i$ is small and use $\langle \spinNucleusVector_{n,_\perp} \cdot \spinNucleusVector_{m,_\perp} \rangle = \delta_{n,m} (2/3) I(I+1)$ corresponding to unpolarized nuclei.

\begin{figure}
\begin{center}
\includegraphics[width=0.6\linewidth]{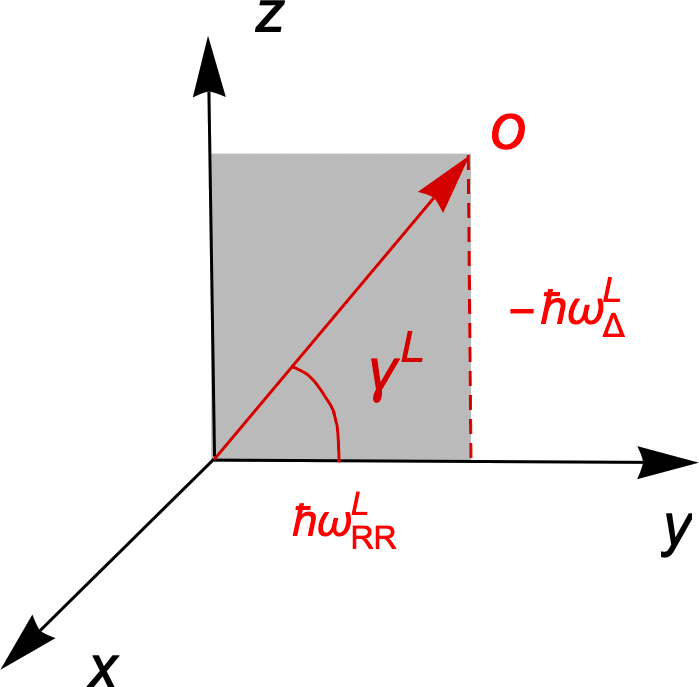}
\end{center}

\caption{\textbf{Rabi oscillations in the Laird mechanism.} The schematic defines the Rabi angle $\gamma^L$ in terms of the detuning and the matrix element given in Eq.~\eqref{eq:Rabi-Laird}. This figure is analogous to Fig.~\ref{fig:Rabi}.
\label{fig:Rabi-Laird}
}
\end{figure}

The transverse field corresponds to the Rabi precession angle (see Fig.~\ref{fig:Rabi-Laird})
\begin{subequations}
\label{eq:Rabi-Laird}
\begin{align}
\sin\gamma^L &= -\frac{\hbar \angularFrequencyDetuning^L}{\hbar \angularFrequencyRabi^L},\\
\cos\gamma^L &= \frac{\hbar \angularFrequencyRabiResonant^L}{ \hbar \angularFrequencyRabi^L},\\
\hbar \angularFrequencyRabi^L  &= \sqrt{ \left( \hbar \angularFrequencyRabiResonant^L \right)^2 + \left( \hbar \angularFrequencyDetuning^L \right)^2 }.
\end{align}
\end{subequations}
The electron spin $z$ component evolves according to
\be \begin{split}
s_z(t) &= s_z(0)\left[ \sin^2 \gamma^L + \cos^2 \gamma^L \cos \angularFrequencyRabi^L t \right] \\
&= s_z(0)\left[ 1 + \cos^2 \gamma^L \left(\cos \angularFrequencyRabi^L t  -1 \right) \right],
\end{split} \ee
an equation analogous to Eq.~\eqref{eq:pzt}.

Finally, since the $z$ component of the total spin of the system is conserved, the change of the electron spin equals the opposite change of the spin of the nuclei,
\be
I_z(t) - I_z(0) =  s_z(0) - s_z(t) =  \frac{p_e}{2}\cos^2 \gamma^L \left( 1- \cos \angularFrequencyRabi^L t \right),
\ee
where we have put $s_z(0)=p_e/2$ in line with the notation in Eq.~\eqref{eq:electronPolarization}.
With the total duration of the driving being $T_\mathrm{pulse}$ (we drop the subscript and use $T$ in the following two equations to improve readability), this change is equivalent to a rate (of polarization the total nuclear spin of isotope $i$)
\be
\Gamma_i^L = \frac{p_e}{2}\cos^2 \gamma^L \frac{1- \cos \angularFrequencyRabi^L T}{T},
\ee
in turn equivalent to the polarization rate
\be
\label{eq:lairdRate}
\begin{split}
\partial_t p_i &= \frac{1}{I_i \fractionIsotope N_\mathrm{tot}}\Gamma^L_i \\
&=\frac{p_e}{2}\frac{1}{I_i \fractionIsotope N_\mathrm{tot}} \frac{\left( \hbar \angularFrequencyRabiResonant^L \right)^2}{\left( \hbar \angularFrequencyRabi^L \right)^2} 
\frac{1- \cos \angularFrequencyRabi^L T}{ T}\\
&=\frac{p_e}{2}\frac{1}{\fractionIsotope N_\mathrm{tot}} \frac{(2/3)(I_i+1)(d/l)^2 \fractionIsotope A_i^2}{ \hbar^2 N_\mathrm{tot}} 
\frac{1- \cos \angularFrequencyRabi^L T}{\left( \angularFrequencyRabi^L \right)^2 T}\\
&=p_e\frac{(I_i+1)}{3 \hbar^2}\frac{A_i^2}{N_\mathrm{tot}^2} \frac{d^2}{l^2} 
\frac{1- \cos \angularFrequencyRabi^L T}{\left( \angularFrequencyRabi^L \right)^2 T}.
\end{split}
\ee
In analogy to Eq.~\eqref{eq:promotionToDeltaFunction}, we interpret the last factor as a (Lorenzian-shaped) spectral density: in the limit of a long evolution time $T\to \infty$ and a small hyperfine matrix element $\angularFrequencyRabiResonant^L \to 0$, it becomes a delta function 
\begin{equation}
\pi\hbar\delta\Big( \hbar\angularFrequencyElectron+\mathrm{sgn}(g_e g_n)  \hbar\angularFrequencyNucleus - \hbar\angularFrequencyRF \Big),
\end{equation}
imposing the conservation of the energy transfer between the electron spin, a nuclear spin, and a microwave photon. With this interpretation, we cast the rate in line with the notation of Eq.~\eqref{eq:rateDNSP},
\begin{subequations}
\label{eq:rateLaird}
\be
\label{eq:rateDNSPL}
\partial_t p_i =  \frac{\pi}{\hbar^2}X_L^2 \alpha_I p_e G_{\Sigma_L} (\angularFrequencyDetuning^L ),
\ee
where
\begin{align}
X_L &= \frac{A_i}{N_\mathrm{tot}}\frac{d}{l}, \label{eq:XL}\\
\Sigma_L^2 &= \frac{2I_i(I_i+1)}{3}\frac{d^2}{l^2} \frac{\fractionIsotope A_i^2}{N_\mathrm{tot}}.\label{eq:gaussianDensityL}
\end{align}
\end{subequations}
We have arrived at a formula analogous to Eq.~\eqref{eq:mainResult}. 
It is interesting to note that, up to an additional factor of 2 in the matrix element $X_L$, Eq.~\eqref{eq:rateLaird} corresponds to Eq.~\eqref{eq:rateDNSP} including only the 'shaking' mechanism in the limit $\frequencyRabiResonant \to 0$ with $\gamma \to \pi/2$. The remaining differences are natural: First, since we assumed unpolarized nuclei, the rate in \eqref{eq:rateDNSPL} contains the factor from Eq.~\eqref{eq:pzI} evaluated at $p_n=0$. Second, the width of the spectral function now refers only to the thermal distribution of the Overhauser field. The latter is similar to the values seen in Sec.~\ref{sec:extensions}.\footnote{Using $d=0.5$ nm, $l_z=10$ nm, and $l=34$ nm, we get 
$\Sigma_L(^{29}\mathrm{Si})=2\pi \times 34$ kHz, $\Sigma_L(^{69}\mathrm{Ga})=2\pi \times 135$ kHz, $\Sigma_L(^{71}\mathrm{Ga})=2\pi \times 140$ kHz, $\Sigma_L(^{75}\mathrm{Ga})=2\pi \times 207$ kHz.} 
However, one also needs to point out substantial differences:

First, the mechanism considered in this section originates in the (reaction) torque that the electron spin exerts in response to the (action) torque from nuclei inducing the electron Rabi rotation. In the main text, this torque was due to the micromagnet and had nothing to do with nuclei. While a stochastic gradient from the Overhauser field will coexist with the one due to a micromagnet, they will have a random mutual orientation (alternatively: random phase). If it is the micromagnet gradient that dominates, the random phase suppresses the `Laird' polarization rate and makes it zero on average in experiments with micromagnets. 

Second, the derivation here applies in the incoherent regime, otherwise the time-dependent factor in Eq.~\eqref{eq:lairdRate} should not be converted to a delta function, but kept as oscillating, leading to an oscillating nuclear polarization. Taking the opposite view, trying to use Eq.~\eqref{eq:mainResult} in the far-off resonant regime, we do not expect to recover Eq.~\eqref{eq:rateDNSPL} from Eq.~\eqref{eq:rateDNSP} upon taking the limit $\frequencyRabiResonant \to 0$. Namely, the assumption that the last term in Eq.~\ref{eq:fourDifferences} is the smallest is not fulfilled far from the resonance and explains the unnatural result $(1+\sin \gamma)/2 \to \theta(\gamma)$ in the limit $\frequencyRabiResonant \to 0$. To correct for this deficiency, one would need to keep both in-phase and out-of-phase frequency components, for example using the technique of Ref.~\cite{shirley_solution_1965}. However, since we are interested primarily in DNSP arising in dots with high-quality single-qubit operations, we do not pursue the off-resonant regime, and the connection between Eq.~\eqref{eq:rateDNSPL} and Eq.~\eqref{eq:rateDNSP}, further.

The most important conclusion of this section is that the DNSP arising as the backaction of the electron `primary' Rabi oscillation on the nuclear spins, that is, the `Laird' mechanism, can be neglected if nuclei are not the dominant source of the primary Rabi oscillations, that is, in experiments employing micromagnets or spin-orbit coupling. The nuclear contribution to the `primary' Rabi oscillations of the electron spin was neglected in the main text, attributing it to the micromagnet entirely. While nuclei also contribute, the corresponding DNSP rate is going to be much smaller than Eq.~\eqref{eq:rateDNSPL}, the latter comparable to one of the mechanisms included in Eq.~\eqref{eq:rateDNSP}.

\section{Effective parameters of bounded diffusion}

\label{app:randomWalk}

The Overhauser field acting on the electron spin in a quantum dot fluctuates because of diffusive thermal fluctuations of nuclear spins mediated by dipolar nuclear spin-spin interactions. The diffusion results in the Overhauser-field variance growing linearly over short times and saturating at long times: The long-time average (probability distribution) of the Overhauser-field components is a Gaussian with a finite variance centered at zero. A simple model of such stochastic quantity is a random walk with a harmonic restoring force \cite{uhlenbeck_theory_1930,rabenstein_qubit_2004}. Using the notation of Ref.~\cite{gutierrez-rubio_optimal_2020}, with $K(\omega_0, \omega, \delta t)$ being the conditional probability of the electron Larmor frequency having value $\omega$ at time $t$ if it had value $\omega_0$ at time $t_0=t-\delta t$, the model gives
\begin{subequations}
\begin{align}
    K(\omega_0, \omega, \delta t)
        = \frac{1}{\sqrt{2\pi\sigma_{\delta t}^2}}
    \exp
    \left[
        -\frac{(\omega-\omega_0
               e^{-\delta t/\kappa})^2}
              {2\sigma_{\delta t}^2}
    \right],
    \label{eq:kernel}
\end{align}
where 
\be
    \sigma_{\delta t}^2
    = \sigma_\Omega^2 (1-e^{-2\delta t/\kappa}).
\ee
\end{subequations}
Hence, the model has two parameters, $\sigma_\Omega^2$ and $\kappa$. The first parameter is the variance long-time saturation value. For the Overhauser field contribution to the electron Larmor frequency,
\be
\sigma_\Omega^2 = \frac{1}{\hbar^2}\langle \left( \delta \hbar \angularFrequencyElectron \right)^2 \rangle,
\ee
it is, by a calculation analogous to Eq.~\eqref{eq:LRR},
\be \label{eq:sigmaOv}
\begin{split}
\sigma_\Omega^2 &= 
\frac{1}{\hbar^2}\langle 
\sum_{n,m} v_0^2 A_n A_m |\Psi_n|^2 |\Psi_m|^2
\spinNucleusVector_{n,z} \cdot \spinNucleusVector_{m,z} \rangle\\
&= \frac{1}{\hbar^2}\sum_i\frac{I_i(I_i+1)}{3} \fractionIsotope A_i^2 \frac{v_0}{V_D}.
\end{split}\ee
The parameter $\kappa$ has the units of time and describes the restoring force that keeps the random walk bounded. Specifically, the expectation value of the distribution in Eq.~\eqref{eq:kernel} is
\be
\overline{\omega} \equiv \int \mathrm{d} \omega K(\omega_0, \omega, \delta t) = \omega_0 e^{-\delta t/\kappa}.
\ee
The Taylor expansion at short time $\delta t \ll \kappa$ gives
\be
\label{eq:firstMoment1}
\overline{\omega - \omega_0} = -\frac{\omega_0}{\kappa} \delta t.
\ee
Since the expected average change is proportional to the time interval, the proportionality factor corresponds to a polarization rate. Further, taking the limit $\delta t \to 0$, 
\be
\label{eq:firstMoment2}
\overline{\partial_t \omega} = -\frac{\omega}{\kappa},
\ee
the equation expresses a restoring force, since the detuning frequency is pulled back to the `equilibrium' $\omega=0$ in proportion to its instantaneous deviation from the equilibrium.

Similarly, the variance growth at short times,
\be
\label{eq:secondMoment}
\overline{(\omega-\omega_0)^2} = \sigma_{\delta t}^2 + \omega_0^2(e^{-2\delta t/\kappa}-1)^2 \approx \sigma_\Omega^2 \frac{2\delta t}{\kappa},
\ee
shows that the process is a diffusion with the diffusion constant
\be
\label{eq:DOmega}
D_\Omega = \frac{2\sigma_\Omega^2}{\kappa}.
\ee
With these results, we can convert the `intrinsic' thermal fluctuations of the Overhauser field, which are bounded and well described by a Gaussian distribution at long times, into the corresponding parameters of the above model. 
As we already noted, there are several experimental measurements of $D_\Omega$ in gated GaAs quantum dots. Based on the values given in Footnote \ref{fnt:diffusion}, we take $D_\Omega = (2\pi \times 7\, \mathrm{kHz})^2/1\,\mu$s as a representative value. Equation \eqref{eq:DOmega} then gives $\kappa = 0.2$ s (we evaluated $\sigma_\Omega^2$ from  Eq.~\eqref{eq:sigmaOv} using our parameters), predicting the Overhauser field equilibration scale in seconds.

In silicon, we are not aware of a direct experimental measurement of the diffusion constant of the quantum dot Overhauser field, $D_\Omega$. To arrive at an estimate, we use the result of Ref.~\cite{rojas-arias_spatial_nodate}, which, using the methods of Refs.~\cite{taylor_hyperne_2006,reilly_measurement_2008}, derives the time-correlation of the Overhauser field, converted to angular frequency units as
\be
\overline{\omega(t)\omega(t+\delta t)} = \frac{\sigma_\Omega^2}{\sqrt{2\pi}} \prod_{\alpha\in \{x,y,z\}} \left(1+2D l_\alpha^{-2} |\delta t| \right)^{-1/2},
\ee
where $D$ is the material bulk nuclear spin diffusion constant. Taylor-expanding for short times $\delta t \to 0$, we get
\be
D_\Omega = \frac{\sigma_\Omega^2}{\sqrt{2\pi}} D (l_x^{-2}+l_y^{-2}+l_z^{-2}),
\ee
and finally
\be
\label{eq:kappa2}
\kappa = \frac{\sqrt{8\pi}}{D} \left( l_x^{-2}+l_y^{-2}+l_z^{-2} \right)^{-1}.
\ee
Cross-checking the value for GaAs, using the bulk diffusion $D=7$ nm$^2$/s estimated theoretically \cite{deng_nuclear_2005,gong_dynamics_2011} gives $\kappa=0.001$ s, implying equilibration time of the order of a minute. The two values delimitate the range for the expected value of the intrinsic rate $\Gamma_0$, which we use in the caption of Fig.~\ref{fig:feedback-1-GaAs} as $0.01$-$0.1$ s$^{-1}$.
Assuming that in silicon the spin diffusion is slower, with $D=2$ nm$^2$/s measured in Ref.~\cite{hayashi_nuclear_2008}, we use an order of magnitude smaller rates, $\Gamma_0 \sim 0.001$-$0.01$ s$^{-1}$, as an orientation value\footnote{Using bulk diffusion constant for a quantum dot has its limits. Compared to a bulk crystal, the diffusion in a dot can be, on the one hand, slowed down by the potentially large inhomogeneities of the magnetic \cite{budakian_suppression_2004}, Knight \cite{madzik_controllable_2020}, or quadrupolar \cite{latta_hyperfine_2011, chekhovich_suppression_2015} fields, and, on the other, boosted by electron-mediated nuclear flip flops \cite{wust_role_2016}.} in Figs.~\ref{fig:feedback-1-Si} and \ref{fig:feedback-1-SiGe}.

\section{DNSP in hole qubits}

\label{app:hole}

We now apply the results of Table~\ref{tab:ManisotropicJ} to quantum dots with holes.  Aiming at rough estimates, we consider the heavy-hole (HH) limit with the hyperfine interaction of the Ising form\cite{fischer_spin_2008}\footnote{Going beyond this simplest limit might require numerics to evaluate the hole wave function. The hyperfine interaction tensor is non-generic, given by the details of the confinement potential  \cite{bosco_fully_2021}.}
\be
H_\mathrm{hf} = \sum_n A_{||,n} v_0 |\Psi(\mathbf{r}_n,z_n)|^2  \spinNucleus_z \spinElectron_z.
\label{eq:hyperfineHH}
\ee
This limit leads to a simple result for the matrix element $X$. Namely, with $J_{zz}$ being the only non-zero element of the hyperfine tensor,\footnote{
The light-hole--heavy-hole mixing results in further elements in the hyperfine tensor. Treating the mixing perturbatively, some of these elements arising in the first and second order were given in Refs.~\cite{xu_optically_2009,eble_holenuclear_2009}.}
Table~\ref{tab:ManisotropicJ} gives $M^{(0)}_{+-} = 0 = M^{(-2)}_{+-}$ and $M^{(-1)}_{+-} =  J_{zz} \cos\delta \sin \gamma /4$. It means, first of all, that in the Ising limit the `shaking' mechanism is not effective, only the `deflection' one contributes,
\begin{align}
\label{eq:XHH}
X_\mathrm{df}^\mathrm{HH} &= \frac{A_{||}}{4N_{\mathrm{tot}}} \sin \delta \cos \gamma,\\
X_\mathrm{sh}^\mathrm{HH} &= 0.
\end{align}
The next difference to electrons is that for holes, apart from the hyperfine tensor, the $g$-tensor is also strongly anisotropic and the confinement has strong effects on the hole spin. Important here, the deflection of the quantization axes of the hole spin and nuclear spins will be most often dominated by the quantum dot confinement rather than the small gradients of the magnetic field. The factor $\sin \delta$ is then not necessarily small for holes. Specifically, consider a quasi-two-dimensional quantum dot with the strong confinement along the $z$ axis, what fixes the heavy-hole spin along z. With the magnetic field in the plane, the deflection angle $\delta$ is $\pi/2$. The factor $\sin \delta$ in Eq.~\eqref{eq:XHH} is then 1, rather than $l \nabla_\perp B/B \approx 0.03$ in Eq.~\eqref{eq:XM}, boosting the rate by orders of magnitude. On the other hand, the nuclei are still polarized in the plane, so that the arising polarization is not visible as a change in the hole Larmor frequency. Additional NMR pulses would be needed to detect this polarization through the hole.

Concerning the material, recent progress with hole qubits \cite{hendrickx_four-qubit_2021} motivates us to consider silicon and germanium atoms for a possible DNSP. The hyperfine constants in the valence band for the two are similar,\footnote{Our Eq.~\eqref{eq:hyperfineHH} corresponds to Eq.~(17) of Ref.~\cite{fang_recent_2023} with $A_n=A_{||}$ and the perpendicular components $A_\perp$ neglected. The reference gives $A_{||}=-2.5$ neV for $^{29}$Si and $A_{||}=-1.1$ neV for $^{73}$Ge, with $A_\perp$ two orders of magnitude smaller.} while the 9/2 nuclear spin of $^{73}$Ge is much larger than the $^{29}$Si spin 1/2. These numbers would suggest germanium as more perspective to search for the DNSP signal. However, its low g-factor makes the nuclear Larmor frequency low, in turn the detection of the Hartmann-Hahn resonance challenging.

For a SiGe hole qubit, we summarize as follows. Since the hyperfine constant in the silicon valence band is similar to the one in the conduction band (see Table~1 in Ref.~\onlinecite{fang_recent_2023}), taking a heavy hole with spin along z and the magnetic field also along z, the resulting DNSP rate is similar to the one for an electron qubit in silicon. It was plotted in Fig.~\ref{fig:Si}, where only the deflection mechanism applies for a hole. The resulting rate is low. A somewhat larger rate arises in germanium atoms, because of the larger nuclear spin. However, the resonance frequency is low (below 1 MHz for $B=1$ T). Concerning a possible observation of the DNSP with holes, the most favorable scenario then looks to be searching for it in silicon atoms with a heavy-hole quantum dot in an in-plane magnetic field. 

\begin{figure}
\begin{center}
\includegraphics[width=0.99\linewidth]{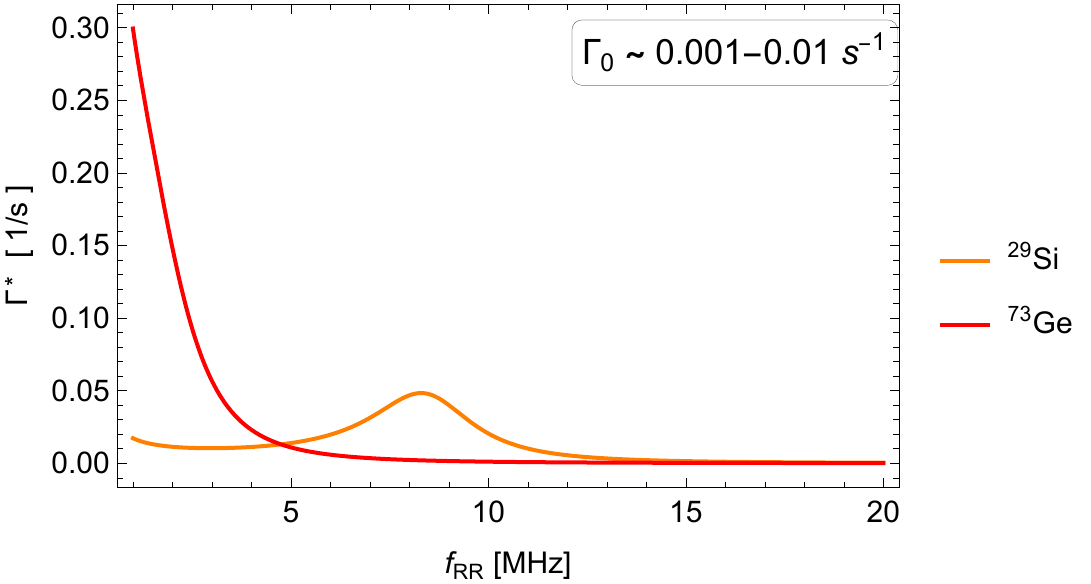}
\end{center}
\caption{\textbf{Stabilization by feedback in a hole qubit in SiGe}. The plot is analogous to Fig.~\ref{fig:feedback-1-GaAs} (see its caption for the description) and adopts the same parameters except for the atomic parameters of Si and Ge.}
\label{fig:feedback-1-SiGe}
\end{figure}

We illustrate this case with Fig.~\ref{fig:feedback-1-SiGe}, plotting the induced rate $\Gamma^*$. For $^{29}$Si atoms, the rate can be compared to the analogous plot for an electron quantum dot, Fig.~\ref{fig:feedback-1-Si}. The lack of suppression due to the factor $\sin \delta$ (with $\sin \delta=1$ for the hole) boosts the rate by three orders of magnitude. On the other hand, one order of magnitude is offset by a smaller size of the electron-qubit quantum dot, due to a larger effective mass. As a result, the difference between the curves for Si in Figs.~\ref{fig:feedback-1-Si} and \ref{fig:feedback-1-SiGe} is approximately two orders of magnitude.

As seen in Fig.~\ref{fig:feedback-1-SiGe}, the rate for $^{73}$Ge atoms can become larger than for $^{29}$Si. It is due to a larger nuclear spin of germanium. However, the resonance happens at a low frequency, so that the resonant peak is not discernible for Ge in Fig.~\ref{fig:feedback-1-SiGe}, overwhelmed by the rate behavior at zero frequency.\footnote{We note that Eq.~\eqref{eq:backforce} diverges in the limit $\frequencyRabiResonant \to 0$. This divergence is spurious, and stems from the assumption $\frequencyRabiResonant \gg \frequencyDetuning$, which we adopted in deriving Eqs.~\eqref{eq:backforce} and \eqref{eq:feedback}. At detunings larger than $\frequencyRabiResonant$, the assumption is violated, Eq.~\eqref{eq:feedback} does not hold, and the quantity $\Gamma^*$, though still well defined, is of little use. For this reason, we limit the lowest frequency on the horizontal axis in Figs.~\ref{fig:feedback-1-GaAs}, \ref{fig:feedback-1-Si}, and \ref{fig:feedback-1-SiGe} to an \textit{ad hoc} value of 1 MHz.}
A discernible peak appears for $B=2$ T or higher (not shown), but such fields might be too high for holes in SiGe to be useful as spin qubits.

\section{Collective enhancement?}
\label{app:collective}

Here we consider the possibility of an enhancement of the polarization rate due to collective effects. We have considered a single nuclear spin in all our derivations of the polarization rate. However, the coupling to a system with many spins can be coherently enhanced (known as `superradiance' \cite{kaluzny_observation_1983}), observed as an increase of the Rabi frequency by the factor \cite{fink_dressed_2009} $\sqrt{N}$ where $N$ is the number of spins. Therefore, one can wonder whether such effects, absent in our single-nuclear-spin calculations, could boost the polarization rate compared to our estimates. 

We find that this is not the case, and \emph{concerning the rates}, calculations within a single spin or many spin basis are \emph{exactly} equivalent. To show the essence of this somewhat surprising equivalence, we consider here only the dependence of the polarization rate on the matrix element of the spin-rasing operator $\spinNucleus_+$. The rate is proportional to a squared matrix element of it, see for example Eq.~\ref{eq:gamman}, with $\mathcal{I}_+ $ defined in Eq.~\ref{eq:YXI}. We calculate the squared matrix element of the total (`collective') spin-raising and lowering operators in a many-spin system
\be
J_\pm = \sum_{n=1}^{2N} I_{n,\pm},
\ee
with $n$ labeling the individual spins, the total number of which is $2N$. We consider nuclear spins 1/2 for simplicity in this section. 

We consider the basis composed of many-spin states with the quantum numbers being the total spin $j$ and its component along the z axis $m$,
\be
|j,m\rangle.
\ee
The admissible values are  $m \in \{-N, -N+1, \ldots, N-1, N\}$, and $j \in \{0, 1, \ldots, N\}$. 
The matrix elements of the total-spin operators are
\begin{subequations}
\begin{align}
\mathbf{J}^2 |j,m\rangle &= j(j+1) |j,m\rangle,\\
J_z |j,m\rangle &= m |j,m\rangle,\\
J_\pm |j,m\rangle &= \sqrt{j(j+1)-m(m\mp1)} |j,m\pm1\rangle.
\end{align}
\end{subequations}

One example of a collective basis state is the totally polarized one,
\be
|j=N,m=N\rangle = |\uparrow\rangle \otimes |\uparrow\rangle \otimes \cdots \otimes |\uparrow\rangle,
\ee
where there are $N_+ =2N$ spins up and $N_-=0$ spins down. Except for the fully polarized one, other collective states are coherent superpositions of several tensor-product states all having the same up and down individual spins, given by $N_\pm = N \pm m$. This property gives the recurrence relation for $C_{jm}$, the degeneracy of the basis state $|j,m\rangle$, as
\be
\left( \begin{tabular}{c} $N_+ + N_-$\\$N_+$ \end{tabular} \right) = \left( \begin{tabular}{c} $N_+ + N_-$\\$N_-$ \end{tabular} \right) = \sum_{j=m}^{N} C_{jm},
\label{eq:recurrence}
\ee
where the bracket denotes a binomial coefficient.
The recurrence is solved by
\be
C_{jm} = \left( \begin{tabular}{c} $2N$\\$j+N$\end{tabular} \right) -\left( \begin{tabular}{c} $2N$\\$j+N+1$\end{tabular} \right),
\label{eq:Cjm}
\ee
valid for any $j\geq0$ including $j=N$.

We now proceed to main calculation of this section, the average total squared matrix element (called in short `rate' in further) in the subspace with a fixed value of the quantum number $m$,
\be
\overline{R}_m^{\pm} \equiv \sum_j p_j R^{\pm}_{jm}.
\ee
The definition comprises the rate in the state $|j,m\rangle$,
\be
R^{\pm}_{jm} \equiv |\langle j,m| J_\pm |j,m\rangle|^2 = j(j+1)-m(m\ps{\pm}1),
\label{eq:Rjm}
\ee
and the probability that the system is in state $|j,m\rangle$,
\be
p_j = \frac{C_{jm}}{\sum_j C_{jm}}.
\ee
In this equation, the normalizing denominator is the number of states with a fixed value of $m$, which is given in Eq.~\eqref{eq:recurrence},
\be \begin{split}
C_m &= \sum_j C_{jm} =  \left( \begin{tabular}{c} $2N$\\$N+m$\end{tabular} \right) = \left( \begin{tabular}{c} $2N$\\$N-m$\end{tabular} \right) \\
&=\left( \begin{tabular}{c} $2N$\\$N+|m|$\end{tabular} \right),
\end{split}
\label{eq:sumCjm}
\ee
where the first two binomial coefficients evaluate to the same value and can thus be written as the third one. It remains to evaluate the following sum
\be
S=\sum_{j=|M|}^N C_{jm} R^{\pm}_{jm}.
\ee
Inserting the definitions from Eqs.~\eqref{eq:Cjm} and \eqref{eq:Rjm} we get
\begin{align}
S&=\left( \begin{tabular}{c} $2N$\\$N+j$\end{tabular} \right) R^{\pm}_{jm} \Big|_{j=|m|} \nonumber \\
& \qquad + \sum_{j=|m|+1}^N \left( \begin{tabular}{c} $2N$\\$N+j$\end{tabular} \right) \left[ R^{\pm}_{jm} - R^{\pm}_{j-1,m} \right] \nonumber \\
&=\left( \begin{tabular}{c} $2N$\\$N+|m|$\end{tabular} \right)(|m| \ps{\mp} m) + \sum_{j=|m|+1}^N \left( \begin{tabular}{c} $2N$\\$N+j$\end{tabular} \right) 2j.
\label{eq:sumS}
\end{align}
Using the identity [see Eq.~(5.18) in Ref.~\onlinecite{graham_concrete_1994}] 
\be
\sum_{k\leq m} \left( \begin{tabular}{c} $r$\\$k$\end{tabular} \right)\left( \frac{r}{2} - k \right) =  \frac{m+1}{2} \left( \begin{tabular}{c} $r$\\$m+1$\end{tabular} \right),
\ee
the sum in Eq.~\eqref{eq:sumS} can be brought to
\be
\left( N+|m|+1\right) \left( \begin{tabular}{c} $2N$\\$N+|m|+1$\end{tabular} \right),
\ee
which, on using Eq.~(5.6) of Ref.~\onlinecite{graham_concrete_1994} twice, equals
\be
\left( N-|m|\right) \left( \begin{tabular}{c} $2N$\\$N-|m|$\end{tabular} \right).
\ee
Collecting the expression in Eqs.~\eqref{eq:sumCjm} and \eqref{eq:sumS}, we get a simple result
\be
\overline{R}_m^{\pm} = N \ps{\mp} m.
\ee
This result is exact, following from identities for binomial coefficients. Importantly, the average rate within a fixed-$m$ subspace is \emph{linear} in $m$. Therefore, the average rate in the total (considering all $m$-subspaces) system, which might be spin-polarized, can be obtained by replacing the spin polarization on the right-hand side of the last equation with its statistical average $m\to\langle m\rangle$. The proof is as follows:
\be
\begin{split}
\overline{R}^{\pm} & \equiv \frac{\sum_{j,m} q_m C_{jm} R_{jm}^{\pm}}{\sum_{j,m} q_m C_{jm} } \\
& = \sum_{m}  \frac{q_m C_m} {\sum_{m} q_m C_m }  \overline{R}_{m}^{\pm}\\
&= \sum_m p_m (N \ps{\mp} m)\\
&=N \ps{\mp} \langle m \rangle,
\end{split}
\ee
where we denoted $p_m$ as the (spin-polarization defining) probabilities of occupation of the subspace $m$. Again, this result is exact and the only assumption it requires is that the probabilities of individual states, denoted $q_m$ in the above, depend only on $m$ (and not $j$ or other, exchange-symmetry related quantum numbers). 

Introducing the spin polarization $p_\mathrm{nuc} = \langle m \rangle / N$, we get the polarization rate evaluated in collective-state basis as
\be
\overline{R}^{\pm} = \frac{1}{2} \left( 1 \ps{\mp} p_\mathrm{nuc} \right) \times 2N.
\ee
On the other hand, using Eq.~\eqref{eq:Rjm} for $j=1/2$ gives the single-spin-increasing and decreasing rate as
\begin{align}
\Gamma_{+\equiv \downarrow \to \uparrow}^\mathrm{single} &= p^\mathrm{nuc}_\downarrow \times R^+_{(j=1/2,m=-1/2)} = p^\mathrm{nuc}_\downarrow \times 1, \\
\Gamma_{-\equiv\uparrow \to \downarrow}^\mathrm{single} &= p^\mathrm{nuc}_\uparrow \times R^-_{(j=1/2,m=1/2)} = p^\mathrm{nuc}_\uparrow \times 1.
\end{align}
Upon introducing single-spin polarization for nuclear spins (here being 1/2 spins) analogously to Eq.~\eqref{eq:electronPolarization}, we thus get
\be
\overline{R}^{\pm} = 2N \times \Gamma_\pm^\mathrm{single},
\ee
the rate for a collection of spins equals their number times the rate of a single-spin.

We thus conclude that there is no `collective enhancement' of the polarization rate. The single-spin calculation gives exactly the same as the many-spin calculation, even if the system is spin-polarized, including fully spin-polarized ($\langle m \rangle =N$). This conclusion seems paradoxical taking into account the superradiance effects of a polarized many-spin system. For example, the coupling (that is, the matrix element of the many-spin operator $J_+$) of a fully spin-polarized system is proportional to $\sqrt{2 N}$.\footnote{This enhancement has been demonstrated experimentally. For example, Ref.~\cite{fink_dressed_2009} has confirmed the increase of the Rabi frequency with the predicted factor $\sqrt{2 N}$ for $N=$1, 2, and 3.} The explanation of the paradox is as follows. When considering Rabi oscillations of a many-spin system due to a resonant excitation induced by $J_\pm$, the fully symmetric sum of individual spin operators, the frequency of these oscillations, given by the matrix elements of $J_\pm$, is proportional to $\sqrt{2N}$ and thus `enhanced'. In contrast to this, the frequency of Rabi oscillations of a single spin is not enhanced. The two calculations differ, and to describe the Rabi oscillations of (say, highly polarized) many-spin system, one should use the collective states. However, when calculating the polarization rate, the limit $t\to \infty$ [see, for example, Eq.~\eqref{eq:promotionToDeltaFunction}] effectively means that we evaluate the rate as the curvature of the Rabi-oscillation curve at $t = 0$. The curvature of that curve is equal to the oscillation amplitude times the oscillation frequency squared. In the many-spin calculation, taking the fully polarized system for illustration, the system oscillates between two states, $|j=N, m=N\rangle$ and $|j=N, m=N-1\rangle$ with the frequency enhanced by a factor $\sqrt{2N}$. The oscillation frequency is large and the amplitude is small, $\Delta m=1$. In a single-spin calculation, each spin oscillates with the same (non-enhanced) Rabi frequency, but the amplitude is 2N, since the system oscillates between $m=N$ and $m=-N$. The resulting rate, being the product of the amplitude and the frequency squared, is the same in both pictures,
\be
R = \underbrace{2N \times 1^2}_{\mathrm{single-spin\, calculation}} = \underbrace{1 \times \sqrt{2N}^2}_{\mathrm{collective\, spin\, states}},
\ee 
The two ways are equivalent, justifying our approach of evaluating the rate in a single-spin calculation.

\section{Quantitative treatment of the quadrupolar-interaction induced polarization}
\label{app:quadrupolar}

In Sec.~\ref{sec:extensions} we have encompassed the quadrupolar interaction effects qualitatively, including it in Eqs.~\eqref{eq:frequencyDispersion} and \eqref{eq:BperpChange} among the sources of deflection of the electron and nuclear spin quantization axes. Here we aim at a more quantitative description, motivated by the experimental results mentioned in Footnote \ref{fnt:non-collinear}, especially the resonances of the electron Rabi frequency with twice the nuclear Zeeman energy. They correspond to double nuclear spin flips and were observed in Refs.~\cite{bodey_optical_2019, gangloff_quantum_2019, gangloff_witnessing_2021}. Among others, we examine what the theory predicts for the ratio of double to single nuclear spin-flip rates.

With this goal, we expand the Hamiltonian in Eq.~\eqref{eq:ZeemanNucleus} by the following term
\be
\label{eq:quadrupolar1}
H_Q=e Q_n \frac{3}{2} \frac{V_{n,\alpha \beta}}{6I_n(I_n+1)}\left( I_{n,\alpha} I_{n,\beta}+I_{n,\beta} I_{n,\alpha}  \right).
\ee 
Here, $Q$ is the quadrupolar moment of the nucleus $n$, $V_{n, \alpha \beta}$ is the matrix of electric field gradients at the nucleus position, and $\alpha$ and $\beta$ are Cartesian coordinates indexes. The nuclear index $n$ could be traded for the isotope index $i$ on all quantities. We will omit it entirely from now on for notational simplicity. We also consider axially-symmetric potential, upon which the interaction can be written as due to a tensor $V$ with a single diagonal component [Eq.~(10.60) in Ref.~\cite{slichter_principles_1996}],
\be
\label{eq:quadrupolar2}
H_Q=\hbar \angularFrequencyQuadrupolar (\spinNucleusVector \cdot \axisQuadrupolarVector)^2.
\ee 
We parametrize it by an energy scale $\hbar \omega_Q$ and a unit vector $\axisQuadrupolarVector$. The scale sets the quadrupolar splittings, being of the order of 10 kHz in GaAs. Anticipating its meaning, we denote the angle of the unit vector $\axisQuadrupolarVector$ with the magnetic field direction (the z axis) as $\delta$, the deflection angle.

We now derive the nuclear spin polarization rate in a way alternative to the main text. We start with Eq.~\eqref{eq:electronNuclearPairH} with the quadrupolar term added,
\be
\begin{split}
H = &
-\hbar\angularFrequencyNucleus \spinNucleusVector \cdot \axisEnergyNucleus
-\hbar\angularFrequencyElectron \spinElectronVector \cdot \axisEnergyElectron
+\hbar \angularFrequencyQuadrupolar (\spinNucleusVector \cdot \axisQuadrupolarVector)^2\\
&- 2\hbar\angularFrequencyRabiResonant   \spinElectronVector \cdot \axisBFieldRF\cos(\angularFrequencyRF t - \phi_\rf)
+J(t) \delta \spinNucleusVector \cdot \spinElectronVector.
\end{split}
\label{eq:electronNuclearPairH-alt}
\ee
and transform only the electron spin operator into the rotating frame
\be
U(t) = \exp ( - i \spinElectronVector \cdot \axisEnergyElectron \angularFrequencyRF t).
\label{eq:U-alt}
\ee
Adopting again the rotating-wave approximation in the fourth term of the Hamiltonian we get
\be
\label{eq:Htilde2-alt}
\begin{split}
H^\prime =&  
-\hbar\angularFrequencyNucleus \spinNucleusVector \cdot \axisEnergyNucleus
+\hbar \angularFrequencyQuadrupolar (\spinNucleusVector \cdot \axisQuadrupolarVector)^2
-\hbar\angularFrequencyRabi \spinElectronVector \cdot \axisEnergyElectronO\\
&+J(t) \delta \spinNucleusVector \cdot R^{-1}_{\axisEnergyElectron,\angularFrequencyRF t}\cdot \spinElectronVector.
\end{split}
\ee
Since that effect was already analyzed in the main text, we neglect the electron wave-function oscillations in space, putting $J_n(t) \approx J(0)\equiv J$. As then the Hamiltonian does not contain a term that can compensate for the fast frequency $\angularFrequencyRF$, we may drop the terms oscillating with this frequency in the last term and get a time-independent Hamiltonian
 \be
\label{eq:Htilde3-alt}
\begin{split}
H^\prime =&  
-\hbar\angularFrequencyNucleus \spinNucleusVector \cdot \axisEnergyNucleus
+\hbar \angularFrequencyQuadrupolar (\spinNucleusVector \cdot \axisQuadrupolarVector)^2
-\hbar\angularFrequencyRabi \spinElectronVector \cdot \axisEnergyElectronO\\
&+J \,  (\delta \spinNucleusVector \cdot \axisEnergyElectron) (\spinElectronVector \cdot \axisEnergyElectron).
\end{split}
\ee
Here one can see the relation to the two effects analyzed in the main text: Had we retained the oscillating part of $J(t)$, it would compensate the oscillating phase of transverse components in the last term, such as $\delta \spinNucleus_+ \spinElectron_-$, and thus provide a channel for nuclear polarization. Alternatively, polarization can arise if the Zeeman terms are not collinear $\axisEnergyNucleus \neq \axisEnergyElectron$. Without either of the two sources, the Hamiltonian without the quadrupolar term can not lead to nuclear spin polarization (as we concluded Sec.~\ref{sec:DNSPorigin}), since it contains only a diagonal operator $\spinNucleus_z$. Examining here the effects of the quadrupolar term, we neglect both of the polarization sources already analyzed and set also $\axisEnergyNucleus = \axisEnergyElectron \equiv \mathbf{z}$, getting
 \be
\label{eq:Htilde4-alt}
H^\prime =
-\hbar\angularFrequencyNucleus \spinNucleus_z
-\hbar\angularFrequencyRabi \spinElectronVector \cdot \axisEnergyElectronO
+\hbar \angularFrequencyQuadrupolar (\spinNucleusVector \cdot \axisQuadrupolarVector)^2
+J \, \delta \spinNucleus_z\, \spinElectron_z.
\ee

Since the Hamiltonian is time-independent, we evaluate the polarization rate using the Fermi's Golden Rule (FGR). During the derivation, we will reuse some of the results of the main text. We first assume that the quadrupolar term is smaller than the nuclear Zeeman energy, so that we can treat it (together with the hyperfine term) perturbatively. We thus define the unperturbed system with the first two terms of Eq.~\eqref{eq:Htilde4-alt}, resulting in the basis states $|sj\rangle$ given in the main text in Eq.~\eqref{eq:basisStates} with $s$ representing the z component of the electron spin and $j$ the nuclear spin. 

We first consider a single-spin-flip resonance, meaning $\hbar\angularFrequencyRabi \approx 1 \times \hbar\angularFrequencyNucleus$. The FGR then gives for the nuclear-spin-increasing transition rate
\be
\Gamma_+ = \frac{2\pi}{\hbar}  \left|  \langle -,j+1 | H_\mathrm{eff} |+, j \rangle \right|^2 p_{j,+} G_\Sigma\left(\angularFrequencyRabi - \omega_n \right).
\ee
Here, we have identified the density of states in the FGR with Eq.~\eqref{eq:gaussianDensity}, $p_{+,j}$ is the occupation probability of the initial state $|+,j\rangle$, and the matrix element between the quasi-degenerate states $|+,j\rangle$ and $|-,j+1\rangle$ should be evaluated by the Hamiltonian $H_\mathrm{eff}$ describing the quasi-degenerate subspace. We get this effective Hamiltonian in the second-order of the degenerate perturbation theory (\cite{bir_symmetry_1974}; see Footnote 1 in Ref.~\cite{stano_orbital_2019}),
\be
\langle m| H_\mathrm{eff} |m^\prime\rangle = \sum_{l \neq m, m^\prime} \langle m| H^\prime_1 |l\rangle\langle l| H^\prime_1 |m^\prime \rangle\left( \frac{1}{E_{ml}}+\frac{1}{E_{m^\prime l}} \right).
\ee
where $m$ and $m^\prime$ are the two quasi-degenerate states, $l$ are other basis states, and we have denoted the third and fourth term of Eq.~\eqref{eq:Htilde4-alt} as $H^\prime_1 = H_Q+H_J$, the perturbation part of the Hamiltonian. Since the two terms of $H_z^\prime$ have simple matrix elements ($H_Q$ being identity in the electron sector and $H_J$ being diagonal in the nuclear sector), one can simplify the effective Hamiltonian to
\be
H_\mathrm{eff} = \frac{1}{\hbar\angularFrequencyNucleus} [H_Q,H_J].
\ee
The simplification shows that the expression for the rate contains the following matrix element of nuclear spin operators
\be
\label{eq:IIII1}
\tilde{\alpha}_I = \frac{1}{I}\mathrm{Tr}\left( \{ \spinNucleus_+ ,\spinNucleus_z\} \rho \{ \spinNucleus_- ,\spinNucleus_z\} \right),
\ee
where $\{.,.\}$ is the anticommutator and $\rho$ is the system density matrix. For unpolarized nuclear spins, a limit that we restrict to, the matrix element can be calculated exactly giving
\be
\label{eq:alphaTilde}
\tilde{\alpha}_I = \frac{2}{15} \left( 4I^3+8 I^2 +I -3\right).
\ee
Introducting $\axisQuadrupolar_\pm = \axisQuadrupolar_x \pm i \axisQuadrupolar_y$ as the complex components of the unit vector $\axisQuadrupolarVector$, the transition rate takes the form
\be
\label{eq:GammaPlus}
\Gamma_+ = \frac{2\pi}{\hbar}  \left|  \frac{\hbar\angularFrequencyQuadrupolar J}{\hbar\angularFrequencyNucleus} \langle -|\spinElectron_z |+\rangle \axisQuadrupolar_- \axisQuadrupolar_z\right|^2 I \tilde{\alpha}_I p_{+} G_\Sigma\left(\angularFrequencyRabi - \omega_n \right),
\ee
where $p_+ \equiv p^e_\uparrow$ introduced in Eq.~\eqref{eq:electronPolarization}. The rate for the opposite (nuclear-spin decreasing) transition takes the same form upon swapping all `+' and `-' indexes, resulting in the only consequential change being $p_{+} \to p_{-}$. For the polarization rate $\partial_t p_i = (\Gamma_+ - \Gamma_-)/I$ we get
\begin{subequations}
\label{eq:rate1f}
\begin{align}
\partial_t p_i^\mathrm{1f} &= \frac{\pi}{\hbar} X_{Q,\mathrm{1f}}^2 \tilde{\alpha}_I p_e G_\Sigma\left(\angularFrequencyRabi - \omega_i \right),\\
X_{Q,\mathrm{1f}} &= \frac{A_i}{4 N_\mathrm{tot}} \frac{\hbar\angularFrequencyQuadrupolar }{\hbar\angularFrequencyNucleus} \frac{\sin 2\delta}{2\sqrt{2}} \cos\gamma,
\end{align}
\end{subequations}
where we introduced the matrix element $X_Q$ as the effective `deflection' matrix element induced by the quadrupolar interation. It should be compared to Eq.~\eqref{eq:XM} and we find an explicit prescription for the quadrupolar-induced effective axes deflection, anticipated in Eq.~\eqref{eq:BperpChange}, as
\be
\label{eq:BperpChangeQ}
\frac{l \nabla_\perp B}{B} \to \frac{\hbar\angularFrequencyQuadrupolar }{\hbar\angularFrequencyNucleus} \frac{\sin 2\delta}{2\sqrt{2}}.
\ee
This is the first main result of this section.

We next consider the double-spin-flip transitions, assuming $\hbar\angularFrequencyRabi \approx 2 \times \hbar\angularFrequencyNucleus$. Since the calculation is analogous, we point out only the differences. The nuclear-spin operator that induces the transitions in Eq.~\eqref{eq:IIII1} is changed to
\be
\label{eq:IIII2}
\frac{1}{I}\mathrm{Tr}\left( \{ \spinNucleus_+ ,\spinNucleus_+\} \rho \{ \spinNucleus_- ,\spinNucleus_-\} \right).
\ee
Interestingly, its average over an unpolarized ensemble is exactly four times the one given in Eq.~\eqref{eq:alphaTilde}. The denominator of the transition rate in Eq.~\eqref{eq:GammaPlus} is now $2\hbar \angularFrequencyNucleus$ instead of $\hbar \angularFrequencyNucleus$ and the factor $\axisQuadrupolar_-\axisQuadrupolar_z$ changes to $\axisQuadrupolar_-^2$. These changes result in an expression basically identical to Eq.~\eqref{eq:rate1f} up to a change in the dependency on the quadrupolar deflection angle $\delta$: 
\begin{subequations}
\label{eq:rate2f}
\begin{align}
\partial_t p_i^\mathrm{2f} &= \frac{\pi}{\hbar} X_{Q,\mathrm{2f}}^2 \tilde{\alpha}_I p_e G_\Sigma\left(\angularFrequencyRabi - 2\omega_i \right),\\
X_{Q,\mathrm{2f}} &= \frac{A_i}{4 N_\mathrm{tot}} \frac{\hbar\angularFrequencyQuadrupolar }{2\hbar\angularFrequencyNucleus} \frac{\sin^2 \delta}{\sqrt{2}} \cos\gamma.
\end{align}
\end{subequations}
We then arrive at the second main result here, the ratio of the single-flip to double-flip polarization rates (at their respective resonances, assuming the density of states are the same):
\be
\frac
{X^2_{Q,\mathrm{1f}}}
{X^2_{Q,\mathrm{2f}}} = 4 \coth^2\delta.
\ee
Interestingly, the double-flip process is not necessarily weaker than a single-flip one. The ratio of the two rates can reach any value, depending on the angle $\delta$, the orientation of the electric field gradient with respect to the magnetic field.

In the preceding calculation, we have considered the limit where the nuclear quadrupolar interaction is smaller than the nuclear Zeeman energy. We finish with a short comment on the opposite limit. The above procedure could be performed similarly, swapping the roles of the quadrupolar and Zeeman term in defining the basis and providing the perturbation allowing transitions. However, if the quadrupolar interaction dominates, only a `single-flip' resonance occurs, when the electron Rabi frequency matches the nuclear Zeeman energy, the energy difference between the spin $\pm 1/2$ nuclear states. Other energy resonances are given the quadrupolar energy, rather than the Zeeman energy. Since in the experiments, clear single as well as double spin-flip resonances were observed in Refs.~\cite{bodey_optical_2019, gangloff_quantum_2019, gangloff_witnessing_2021}, we do not pursue the calculation in this limit.

\section{Notation: list of defined quantities}
\label{app:tables}

We collect the definitions of the main symbols used throughout the text for easier reference and lookup. We group them in the three parts of Table~\ref{tab:tables}.

\newcounter{subtable}

\refstepcounter{table}\addtocounter{table}{-1}\label{tab:tables}
\renewcommand\thetable{\Roman{table}-\arabic{subtable}}
\setcounter{subtable}{1}

\begin{widetext}

\begin{table*}[b]
\begin{tabular}{@{}c@{\quad}c@{\quad}c@{\quad}c@{\quad}c@{}}
\toprule
\multicolumn{5}{c}{Zeeman energies, Larmor frequencies, and related quantities}\\
\midrule
electron & nucleus & description & sign & definition\\
\midrule
$q_e$ & $q_n$ & sign of electric charge & signed & $q_e=-1$, $q_n=+1$\\
$\gFactorElectron$ & $\gFactorNucleus $ & $g$ factor & signed & material parameter\\
$\mu_B$ & $\mu_N$ & magneton & positive & nature constant\\
$\spinElectron$ & $\spinNucleus $ & spin magnitude & positive & $\spinElectron=1/2$, $\spinNucleus = \mathrm{(half)integer}$  \\
$\spinElectronVector$ & $\spinNucleusVector $ & spin operator& vector &
\multicolumn{1}{c}{$\spinNucleusVector^2 = \spinNucleus(\spinNucleus+1)$, $\spinElectronVector^2 = \spinElectron(\spinElectron+1)$}\\
$\bFieldElectron$ & $\bFieldNucleus $ & magnetic field & vector & tunable parameter\\
$\axisEnergyElectron$ & $\axisEnergyNucleus $ & spin ground-state direction& unit vector &\multicolumn{1}{c}{$\axisEnergy = \mathrm{sgn}(q \gFactor)\bField / B$}\\
$\hbar \angularFrequencyElectron$ & $\hbar \angularFrequencyNucleus$ & Zeeman energy & positive & $\hbar \omega_n = | g_n \mu_N B |$; for $\hbar \angularFrequencyElectron$, see Eq.~\eqref{eq:frequencyElectron}\\
$\frequencyLarmorElectron$ &  $ \frequencyLarmorNucleus$ & Larmor frequency & positive & $f = \omega /2\pi $\\ 
\multicolumn{2}{c}{$\delta, \delta^\prime$} & angles relating $\axisEnergyElectron$ and $\axisEnergyNucleus$ & signed & see Fig.~\ref{fig:angles}\\
\bottomrule
\end{tabular}
\caption{\label{tab:Zeeman}
Quantities related to the Larmor precession speed and the spin orientation.}
\end{table*}
\addtocounter{table}{-1}
\addtocounter{subtable}{+1}

\begin{table*}
\begin{tabular}{@{}c@{\quad}c@{\quad}c@{\quad}c@{}}
\toprule
\multicolumn{4}{c}{Atomic and nuclear quantities}\\
\midrule
quantity & description & sign & definition\\
\midrule
$a_0$ & lattice constant & positive & material parameter\\
$v_0$ & volume per atom & positive &$v_0 = a_0^3 / 8$\\
$V_D$ & quantum dot volume & positive & $V_D = 1/\int \mathrm{d}V |\Psi|^4$\\
$N_{\mathrm{tot}}$ & number of atoms in the dot & positive &$N_{\mathrm{tot}} = V_D / v_0$\\
$\fractionIsotope$ & isotopic fraction& positive & material parameter\\
$N_i$ & number of atoms for $i$-th isotope & positive & $N_i = N_{\mathrm{tot}} \fractionIsotope$\\
$A_i$ & hyperfine constant for $i$-th isotope & signed & $\frac{4}{3 v_0} \mu_N \mu_B g_i |\eta_i|^2 $ \\
$J_n$ & hyperfine coupling strength for nucleus $n$ & signed & $J_n = A_n v_0 |\Psi_n|^2$\\
$J$ & average hyperfine coupling strength & signed & $J \equiv \langle J_n \rangle = A_i / N_\mathrm{tot}$\\
$\hbar \angularFrequencyQuadrupolar$ & quadrupolar interaction strength & signed & $3eQ ( \axisQuadrupolarVector\cdot \mathbf{V} \cdot \axisQuadrupolarVector)/6I(I+1)$\\
\bottomrule
\end{tabular}
\caption{\label{tab:nuclear}
Quantities related to nuclear spins. The relation $v_0=a_0^3/8$ applies for zinc-blende and diamond crystals. For our wave-function choice in Eq.~\eqref{eq:wavefunction}, the quantum dot volume is $V_D=2\pi l^2 l_z$, see Appendix~\ref{app:nuclearDensity}. The amplitude of the electron Bloch wave function at the atomic nucleus is $\eta_i$. For the electric-field gradient tensor is $\mathbf{V}$ the quadrupolar interaction magnitude $\hbar \angularFrequencyQuadrupolar$, see Eqs.~\eqref{eq:quadrupolar1}-\eqref{eq:quadrupolar2}.}
\end{table*}
\addtocounter{table}{-1}
\addtocounter{subtable}{+1}

\begin{table*}
\begin{tabular}{@{}c@{\quad}c@{\quad}c@{\quad}c@{\quad}c@{\quad}c@{}}
\toprule
\multicolumn{6}{c}{EDSR related quantities}\\
\midrule
\multicolumn{3}{c}{quantity}& description & sign & definition\\
\midrule
\multicolumn{3}{c}{$\dotShift$} & dot shift in space & in-plane vector & $\dotShift = e \mathbf{E}_0 l^2 / (\hbar^2 /m l^2)$\\
\multicolumn{3}{c}{$\axisBFieldRF$} &direction of the EDSR field & unit vector & see Eq.~\eqref{eq:EDSRTerm}\\
\multicolumn{3}{c}{$\angleDetuning$} & detuning angle & signed & $\sin \gamma = - \frequencyDetuning / \frequencyRabi$ \\
\multicolumn{3}{c}{$\angleRF$} & phase shift of the EDSR signal& signed & $\mathbf{E}(t) = \mathbf{E}_0 \cos(\angularFrequencyRF t -\angleRF)$\\
\midrule
 freq. &&ang.~freq.&\multicolumn{3}{l}{$(\omega = 2 \pi f)$}\\
 \midrule
$\frequencyRF$ && $\angularFrequencyRF$ & frequency of the EDSR drive & positive & tunable parameter\\
$\frequencyDetuning$ && $\angularFrequencyDetuning$&detuning frequency & signed & $\frequencyDetuning = \frequencyRF - \frequencyLarmorElectron$\\
$\frequencyRabiResonant$ &&$\angularFrequencyRabiResonant$ & Rabi frequency at resonance & positive & see Eq.~\eqref{eq:EDSRTerm}\\
$\frequencyRabi$ &&$\angularFrequencyRabi$ & Rabi frequency & positive & $\frequencyRabi = \sqrt{(\frequencyRabiResonant)^2 + \frequencyDetuning^2}$\\
\bottomrule
\end{tabular}
\caption{\label{tab:driving}
Quantities related to the EDSR drive.
}
\end{table*}
\addtocounter{table}{-1}
\addtocounter{subtable}{+1}

\end{widetext}

\clearpage

\bibliographystyle{naturemag}
\bibliography{bib/2020-Stano-DNSP}

\end{document}